# Hole-Doped Cuprate High Temperature Superconductors

C. W. Chu, L. Z. Deng and B. Lv

Department of Physics and Texas Center for Superconductivity
University of Houston

## Abstract

Hole-doped cuprate high temperature superconductors have ushered in the modern era of high temperature superconductivity (HTS) and have continued to be at center stage in the field. Extensive studies have been made, many compounds discovered, voluminous data compiled, numerous models proposed, many review articles written, and various prototype devices made and tested with better performance than their nonsuperconducting counterparts. The field is indeed vast. We have therefore decided to focus on the major cuprate materials systems that have laid the foundation of HTS science and technology and present several simple scaling laws that show the systematic and universal simplicity amid the complexity of these material systems, while referring readers interested in the HTS physics and devices to the review articles. Developments in the field are mostly presented in chronological order, sometimes with anecdotes, in an attempt to share some of the moments of excitement and despair in the history of HTS with readers, especially the younger ones.

## 1. Introduction

Hole-doped cuprate superconductors have played an indispensable role in the exciting development of high temperature superconductivity (HTS) science and technology over the last 28 years. They ushered in the era of cuprate high temperature superconductivity and helped create a subfield of physics, namely, “high temperature superconductivity” as we know it today. It all began with the observation of superconductivity up to 35 K in the Ba-doped $La_2CuO_4$ ternary compound by Alex Mueller and George Bednorz of IBM Zurich Laboratory in 1986 [1], followed immediately by the discovery of superconductivity at 93 K in the new self-doped $YBa_2Cu_3O_7$ quaternary compound by C. W. Chu of University of Houston, M. K. Wu of University of Alabama at Huntsville, and their team members in 1987 [2]. The explosion of activities in the ensuing 28 years has led to the discovery of more than two hundred cuprate superconductors that belong to 7 families with $T_c$ up to 134 K in $HgBa_2Ca_2Cu_3O_9$ at ambient pressure [3] and 164 K under 30 GPa [4], the revelation of the anomalous behavior of these compounds in their normal state [5] and unusual flux dynamics in their superconducting state [6], the proposition of various models to account for the observations in both the normal and superconducting states [7], and the construction and demonstration of the prototypes of HTS devices with superior performance to that of their nonsuperconducting counterparts [8]. In this review we shall focus mainly on the results achieved to date in hole-doped cuprate materials that have laid the foundation of HTS science and technology, as well as presenting several of the scaling laws that reveal the systematic and universal simplicity inherent in the complexity of these vast material systems, while leaving the details of the HTS physics and devices to the voluminous review articles [9] for interested readers. Achievements in the field are generally presented in chronological order and sometimes with anecdotes to share moments of both excitement and despair in the development of HTS with readers, especially the younger ones. The future of HTS will also be discussed at the end of this review.

## 2. Materials

More than 200 cuprate superconductors have been found through doping or self-doping their parent compounds, Mott-insulators. The evolution of superconductivity with doping from an antiferromagnetic insulator to a spin glass to a superconductor and finally to a normal metal is represented by the generic phase diagram shown in Fig. 1. It may not be overstated that the generic phase diagram encompasses the gist of physics of HTS. All of the cuprate high temperature superconductors (HTSs) can be represented by the generic formula $A_mE_2R_{n-1}Cu_nO_{2n+m+2}$ designated as Am2(n-1)n as shown in Fig. 2 [10]. It shows the main architecture of the layer cuprate HTSs that reflects some of the intricate physics of HTS.

Almost all cuprate HTSs can be grouped into 7 families with hole- or electron-doping as shown in Tables I and II, which include properties and references. It is one thing to make a layer cuprate sample superconducting once discovered; it is quite another to make it with high quality for scientific studies or with enhanced performance in desired forms for applications. Due to the space limitations of this review, we choose to refer readers to the review articles for cuprate synthesis and processing [11].

The evolution of record-$T_c$ at the time for all superconductors is displayed in Fig. 3. It is evident that superconductors that exhibit a $T_c$ above 77 K are hole-doped cuprates. The current record-$T_c$ of 134 K [3] and 164 K [4] occur in $HgBa_2Ca_2Cu_3O_9$ at ambient pressure and 30 GPa, respectively. Cuprate HTSs display a layer structure and can be represented by a generic formula $A_mE_2R_{n-1}Cu_nO_{2n+m+2}$, as shown in Fig. 2, where A = Bi, Tl, Pb, Hg, Cu, or rare-earth; E = Ca, Sr, Ba, or vacant; R = Ca , Y, or rare-earth; m = 0, 1, or 2; and n = 1, 2, ... The generic formula $A_mE_2R_{n-1}Cu_nO_{2n+m+2}$ can be rewritten as $[(EO)(AO)_m(EO)] + \{(CuO_2)[R(CuO_2)]_{n-1}\}$, which consists of two substructures: the active block of $\{(CuO_2)[RCuO_2]_{n-1}\}$ and the charge reservoir block of $[(EO)(AO)_m(EO)]$. The space group for compounds with m = 2 or 0 is I4/mmm but changes to P4/nmm when m = 1. The active block comprises n-square-planar-($CuO_2$)-layers per formula interleaved by (n-1)-R-layers and the charge reservoir block contains m-(AO)-layers bracketed by 2-(EO)-layers. Superconducting current flows mainly in the ($CuO_2$)-layers in the active block and doping takes place in the charge reservoir block, which transfers charges without introducing defects into the ($CuO_2$)-layers (similar to modulation-doping in semiconducting superlattices). Almost all cuprate HTSs can simply be designated as Am2(n-1)n or just 0(n-1)n-E when the two (AO)-layers are absent. The rest can be considered derivatives from them by replacing the AO- or R-layers by more complex layers, such as the oxyfluoride layers [12]. For instance, the first cuprate superconductor Ba-doped $La_2CuO_4$, which is commonly known as La214 or LCO, can be written as $[(LaO)_2](CuO_2)$ and represented by La0201 with $(LaO)_2$ as the charge reservoir and $(CuO_2)$ as the active block without the (EO)-double layers; and the first liquid nitrogen superconductor $YBa_2Cu_3O_7$, which is commonly known as Y123 or YBCO, can be written as $[(BaO)(CuO)(BaO)][(CuO_2)Y(CuO_2)]$ and represented by Cu1212 with $[(BaO)(CuO)(BaO)]$ as the charge reservoir and $[(CuO_2)Y(CuO_2)]$ as the active block.

The basic considerations for the formation of the compounds are charge neutrality and ionic size matching. For doping, one may replace the anion partially with another anion of different valence or vary the oxygen content. All major families of the cuprate HTSs are listed in Table I according to the chronological order of their discoveries. They will be briefly discussed accordingly.

### 2.1. $R_2CuO_4$ (R214 or R0201) with R = La, Pr, Nd, Sm, Eu

$R_2CuO_4$ is known to crystallize in the $K_2NiF_4$ structure with a space group I4/mmm or P4/nmm with three different phases of T, T', and T*, depending on the specific R and dopant [13]. The appearance of the three phases can be understood in terms of the Goldschmidt rules of the formation of perovskite-like crystal where the ionic radii of the elements and bond-length play a critical role to reduce the mismatch between the square-planar $CuO_2$-layers and the RO-layers [14]. As a result, the T-phase consists of the corner-sharing $CuO_6$ octahedra and the T'-phase consists of only the square-planar $CuO_2$-layers, while the T*-phase consists of the combination of the two as shown in Figs. 4a-c.

#### 2.1.1. The T-phase of $La_2CuO_4$ (or La214) doped with the alkaline-earth (AE)

*The Ba-doped T-phase of $La_2CuO_4$ (La214) with a $T_c$ up to 35 K discovered by George Bednorz and Alex Mueller of IBM Zurich Laboratory in 1986 was the first cuprate HTS that ushered in the era of cuprate high temperature superconductivity. For their seminal work, both were awarded the 1987 Nobel Prize in Physics.*

Before the discovery of a $T_c$ up to 35 K by George Bednorz and Alex Mueller, physicists in general followed two paths to search for superconductors of higher $T_c$: Matthias's enlightened empirical approach [15] and the BCS rational approach [16]. Matthias's rule correlates well the $T_c$ with the valence electron to atom ratio (e/a) and suggests a maximum $T_c$ of an intermetallic system at e/a ~ 4.75 or 6.4. Indeed, the then high $T_c$ superconductors, such as $V_3Si$ ($T_c$ = 17.5 K), $Nb_3Sn$ ($T_c$ = 18 K), and $Nb_3Ge$ ($T_c$ = 23.2 K), were all intermetallics with an e/a = 4.75. The highest $T_c$ of 23.2 K was achieved by John Gavella et al. [17] and Lou Testardi et al. [18] in 1973, respectively, representing only a $T_c$-increase of less than 20 K more than six decades after the discovery of superconductivity in Hg ($T_c$ = 4 K) by Heike Kamerlingh Onnes in 1911 [19]. Disappointed by the slow progression of $T_c$ in the conventional intermetallics, Mueller and Bednorz (M&B) [20] decided to deviate from the conventional intermetallic path and dive into oxides. According to the BCS theory, $T_c = 1.14\Theta_D\exp[-1/N(E_F)V]$, where $\Theta_D$ is the Debye temperature, $N(E_F)$ the electron density of states at the Fermi surface $E_F$, and V the attractive electron-phonon interaction, so M&B reasoned that a higher $T_c$ could be realized in a compound if V and $N(E_F)$ could be further enhanced. Based on their previous studies on perovskite oxides, they concluded that an enhanced V may be achieved in oxides due to the polaron formation as well as the mixed valence states that exist in the oxide superconductors of $(Ba,Pb)BiO_3$ [21] and $Li_{1+x}Ti_{2-x}O_4$ [22] with a relatively high $T_c$ at the time. Encouraged by the suggestion of a possible bipolaronic to superconducting transition [23], they started the search for superconductivity in 1983 in $LaNiO_3$, $LaAlO_3$, $LaCuO_3$, and their mixtures. They finally detected superconductivity up to 35 K in their multiphase samples with a nominal composition of $Ba_xLa_{5-x}Cu_5O_{5(3-y)}$ [(La,Ba)-Cu-O] by varying the doping x and synthesis temperature, as shown in Fig. 5. The resistive results were published in the September issue of *Zeitschrift für Physik* in 1986 [1]. The appearance of superconductivity at such a high temperature in an oxide surprised many experts. This was against the conventional wisdom that most oxides are usually insulating and not even metallic, let alone superconducting, especially with a high $T_c$. In addition, with the many previous false alarms of high $T_c$, the absence of magnetic data in the report, and the dwindling worldwide effort in the search for higher $T_c$, coupled with their modest title "Possible High $T_c$ in the La-Ba-Cu-O System," the report did not initially attract the attention it deserved. However, its significance was immediately recognized, respectively, by three groups in Tokyo, Houston, and Beijing who had been continuously searching for higher $T_c$ in oxides throughout the dog days of superconductivity. It was the announcement of reproduction of the Zurich observation by Chu of Houston at the 1986 Materials Research Society Fall Meeting on December 4 in Boston at the end of his presentation on another oxide superconductor, $Ba(Pb,Bi)O_3$, that prompted Koichi Kitazawa of Tokyo to reveal that Tanaka's group had also observed the same superconducting results not just resistively, but magnetically as well. He also told Chu after the session that they had identified the superconducting phase to be the Ba-doped $La_2CuO_4$ [(La,Ba)214]. (It should be noted that in fact, Chu had already mentioned his superconductivity results informally at the banquet talk of a meeting organized by Gary Vezzoli at the Army Picatinny Arsenal in New Jersey two days earlier on December 2.) The genie was finally out of the bottle and set the world of superconductivity on fire. Superconductivity was soon discovered up to the 50s K in the (La,Ba)214 under pressures [24] and up to the high 30s K in the Sr-doped La214 [(La,Sr)214 [25] at ambient. This triggered the search for the liquid nitrogen superconductivity to be discussed later in Section 2.1.3.

$La_2CuO_4$ crystallizes in the $K_2NiF_4$ tetragonal structure with a space group of I4/mmm, known as the T-phase with apical oxygen atoms associated with the $CuO_6$-octahedra that form the square planar $CuO_2$-layers, as shown in Fig. 4a. The $CuO_2$-layers and the LaO-layers stack on one another very much like in the cubic perovskite $ABO_3$, except that one slab of $La_2CuO_4$ is shifted by (a/2)(1,1,0) with respect to the one below, where a is the lattice parameter along the layer. $La_2CuO_4$ is an antiferromagnetic insulator and the parent compound of the superconducting 214 family. Superconductivity is induced only upon the introduction of charge carriers through hole-doping by partial replacement of the trivalent La by the divalent alkaline earth (AE) Ba, Sr, or Ca or by increasing the O-content of the T-La214.

The first high pressure experiment on our multiphase (La,Ba)-Cu-O sample revealed a $T_c$ up to 52.5 K under an unprecedentedly large positive pressure effect of ~ 0.9 x $10^{-2}$ $KGPa^{-1}$. The new record-$T_c$ exceeded unambiguously the 30s K-limit suggested by theory [26]. Such an observation has reinforced our faith in higher $T_c$ and has subsequently led us on the right path to bring down the liquid nitrogen barrier.

2.1.2. The T'-phase of $R_2CuO_4$ with R = $(R'_{1-x}Ce_x)$, where R' = Pr, Nd, Sm, or Eu; and T*-phases of $R_2CuO_4$ with R =$(La_{1-x}Sr_x)(R'_{1-x}Ce_x)$, where R' = Nd, Sm, Eu, or Gd

*T*-(Nd,Ce,Sr)$_2$CuO$_4$ with a $T_c$ of 28 K and T'-(Nd,Ce)CuO$_4$ with a $T_c$ of 24 K, discovered by J. Akimitsu et al. of Aoyama-Gakuin University in late 1988 [27] and by Y. Tokura, H. Takagi and S. Uchida of University of Tokyo in early 1989 [28], respectively, were the first electron-doped superconductors discovered, opening up an interesting chapter in HTS to study electron-doped superconductivity and to contrast it with the hole-doped superconductivity in T-La214.*

The crystal structure of the T'- and T*-phases are shown in Figs. 4b-c. The parent compounds are antiferromagnetic and become superconducting upon electron doping the anions or varying the O-content as shown in Table I. In spite of the gross structural similarity of the hole-doped and electron-doped cuprate superconductors, the $T_c$ of the electron-doped cuprate superconductors has not been able to surpass 30s K, in contrast to the record of 134 K for the hole-doped cuprate superconductors at ambient. However, interesting physics emerges from this class of compounds. Both similarities and differences have been found between the electron-doped and hole-doped cuprate superconductors. Details of the former are discussed in the chapter by Fournier in this volume.

2.2. $RBa_2Cu_3O_7$ or Cu1212 (RBCO, 123, or R123) with R = Y, La, Nd, Sm, Eu, Gd, Tb, Dy, Ho, Er, Tm, Yb, or Lu

*RBa$_2$Cu$_3$O$_7$, where R = Y or rare-earth (RBCO, R123, or Cu1212) – the first (self-doped) cuprate family that displays a $T_c$ in the 90s K above the liquid nitrogen boiling point was discovered by Paul C. W. Chu of University of Houston (UH), Maw-Kuen Wu of University of Alabama at Huntsville (UAH), and their team members, representing a giant advancement in modern science and drastically changing the psyche of superconductivity research. In spite of the many cuprate HTSs subsequently discovered, RBCO remains the most desirable HTS material for applications to date due to its physical robustness and superior superconducting behavior in high magnetic field. An efficient hybrid RBCO/Liq. N$_2$ transmission system for the delivery of electrical and chemical energies was subsequently proposed.*

With RBCO being the first HTS family to bring down the liquid nitrogen temperature barrier of 77 K and the HTS of choice for applications, as well as the heavy personal involvement by one of us (CWC), we choose to indulge ourselves in a slight digression in this section to recall selected events, some by design and others by coincidence, before and during its discovery, although some of them have been presented previously [29].

2.2.1. $RBa_2Cu_3O_7$ with R = Y, La, Nd, Sm, Eu, Gd, Tb, Dy, Ho, Er, Tm, Yb, or Lu

As mentioned previously in Section 2.1.1., BCS theory gives $T_c = 1.14\Theta_D \exp[-1/N(E_F)V]$ [16]. In principle, $T_c$ can be raised simply by increasing one or more of the three parameters, i.e. $\Theta_D$, $N(E_F)$, and V.

Unfortunately, not all the three parameters in the BCS-formula are completely independent. For example, a compound with a large $\Theta_D$ tends to be very hard mechanically and thus has a very small V. Because of the negative inverse exponential dependence of $T_c$ on V, the gain in $T_c$ by a modestly enhanced $\Theta_D$ can be overcompensated by the $T_c$-reduction due to a reduced V. Additionally, the increase of N and/or V may lead to instabilities such as the formation of charge- or spin-density waves, magnetic ordering, Peierls instabilities, a structural transformation, or even a collapse of the structure. We learned that optimization rather than maximization of the parameters is the best way to achieve higher $T_c$, i.e. to increase the parameters without triggering the onset of serious instabilities, and to adjust the parameters in such a way that highest $T_c$ is obtained.

In the 1970s and 1980s, one of the major concerns for higher $T_c$ was structural instabilities often observed in superconductors with a relatively high $T_c$. To determine the correlation of lattice instabilities with superconductivity, we developed an ac calorimetric high pressure technique [30] to vary and detect simultaneously the structural and superconducting transitions without introducing any chemical complexity. We found that lattice instabilities do affect the $T_c$ of the A15 superconductors, but only slightly by no more than a few tenths of one degree [31]. The observation gave Chu the confidence that superconductivity at higher $T_c$ might be achievable. During the same period of time, we had also studied the oxide superconductors $Ba(Pb_{1-x}Bi_x)O_3$ and $Li_{1+x}Ti_{2-x}O_4$ with unexpected high $T_c$ ~ 13 K under pressures [32], trying to see whether a novel superconducting mechanism was in operation and giving us experience in preparing and handling perovskite-type oxides. This is why Chu paid great attention to the observation of a possible $T_c$ in [(La,Ba)-Cu-O] by B&M, reproduced it quickly, and announced it at the Fall MRS Meeting on December 4, 1986, in Boston (and informally on December 2 at the Army Picatinny Arsenal in New Jersey).

Our groups worked day and night on the search in the ensuing critical months. Even though Chu officially co-directed the NSF Solid State Physics Program with Joe Trivisonno, the flextime scheme of NSF and the help from Trivisonno allowed him to spend on the average about 42% (and 75% in February 1987) of his time in Houston personally engaging in and directing the work.

After we successfully reproduced the results of M&B, we subjected the (La,Ba)-Cu-O mixed-phase samples to high pressure, focusing only on the superconducting phase in an attempt to reveal the nature of such an unusually high $T_c$ in oxides. The $T_c$ was unexpectedly raised to 40 K [33] and then to 52 K [24] at a rate more than ten times that of the intermetallic superconductors. The observation of a $T_c$ higher than 40 K clearly exceeded the then-theoretical limit of 30s K [26] and raised serious questions about the theory at the time. The unusually large positive pressure effect on $T_c$ observed further suggested to Chu that higher $T_c$ might be achievable through chemical pressures by replacing anions in (La,Ba)-Cu-O with those of smaller ionic radii and the same valences, such as Ba by Sr or Ca and La by the non-magnetic Y or Lu. The Ba replacement by Sr to raise $T_c$ was quickly confirmed, but the Ba to Ca substitution was unfortunately found later to lower the $T_c$. At the urging of the Dean of the UH College of Natural Sciences and Mathematics, Roy Weinstein, and the university legal counsel, Scott Chaffin, Chu prepared a patent disclosure at the beginning of January and filed a patent application [34] with the US Patent Office on January 12, 1987, which happened to contain the nominal $Y_{1.2}Ba_{0.8}CuO_4$, in which the stable 90 K superconductivity was first observed less than three weeks later by Wu and students in Huntsville. It should be noted that attempts to make nominal $Y_{1.2}Ba_{0.8}CuO_4$ in Houston failed, as shown in the January 13, 1987, entry to our lab book.

As mentioned previously in Section 2.1.1., Chu learned from Kitazawa in Boston that La214 was the superconducting phase in the (La,Ba)-Cu-O mixed-phase samples of M&B. It was only natural for Chu to decide to make pure 214-phase samples in his group, preferably single crystals, and to examine the origin of the 30s K-$T_c$ before contemplating the next step to raise $T_c$. After receiving the paper by Raveau et al. [35] on the structure data of 214 from Wu, we tried but failed to grow La214 single crystals, following the destruction of two of our three expensive crystal-growing Pt-crucibles. This turned out to be a blessing in disguise or “a kick of luck” in Mueller’s words. Subsequently, Chu decided to turn the team’s attention to stabilizing the high temperature anomalous resistance drops at ~ 70 K as shown in Fig. 6, indicative of superconductivity, which were detected sporadically in the multiphase samples above 70 K as early as November 25, 1986, but not in the pure 214 ones. This suggested to us that superconductivity above the $T_c$ of 214 at ambient could occur only in phases other than 214, if found. Regardless of its unstable nature, Chu showed the ~70 K preliminary data to his former student M. K. Wu, who was then an Assistant Professor at UAH, during lunch on December 4 at the 1986 Fall MRS Meeting in Boston and successfully convinced him to join our team for the search. On January 12, 1987, we observed a large diamagnetic shift or Meissner signal up to ~ 96 K in one of our mixed-phase samples, as displayed in Fig. 7, representing the first definitive superconducting sign detected above the liquid nitrogen temperature of 77 K. Unfortunately, the sample degraded and the diamagnetic signal disappeared the following day. No effort of ours in the ensuing two weeks succeeded to reproduce and stabilize this high temperature superconducting signal. Chu decided to write up and report details of the experiment and let other better equipped groups stabilize and identify the high temperature superconducting phase. No sooner than half of the paper was drafted, Wu called from Alabama in the afternoon of January 29, 1987, with the exciting news that a resistive drop indicative of a superconducting transition above 77 K was detected in the mixed-phase samples $Y_{1.2}Ba_{0.8}CuO_4$. On January 30, Wu brought to Houston the sample in which a resistive transition was immediately reproduced and the Meissner effect observed. Stable superconductivity at ~ 93 K was finally achieved as shown in Fig. 8, nearly tripling the $T_c$ of (La,Ba)214. The excitement preempted Chu’s desire to complete the manuscript on the unstable 90 K LBCO, of which XRD data taken at the time was later identified to be the La123 phase. The La123 results were mentioned as a footnote in the YBCO paper. It took Chu less than 24 hours to complete drafts of the manuscripts entitled: “Superconductivity at 93 K in a New Mixed Phase Y-Ba-Cu-O Compound System at Ambient Pressure” and “High Pressure Study on the New Y-Ba-Cu-O Compound System.” The latter was done to demonstrate that the 90 K Y-Ba-Cu-O belonged to a phase different from the 35 K 214 due to its different $T_c$-response to pressure. After they were circulated among collaborators for comments, the manuscripts were FedExed to the *Physical Review Letters* on February 5, were accepted on February 11, and appeared on March 2, 1987 [2,36]. March 2, 1987, was called by some the “super day of physics”, since besides the announcement of the 90 K superconductor, other news of the day included the detection of a supernova and the US contemplating the construction of the Superconducting Supercollider. Contrary to the initial skepticism toward the 30 K (La,Ba)-Cu-O, the 90 K Y-Ba-Cu-O results were accepted after the news broke because of their quick duplication worldwide. The world of HTS was on fire and YBCO ushered in the modern era of liquid nitrogen HTS.

Being the first to claim superconductivity with a $T_c$ above 77 K laid upon us a tremendous psychological burden. In spite of Chu’s confidence in our results, the thought, “Could it be too good to be true?”, did enter his mind occasionally after submitting the manuscripts. A bungle of this magnitude would have meant the abrupt end of his career in superconductivity. More than once, he dropped by the offices of his colleagues and asked, “Can there be phenomena other than superconductivity that can account for our observations? Please think and think hard!” Even in the cover letter to Myron Strongin, the *Physical Review Letters* editor, Chu put in writing the earlier verbal agreement with him over the phone that “information in these manuscripts should not be released nor experimentally tested prior to their formal publications by the journal.” To do so was to let us have more time to correct possible mistakes and to crack the mystery.

The next challenge was to isolate and identify the 93 K superconducting phase from the greenish-looking mixed phase YBCO samples and to determine its composition and structure. After failing to resolve the problems with colleagues at Houston, Chu sought help from Dave Mao and Bob Hazen at the Geophysical Lab at Washington, D.C., who had expertise in determining the structures and compositions of tiny crystals from rocks. They graciously agreed. They started working feverishly on the sample Chu brought them the third week of February 1987. Within a week, they determined the compositions and symmetries of the black and green phases closely intertwined in the small grains of the crushed sample. At the same time, we determined that the black phase was superconducting by correlating the sample color with the superconducting volume of samples of different compositions and synthesis conditions in Houston. Through continued and instant exchanges of information between Hazen and Chu, within a few days Hazen solved the structure of the first liquid nitrogen HTS as $YBa_2Cu_3O_{7-\delta}$ = $(BaO)(CuO_{1-\delta})(BaO)(CuO2)(Y)(CuO2)$ [37], except for the oxygen vacancy locations. The orthorhombic crystal structure of R123 (space group Pmmm) with corrugated $CuO_2$-layers is shown in Fig. 9.

Before the superconducting Y123 phase was identified, we were puzzled by the grossly different $T_c$s between the nominally similar samples of ($Y_{1.2}Ba_{0.8}CuO_4$) and ($La_{1.2}Ba_{0.8}CuO_4$). We decided to probe the role of Y in the superconducting YBCO mixed phase samples, while pursuing complete replacement of La by the non-magnetic Lu as conjectured in the patent application. We adopted the standard practice by examining the effect of partial replacement of Y by Gd on the $T_c$ of ($Y_{1.2}Ba_{0.8}CuO_4$). To our great surprise, several percent of Y-replacement by Gd, the strongest magnetic element among rare earths, did not suppress the $T_c$ of ($Y_{1.2}Ba_{0.8}CuO_4$) at all, in contrast to the case of La, in which superconductivity is completely destroyed by ~ 1% doping by Gd [29,38]. The observation suggested that Y in YBCO is electronically isolated from the superconducting electron system of the compound and serves only as a stabilizer to hold the crystal structure of the superconducting phase together. The information led us to the subsequent discovery of the series of RBa2Cu3O7 (RBCO, R123, or Cu1212) with a $T_c$ varying between 93 and 100 K where R = Y and all rare-earth elements except Ce and Pr. Chu discussed part of the results at a special seminar at Harvard on March 16 before rushing back to Houston to prepare the paper. The manuscript was submitted to *PRL* on March 17. The results were presented at the APS March Meeting on March 18 in New York City. The paper appeared in the May issue of *PRL*, entitled "Superconductivity above 90 K in the Square-Planar Compound System $ABa_2Cu_3O_{6+x}$ with A = Y, La, Nd, Sm, Eu, Gd, Ho, Er, and Lu" [39]. Later work added Tb, Dy, Tm, and Yb but not Pr or Ce to the series [40-42]. Different reports appeared concerning the absence of superconductivity in $PrBa_2Cu_3O_7$. One claimed a $T_c$ in the 90s K in a single crystal sample. Unfortunately, no effort has reproduced the claim to date.

The discovery has brought superconductivity applications a giant step closer to reality and at the same time poses serious challenges to physicists concerning the origin of HTS. The excitement at the time was amply demonstrated by the Special Panel Discussion on Novel Materials and High Temperature Superconductivity initiated by Chu and organized by the American Physical Society (APS) on March 18, 1987, at the New York Hilton attended by thousands of fellow physicists and by the Federal Conference on High Temperature Superconductivity Science and Technology in July 1987 at Washington, D.C., attended by President Reagan and his cabinet members. The special APS discussion panel started at 7:30 p.m. with short presentations by five panelists: Alex Mueller (https://www.youtube.com/watch?v=1IJoikdY5dA&list=PLDc_NxcK1lnwqiynxBCiWy2AZA-avfKoC&index=8), Shoji Tanaka (https://www.youtube.com/watch?v=k0SqSgGRpl0&index=3&list=PLDc_NxcK1lnwqiynxBCiWy2AZA-avfKoC), Paul Chu (https://www.youtube.com/watch?v=stVkHt4qox8&index=12&list=PLDc_ NxcK1lnwqiynxBCiWy2AZA-avfKoC), Bertrum Batlogg (https://www.youtube.com/watch?v=0xJI_VMevbI&list=PLDc_NxcK1lnwqiynxBCiWy2AZA-avfKoC&index=13), and Zhongxian Zhao (https://www.youtube.com/watch?v=JGQfdXeg2Gs&list=PLDc_NxcK1lnwqiynxBCiWy2AZA-avfKoC&index=1), followed by short contributions and discussions that lasted until the wee hours of the next morning. The 1,200-seat meeting room was packed with more than 2,000 people. Many more who could not get into the room watched on TV screens outside the room to witness this exciting event, coined as the "Woodstock of Physics" (Fig. 10) by the late Bell Labs physicist Mike Schluter, referring to the legendary 1969 Woodstock Music and Art Festival against the Vietnam war in upstate New York. In spite of later discoveries of many other cuprate HTS families, some of which have higher Tc, YBCO remains to be the most desirable material for HTS science and technology due to its superior sample quality, current carrying capacity in the presence of high magnetic fields, and physical robustness in thin-film form. A YBCO puck was chosen as an entry for the White House's National Millennium Time Capsule in 2000, which was created in the spirit of “honor the past—imagine the future" to contain discoveries and achievements in all areas by Americans over the previous 100 years considered to be significant (Fig. 11). It will be opened in 2100 to communicate to the future generations about US accomplishments and visions made in the 20th Century.

2.2.2. $YBa_2Cu_4O_8$ (Y124) and $Y_2Ba_4Cu_7O_{15}$ (Y247)

*$YBa_2Cu_4O_8$ (Y124) and $Y_2Ba_4Cu_3O_7$ (Y247) – cuprates closely related to Y123 with $T_c$s around 80 K were identified and characterized in 1988-1989 by P. Marsh et al. of Bell Labs and J. Karpinski et al. of ETH. The compounds fit the generic formula $Y_2Ba_4Cu_{6+n}O_{14+n}$ proposed by R. J. Cava et al. of Bell Labs in 1989. They are more stable against the loss of oxygen above 800 °C than Y123 and do not exhibit the orthorhombic structure transition with oxygen-content like the Y123, making them better candidates for devices.*

$YBa_2Cu_4O_8$ (Y124) and $Y_2Ba_4Cu_7O_{15}$ (Y247) were first identified by Marsh et al. [43] and Karpinski et al. [44], respectively, as distinct impurity phases in decomposed Y123 samples. After fine-tuning the synthesis conditions, pure bulk Y124 and 247 samples were prepared and characterized by Fischer et al. [45], Karpinski et al. [46], and Cava et al. [47]. In contrast to Y123, which has single CuO-chains in the b-direction with the $CuO_5$-pymids, Y124 has double CuO-chains along the b-direction while Y247 possesses a mixture of single and double chains. The reported $T_c$s for the compounds are around 80 K. Later studies show that they are stable against the loss of oxygen above 800 K that can result in the degradation of their superconducting properties. They also do not exhibit the orthorhombic-tetragonal transition that creates defects and reduces their performance. Unfortunately, the compounds are difficult to synthesize and have to be made with special care, making their utilization less practical regardless of the above-mentioned beneficial properties.

2.2.3. $RSr_2Cu_3O_x$ or Cu1212 (RSCO, 123 or R123) with R = Y, Eu, Gd, Tb, Dy, Ho, Er, Tm, or Yb

*$RSr_2Cu_3O_7$, where R = Y or rare-earth (RSCO, R123, or Cu1212) – the cuprate family that is closest to YBCO displays a $T_c$ up to 80 K and was discovered by Maw-Kuen Wu et al. at the University of Alabama at Huntsville in 1988. In spite of the gross chemico-structural similarity between YSCO and YBCO, YSCO is unstable at ambient and always shows a lower $T_c$ than YBCO when stabilized by slight doping into the CuO-chains. R exhibits a large negative effect on the $T_c$ of RSCO, in contrast to RBCO.*

Since the $T_c$ of (La,Sr)214 is higher than that of (La,Ba)214, a search for an expected higher $T_c$ in $YSr_2Cu_3O_7$ (YSCO) than that of $YBa_2Cu_3O_7$ (YBCO) was immediately launched after the discovery of YBCO. Unfortunately, YSCO does not form under the synthesis conditions of YBCO. In 1988, M. K. Wu et al. detected superconductivity only up to 80 K in mixed-phase YSCO samples fast-quenched from 1300 °C [48]. This is because YSCO does not form at ambient pressure below 1250 °C. Later it was successfully synthesized under high pressures at high temperatures [49]. YSCO so-formed shows a $T_c$ of ~ 60 K with a tetragonal structure and an oxygen deficiency. However, the YSCO can be stabilized at ambient by the slight replacement of the Cu-atoms in the chains of YSCO by a large number of the metal elements, such as Fe, Co, W, Mo, Re, etc. [50] For instance, $YSr_2(Cu_{2.85}Re_{0.15})O_x$ is stable at ambient with a $T_c$ ~ 41 K. This provides the opportunity to determine whether Y in YSCO is isolated from the electron system responsible for its superconductivity as in YBCO and whether a family of RSCO with high $T_c$ can be found. An experiment to replace Y in $YSr_2(Cu_{2.85}Re_{0.15})O_x$ by R = Eu, Gd, Tb, Dy, Ho, Er, and Tm was carried out [51]. It was found that the $T_c$ is drastically influenced by the partial R-replacement for Y, in strong contrast to YBCO. There appears to be a strong correlation of the decrease of $T_c$ with the decrease of orthorohmbicity. It is known that orthorhmbicity of the 123 phase is related to the oxygen-deficiency and R-substitution for Cu-created defects, which are known to suppress $T_c$.

2.2.4. $RuSr_2RCu_2O_8$ (Ru1212) and $RuSr_2(R,Ce)_2Cu_2O_{10}$ (Ru1222) with R = Sm, Eu or Gd

*$RuSr_2RCu_2O_8$ (Ru1212) and $RuSr_2(R,Ce)_2Cu_2O_{10}$ (Ru1222) with R = Sm, Eu or Gd – the first ferromagnetic superconductors with a $T_c$ ~ 40s K below their ferromagnetic transition temperature $T_m$ ~ 120-180 K were first synthesized and found by L. Bauernfeind, L. Widder, and H. T. Braun of University of Bayreuth to be superconducting in 1995 and later discovered to be weak ferromagnets by I. Felner et al. of Hebrew University who coined the name of "superconducting ferromagnet."*

In an attempt to broaden the HTS material base and to further improve the flux pinning force, Braun et al. modified the RO-layers between the two $CuO_2$-layers in YBCO [52]. In 1995, they succeeded in synthesizing two general groups of ruthenocuprate compounds, $RuSr_2RCu_2O_8$ (Ru1212) and $RuSr_2(R,Ce)_2Cu_2O_{10}$ (Ru1222), where R = Sm, Eu, or Gd, and found them to be superconducting with a $T_c$ up to ~ 40s K by introducing O-deficiency or Ce-doping. Both Ru1212 and Ru1222 compounds have a tetragonal symmetry with a space group P4/mmm and I4/nmm, respectively. Ru1222 can be regarded as the result of substituting the CuO linear-chain layer, the BaO layer, and the Y layer in YBCO with the $RuO_2$ square-planar layer, the SrO layer, and the $(R,Ce)_2O_2$ fluorite-type layer block, respectively, while Ru1212 can be regarded as the result of replacing the CuO linear-chain layer, the BaO layer, and the Y layer with the $RuO_2$ square-planar layer, the SrO layer, and the R layer, respectively, as shown in Fig. 12. They were later found to be weak ferromagnets with a magnetic onset temperature $T_m$ above $T_c$, i.e. up to ~ 180 K for Ru1222 and ~ 133 K for Ru1212 as displayed in Fig. 13. These compounds with their $T_m > T_c$ were first called superconducting ferromagnets (SCFMs) by Felner [55] to differentiate them from the previously known ferromagnetic superconductors with their $T_m < T_c$. The discovery of these SCFMs has attracted intense interest over the past few years. The appearance of magnetism and superconductivity in Ru1212 and Ru1222 does not seem to be surprising in view of the layer blocks of [$(RuO_2)$(SrO) = $RuSrO_3$] and [$(CuO_2)$(R)$(CuO_2)$] in the compounds, since the former represents the itinerant ferromagnet $RuSrO_3$ with a $T_m$ ~ 160 K and the latter is the active component of the superconductor YBCO with a $T_c$ ~ 93 K. While the simultaneous occurrence of superconductivity and magnetism in these compounds has been established, many issues remain open. They include the absence of a bulk Meissner effect, the large variation in $T_c$, the nature of the magnetic state, the nature of the superconducting state, and the possible occurrence of the spontaneous vortex state in this fascinating class of compounds [54].

2.3. $Bi_2Sr_2Ca_{n-1}Cu_nO_{2n+4}$, where n = 1,2,3,….[BSCCO or Bi22(n-1)n]

*$Bi_2Sr_2Ca_{n-1}Cu_nO_{2n+4}$, where n = 1,2,3,….[BSCCO orBi22(n-1)n] - the first cuprate family without rare-earth element that displays a $T_c$ up to 110 K at ambient and ~ 135 K at ~ 35 GPa was discovered by Hirosh Maeda et al. [55]of the National Institute for Metal in January 1988, suggesting that HTS had a broader material base than R123 with rare earth, as originally thought, and that more HTSs with a higher $T_c$ were possible. BSCCO has been the material used for the first generation HTS-wires and for extensive ARPES studies, due to its graphitic-like behavior and the ease to obtain quality samples for the studies.*

Euphoria permeated the field following the announcement of superconductivity above 90 K in YBCO and RBCO, and the sky seemed to be the only limit to $T_c$. As 1987 was drawing to an end, an impatient physicist of an industrial lab told the *Wall Street Journal* that the accumulated man-hours devoted to cuprate-HTSs in 1987 had exceeded all previous effort devoted to LTSs in the preceding 75 years since its discovery and that any cuprate with a $T_c$ above 90s K should have been found. He went on to propose that one should search outside the cuprates for superconductors with higher $T_c$. The fallacy of prophecy based on past statistics immediately faced the truth on January 22, 1988, when *Nippon Keizai Shimbun* briefly reported the observation of superconductivity between 75 and 105 K in the Bi-Sr-Ca-Cu-O system without any information about the stoichiometry, structure, or processing condition. The observation was easily reproduced by us in three days with a $T_c$ up to 114 K [56]. The paper by Hirosh Maeda et al. on superconductivity above 100 K in their mixed phase samples of Bi-Sr-Cu-O without rare-earth later appeared in the February 20 issue of *Japanese Journal of Applied Physics*, entitled “A New High-$T_c$ Oxide Superconductor without a Rare Earth Element" [55]. The $T_c$ was later raised to ~ 135 K under 35 GPa [57].

In their attempt to broaden the material base for HTSs, Maeda et al. examined elements in the VB group of the periodic table, such as Bi, which is trivalent and has a similar ionic radius to those of the trivalent rare earth elements, along the lines of 214 and RBCO. They found superconductivity in samples of nominal $Bi_1Sr_1Ca_1Cu_2O_x$ above 105 K, as shown in Fig. 14 [55]. The multi-transitions in their resistivity data suggested that the sample must consist of multiple phases. Indeed, shortly afterward, three members of the Bi22(n-1)n family with n = 1, 2, and 3 were isolated and the structures determined by several groups with a distinct modulation in the BiO-double layers [58], including the Carnegie Institute Geophysical Lab and our own [59-61]. The layer stacking sequence of the highest $T_c$ member of the family is Bi2223 or $Bi_2Sr_2Ca_2Cu_3O_{10}$ =[(SrO)$(BiO)_2$(SrO)][$(CuO_2)$][Ca$(CuO_2)$Ca$(CaO_2)$], as also shown in Fig.15. Layer structural modulation appears in the (BiO) double layers for all n. The maximum $T_c$s for members of the homologous series at ambient increase with n and are ~22, ~80, and ~110 K for n = 1, 2 and 3, respectively. The $T_c$ of Bi2223 was later reported to reach 135 K without sign of saturation at ~ 35 GPa by Chen et al. in August 2010 [57]. However, for n > 3, $T_c$ starts to drop and the drop is attributed to the combined influence of electrostatic shielding and proximity effects of the $CuO_2$-layers. Aided by the experience on Y123, it took less than two days for Hazen et al. of Carnegie Geophysical Lab and Veblen et al. of Johns Hopkins [59] to crack the structure of BSCCO after receiving the samples from us, in contrast to the more than 10 days taken by Hazen et al. to do the same for Y123 in 1987. From their structures, it was immediately evident that the weak van der Waal force between the (BiO)-double-layers in BSCCO is responsible for the graphitic-behavior of the compound, making cleaving the sample in vacuum easy for spectroscopic studies and mechanical rolling possible for aligning the (CuO2)-layers in the first generation-HTS-wire processing but with weak flux pinning and thus low $J_c$. Unfortunately, the softness and structural modulation nature of the compound make thin film synthesis of BSCCO for devices a challenge.

By introducing complexity to the charge reservoir blocks and the R-layers, many other layer cuprate superconductors have been subsequently discovered, including the Pb-based cuprates. While their $T_c$s never exceed 70 K, the local atomic configurations display different interesting features [62].

It is interesting to note that in the summer of 1987 Bernard Raveau et al. of University of Caen [63] and Jun Akimutsu et al. of Aoyama-Gakuin University [64] reported an 8 K superconducting transition in the Bi-Sr-Cu-O, which was later found to be associated with the n = 1 member of the BSCCO-family. This important piece of information was lost at the time due to the mad rush for superconductors with a higher $T_c$ than that of YBCO. Chu even marked their preprints with “exciting, more study needed!" in August 1987, but took no immediate action. Should one have paid greater attention to these early results, the discovery of the 110 K superconducting BSCCO-family without rare-earth could have been advanced by at least half a year and by someone other than Maeda. A lesson from this episode is never to get trapped in the turbulence of excitement presented by obvious fashionable pursuits.

2.4. $Tl_2Ba_2Ca_{n-1}Cu_nO_{2n+4}$ where n = 1, 2, 3,… [TBCCO or Tl22(n-1)n]; and $TlBa_2Ca_{n-1}Cu_nO_{2n+3}$ with n = 1, 2, 3, .... [TBCCO, Tl12(n-1)n]

*$Tl_2Ba_2Ca_{n-1}Cu_nO_{2n+4}$, where n = 1, 2, 3,…[TBCCO or Tl22(n-1)n] – the second cuprate family without rare earth that displays a $T_c$ up to 125 K at ambient and 131 K at 7 GPa was discovered by Zhengzhi Sheng and Allen Hermann of University of Arkansas in February 1988 [65,66]. The discovery appeared to justify the early optimism after RBCO and BSCCO that more high HTSs would be on their way to be* *found. The physical softness and complex synthesis procedure of TBCCO make its applications difficult, in spite of the early optimism regarding its possible use for thin film devices due to its higher $T_c$ and greater stability than BSCCO.*

Trying to simulate the role of R in R123, Sheng and Hermann started to replace the trivalent R in R123 with the trivalent Tl, which has an ionic radius similar to R. They found superconductivity up to 90 K in their mixed phase samples of Tl-Ba-Cu-O with a nominal composition of $Tl_2Ba_2Cu_3O_{8+x}$ in early January 1988 as shown in Fig. 16 [65], at about the same time that Maeda et al. announced their discovery of superconductivity up to 105 K in BSCCO. This marked the discovery of the first member, Tl2201, of the TBCCO family. Sheng and Hermann soon partially replaced Ba with Ca and observed superconductivity above 120 K in their multiphase samples of nominal $Tl_2Ca_{1.5}BaCu_3O_{8.5+x}$ during the second week of February 1988 as shown in Fig. 17. The results appeared in the March 10, 1988, issue of *Nature* in an article entitled “Bulk superconductivity at 120 K in the Tl-Ca/Ba-Cu-O system" [67]. The n = 2 and 3 members of the Tl22(n-1)n homologous series were quickly identified in these samples by Hazen et al. of the Carnegie Geophysical Lab [68] and Torardi et al. [69] of Du Pont Research Lab only two days after the announcement of the 120 K superconductivity. Parkin et al. of the IBM Almaden Lab later obtained pure-phase samples of Tl2223 and achieved a $T_c$ of 125 K [70], which was the record-$T_c$ at ambient pressure until the discovery of HBCCO in April 1993. The $T_c$ of Tl2223 was raised to 131 K by Berkeley et al. [71] at the Naval Research Lab by pressures up to 7 GPa in September 1992. The whole homologous series Tl22(n-1)n was discovered and identified to display the layered stacking sequence. For instance, $Tl_2Ba_2Ca_2Cu_3O_{10}$ with n = 3 or Tl2223 exhibits the stacking of $[(BaO)(TlO)_2(BaO)][(CuO_2)(Ca)(CuO_2)(Ca)(CuO_2)]$. Similar to Bi22(n-1)n, the $T_c$ of Tl22(n-1)n increases with n up to 3 and decreases when n > 3, similar to BSCCO. The maximum $T_c$s are 90, 110, and 125 K, respectively, for n = 1, 2, and 3. Depending on the oxygen content, the $T_c$ can be varied by more than 10 K. In spite of the gross similarity between Tl22(n-1)n and Bi22(n-1)n, there exists no structural modulation along the double (TlO)-layers in contrast to that along the double (BiO)-layers, suggesting that such a structural anomaly does not play a role in their high $T_c$ as was initially thought.

The single (TlO)-layer TBCCO homologous series Tl12(n-1)n was later discovered [72,73] (Table I). The layer stacking sequence is similar to that for Tl22(n-1)n except that the double (TlO)-layers in Tl22(n-1)n are replaced by the single (TlO)-layer with an I4/mmm symmetry. The $T_c$ of members of this series is, in general, lower than that for the corresponding members of Tl22(n-1)n. It was found to increase continuously with n up to n = 4 before it decreases [74,75], i.e. $T_c$ = 50, 82, 110, and 120 K for n = 1, 2, 3, and 4, respectively. The reasons for the lower $T_c$s and for the continuous increase of $T_c$ to n = 4 could be interesting for understanding the inner layer atomic structure influence on $T_c$, but unfortunately remain unknown.

It is interesting to note that the news of the TBCCO discovery caught many in the field and in the media by surprise – for such an exciting discovery came from an institute not known for superconductivity study. Sheng took his first 120 K sample to Chu immediately after their discovery and asked him to confirm it. This demonstrated that HTS is a level playing field – there is a chance for everyone with vision and willingness to try.

2.5. $HgBa_2Ca_{n-1}Cu_nO_{2n+3-\delta}$, where n = 1, 2, 3,…[HBCCO or Hg12(n-1)n]

*$HgBa_2Ca_{n-1}Cu_nO_{2n+3-\delta}$, where n = 1, 2, 3,… [HBCCO or Hg12(n-1)n] - the third cuprate family without rare-earth that shows a $T_c$ up to 134 K at ambient and 164 K at ~ 30 GPa was discovered by Andeas Schilling et al. of ETH in mid-April 1993. [76-78]. A $T_c$ of 134 K is above the temperature in the cargo bay of the Space Shuttle when orbiting in Earth's shadow and above the boiling point of liquid natural gas (LNG) on earth. The former makes HBCCO a possible material for HTS devices operable on the Space Shuttle without liquid cryogen and the latter may enable the development of a hybrid HTS/LNG* *transmission system for the efficient delivery of electrical and chemical energies simultaneously as was proposed by Chu and Grant.*

In late 1989, the $T_c$ of cuprates appeared to have stagnated at 125 K since the Spring of 1988. A prominent chemist conjectured that the $T_c$ of cuprates could not exceed 160 K based on his physical chemistry arguments. However, $HgBa_2Ca_2Cu_3O_{7-\delta}$ was discovered in 1993 to display a $T_c$ up to ~ 133 K by Andreas Schilling et al. of ETH at Zurich in their mixed phase samples as shown in Fig. 18. The results appeared in the May 6, 1993, issue of *Nature* in an article entitled “Superconductivity above 130 K in the Hg-Ba-Ca-Cu-O system" [76]. The results took several weeks to reproduce, which was possible only after the sample preparation challenges associated with the complex chemistry of the compound were overcome by us [77,78]. The homologous series of HBCCO, Hg12(n-1)n with n = 1, 2, 3, … was subsequently identified and characterized. The maximum $T_c$s were found to be 97, 128, and 134 K, for n = 1, 2, and 3, respectively [79]. They all exhibit layer structures with the stacking sequence as specified by the generic formula. This is exemplified by the highest $T_c$ member of the HBCCO family, Hg1223, as $HgBa_2Ca_3Cu_{9-\delta}$ = $[(BaO)(HgO_{1-\delta})(BaO)]$ $[(CuO_2)(Ca)(CuO_2)(Ca)(CuO_2)]$, shown in Fig. 19.

It is interesting to note that attempts were made as early as 1991 to substitute the linearly coordinated $Hg^{+2}$ for the similarly coordinated $Cu^{+2}$ in the (CuO)-chain-layer of R123. Compounds of $HgBa_2EuCu_2O_x$ with a structure similar to Tl1212 were first made but found not superconducting by Antipov of Moscow State University in 1991 [80]. A small superconducting signal due to an impurity phase (thought to be Eu123 at 94 K) was detected in 1991 in one of our $HgBa_2EuCu_2O_x$ samples without recognizing that it was associated with the Hg1201. The whole story of HBCCO did not start to unfold until September 1992 when E. Antipov of Moscow State University and M. Marezio of CNRS at Grenoble successfully synthesized the Hg1201 with a $T_c$ = 94 K [81]. Knowing that increasing the number of $(CuO_2)$-layers per cell will lead to an increase of $T_c$, A. Schilling et al. [76] of ETH added Ca to achieve a $T_c$ up to 133 K in their mixed phase sample that consisted of the n = 2 and 3 members of the Hg12(n-1)n with an enhanced $T_c$.

While Hg12(n-1)n has a layer structure similar to Tl12(n-1)n, there exists a subtle difference, presumably arising from the linear oxygen coordination of $Hg^{+2}$-ions in HBCCO as reflected in the relatively short Hg-O bond length along the c-axis and the large number of voids in the $HgO_{1-\delta}$-layer [82]. Higher $T_c$ was therefore expected in HBCCO under pressures. Experiments on optimally doped pure Hg1201, 1212, and 1223 were carried out under pressures. We, in collaboration with Mao et al. at the Carnegie Geophysics Lab, quickly found in 1993 that the $T_c$ of these members grows with pressure in parallel and peaks at ~ 118 K in Hg1201 at ~ 24 GPa, at ~ 154 K in Hg1212 at ~ 29 GPa, and at 164 K in Hg1223 at ~ 30 GPa, as shown in Fig. 20 [78]. The observation strongly suggests that superconductivity of different members of the Hg12(n-1)n homologous series arises from a common origin. A $T_c$ = 134 K at ambient or 164 K under high pressure remains as today’s respective records. However, the unusually large pressure-induced $T_c$-enhancement in their optimally doped states is still unexplained and cannot be accounted for by models commonly used to explain the pressure effect on $T_c$ for other HTSs. A modified rigid band model has been proposed for the observations [83].

2.6. $(Sr_{1-x}\,Ax)CuO_2$ where A = Ba, Sr, Ca or Nd

*$(Sr_{1-x}\,Ax)CuO_2$ where A = Ba, Sr, Ca, or Nd – the infinite layer cuprate and the most basic building element, $CuO_2$, with a $T_c$ between 40 – 110 K discovered by J. Goodenough et al. of University of Texas at Austin and M. Takano et al. of Kyoto University with the hope to resolve the basic problem of HTS. Unfortunately, the origin of the 110 K superconductivity in these infinite layer compounds remains unsettled.*

After the discovery of R123, we suggested that $T_c$ might increase with the number of ($CuO_2$)-layers per unit cell (n). The later observation of $T_c$-increase with n up to 3 in Bi22(n-1)n, Tl22(n-1)n, and Hg12(n-1)n gave support to the suggestion, consistent with some subsequent model calculations. It would be very interesting to investigate the $T_c$ of compounds with very large n, preferably an n approaching ∞, even though we knew at the time that the $T_c$ of the above three families decreases as n becomes greater than 3. For instance, $HgBa_2Ca_{n-1}Cu_nO_{2n+3}$, which has n $CuO_2$-layers per formula may be approximated as a homologous series $Ca_{n-1}Cu_nO_{2n}$ for n >> 2 and becomes $CaCuO_2$ as n → ∞. Therefore, the $ACuO_2$ with A = Ba, Sr, and Ca may be considered the infinite layer member of the layer cuprate system. It has the simplest basic structure of all cuprate HTSs with only $CuO_2$-layers separated by the A-layers as shown in Fig. 21. Therefore it was considered to hold the key to the mystery of cuprate HTS.

In 1988, T. Siegrist et al. of AT&T Bell Labs succeeded in stabilizing the compound $(Ca_{0.85}Sr0_{.15})CuO_2$ [84]. Unfortunately, it was not superconducting and the homogeneity range of the compound was found to be very limited at ambient pressure, i.e. one could not vary the Ca/Sr-ratio without triggering the collapse of the structure. Almost three years later, Goodenough and coworkers of the University of Texas at Austin decided to substitute Nd for Ca on the basis of consideration of Cu-O bond length in the $CuO_2$-planes of the compound. They synthesized $Sr_{1-y}Nd_yCuO_2$ with a $T_c$ ~ 40 K at 2.5 GPa [85]. Other similar compounds with rare earth replacements were also made later with a $T_c$ not higher than 43 K [86]. In 1991 M. Takano et al. of Kyoto University demonstrated that $(Sr_{1-x}Ca_x)_{0.9}CuO_2$ was superconducting up to 110 K as displayed in Fig. 22, when it was prepared at 6 GPa [87]. The high resolution electron micrographs revealed that the superconducting sample was loaded with defect layers with Ca- and Sr-vacancies. In fact, $Sr_{0.9}CuO_2$ was subsequently found to be superconducting with an onset $T_c$ ~ 100 K. Although the Sr-deficient compound is believed to be responsible for the ~ 100 K-$T_c$, the exact superconducting phase is yet to be determined, since it has been shown that there exists an infinite number of Sr-deficient phases in the homologous series $Sr_{n-1}Cu_{n+1}O_{2n}$ with n = 3, 5, … [88].

## 2.7. $Ba_2Ca_{n-1}Cu_nO_x$, where n = 1, 2, 3… [BCCO or 02(n-1)n]

*$Ba_2Ca_{n-1}Cu_nO_x$ [BCCO or 02(n-1)n] – the cuprate superconductors without the charge reservoir block and with only the active block of the cuprate HTSs, BSCCO and TBCCO were synthesized under pressures and discovered to be superconducting with a $T_c$ up to 126 K at ambient in 1997 by C. W. Chu et al. at the University of Houston. It demonstrates that a superconductor with a high $T_c$ can exist through interstitial doping alone without the charge reservoir.*

It has been demonstrated that the layer cuprate HTSs consist of two substructures, namely the active block of *$[(CuO_2)R_{n-1}(CuO_2)_{n-1}]$* and the charge reservoir block of *$[(EO)(AO)_m(EO)]$*. The active block comprises *n-square-planar-($CuO_2$)*-layers interleaved by *(n-1)- R*-layers and the charge reservoir block contains *m-(AO)*-layers bracketed by *2 chemically inert (EO)*-layers. It is known that processing of cuprate HTS materials into practical forms is a challenge due to the chemical and physical complexity of the materials. We therefore tried to synthesize layer cuprate HTSs of simpler structure, *e.g.* without the charge reservoir blocks and thus without the volatile toxic elements such as Tl and Hg. We attempted to deactivate the charge reservoir block by removing the *m-(AO)*-layers and allowing interstitial doping to take place in the two (EO)-layers that usually bracket the *m*-(AO)-layers. The compound does not form at ambient. However, through high pressure synthesis processing in a C- and O-free environment, we succeeded in stabilizing the $Ba_2Ca_{n-1}Cu_nO_x$ *[BCCO or 02(n-1)n-Ba]*. The XRD shows a layer lattice following the I4/mmm space group symmetry. It superconducts with a $T_c$ of 126 K for the n = 4 member and the $T_c$ increases to ~ 150 K under pressure as shown in Fig. 23. The results were published in the August 22, 1997, issue of *Science* in an article entitled "Superconductivity up to 126 Kelvin in Interstitially Doped $Ba_2Ca_{n-1}Cu_nO_x$ [02(n-1)n-Ba]" [89].

Although $Ba_2Ca_{n-1}Cu_nO_x$ *[BCCO or 02(n-1)n]* meet the original goal of achieving a simpler structure without toxic elements and are interesting from chemistry and physics points of view, the compounds are not very stable and degrade in the presence of humid air, and thus are of no practical application value. However, they do show that the structural and chemical instabilities associated with the building blocks of layer cuprates should be taken into consideration in future design of cuprate HTSs with higher $T_c$.

3. Several scaling rules of $T_c$ of hole-doped cuprate high temperature superconductors

There appear to be three ultimate goals in the study of hole-doped cuprate HTSs: to search for a path to further enhance their $T_c$s; to unravel the underlying mechanism to develop a comprehensive microscopic theory of HTS; and to develop practical HTSs with improved performance for applications. Great progress has been made in bringing us closer to our goals over the last 28 years. The continuous broadening of the material base, improvement of the materials, and development of the characterization techniques have removed much confusion that confronted the field in the early days of HTS. Many excellent review articles have appeared [5-9,90], covering almost all aspects of cuprate HTSs, ranging from materials, phenomena, basic physics, pairing mechanisms, symmetry of the order parameter, electronic structures, and flux dynamics to applications. While consensus has emerged concerning certain specific aspects, such as many of the phenomena, and d-symmetry of the order parameter, more diverging views remain about the fundamental issue of HTS, for example, the electron pairing force responsible for the high $T_c$ and even the approach adopted to solve the HTS problem. This is amply evident in the various reviews and the absence of a commonly accepted comprehensive microscopic theory. In this review, we decide to let the experts resolve the different views and to follow instead a holistic approach to summarize some of the empirical scaling rules of $T_c$, which often embed more subtle physics of HTS than it appears. In the absence of a commonly accepted theory of HTS, universal trends and relationships found between $T_c$ and other physical parameters, especially those easily determined ones, will provide useful guides to higher $T_c$ and significant physics of HTS. For example, the physics significance of the successful empirical Matthias rule for low temperature intermetallic superconductors in the early days of superconductivity was not realized until the BCS theory was developed and applied. We shall list a few of them in the chronological order of their propositions:

3.1. $T_c$ - n (the number of square-planar $CuO_2$-layers per formula)

In 1964, W. Little [91] examined the question posed by F. London of whether superconductivity occurs in organic macromolecules within the framework of BCS theory and concluded that superconductivity not only exists, but also that it is not impossible to have a $T_c$ above room temperature in organic molecules with a special design. The proposed design is a one-dimensional (1D) long organic polymer with side chains with proper oscillation of charges to provide the glue (excitons of high characteristic temperature) to pair the electrons in the spine. In the same year, V. Ginzburg [92] noticed the drawbacks of such a 1D system, e.g. fluctuations, instabilities, and lack of Coulomb screening. Instead, he focused on 2D materials systems to alleviate the impasse at least in part. He conjectured that a $T_c$ of ~ $10^3$ K might be possible via the exchange of excitons in metallic thin films, metal surface covered by dielectrics, metal sandwiches with dielectrics in between, or 2D layer compounds. In the ensuing years, extensive effort has been devoted to the search for high $T_c$ in low-dimension materials, especially the transition-metal dichalcogenides where the transition-metal layers are loosely held together by the van der Waal force. Unfortunately, the $T_c$ was low and the superconductivity observed could be explained by the exchange of phonons without invoking excitons [93]. It should be noted that in spite of the disappointing outcome, the search led to the discovery of charge-density-waves [94], which had become one of the most studied topics of condensed matter physics in the ensuing decades. As an extension of the schemes of Little and Ginzburg, in 1973, Allender, Bray, and Bardeen carried out the first analytic analysis on a simple model with a 2D-interface between a metal layer on top of a semiconductor. They showed a possible enhanced pairing interaction between electrons via exciton exchange in the 2D interface, giving a possible $T_c$ up to 800 K, when all stringent requirements of the metal/semiconductor are met [95]. Many experiments to artificially synthesize the metal/semiconductor interfaces were done in the late 1970s without producing a clear exciton-enhanced $T_c$ [96]. Suspecting that the failure might stem from the serious challenges in physically fabricating the ideal interfaces to meet the stringent requirements imposed by the model, we turned to Nature for help by examining some eutectic and immiscible alloy systems where clean and physically coupled interfaces exist. We searched for exciton-induced superconductivity in the immiscible Au/Ge system in the early 1980s [97]. Since neither Au nor Ge is superconducting, the detection of any sign of superconductivity would have provided strong evidence for exciton superconductivity even if the $T_c$ were low. To increase the chance of success, we secured from NASA samples prepared in a microgravity clean environment in Space, thoroughly mixed acoustically, and quenched quickly. Indeed, superconductivity up to 3 K was detected in some of the samples. Unfortunately, the excitement was short-lived when we found out that superconductivity also exists in the metastable Au/Ge alloy during fast quenching on earth. This possibility has been constantly under our watch.

When superconductivity with new record-high $T_c$ was discovered in mixed phase samples first of (La,Ba)-Cu-O [(LB)CO] and then of Y-Ba-Cu-O [YBCO], the thought of interface-enhanced $T_c$ was in our mind, as reflected in some of our early papers [98]. However, after the superconducting phases were identified in these samples, we shifted our attention away from mixed phase systems to pure compound systems with layer structures for possible interfacial enhancement effect on $T_c$. The layer-structure of cuprate superconductors reminds us of the possible self-assembled metal/semiconductor interfaces with the metallic $CuO_2$-layers embedded in the semiconducting environment and raises the question of whether excitons play a role in their high $T_c$. In fact, one of us (Chu) discussed this possibility in YBCO with Bardeen at the International Workshop on Novel Mechanisms of Superconductivity organized by S. Wolf and V. Kresin in Berkeley, California, June 22-26, 1987. Apparently, he and D. Ginsberg also had considerations along these lines at the time. With the rudimentary structural and chemical information about the (LB)CO and YBCO, we ventured to point out "euphorically" that the threefold increase of $T_c$ to 90s K in YBCO from the 30s K in (LB)CO might result from the increase of $CuO_2$-layers per formula (n) to "3" in YBCO from 1 in (LB)CO [although the correct n for YBCO should be 2 as shown later]. We also conjectured to raise the $T_c$ further by increasing n associated with the possible enhanced electron-density of states and arrested instability. Many models proposed seemed to support such a view. Indeed, to raise the $T_c$ by increasing n was one main guide used, although later it was found to work only to n = 3 before $T_c$ starts to decrease with larger n, as shown in Fig. 24.

As mentioned above, almost all early models seemed to suggest that $T_c$ should continue to increase with n beyond 3 [99]. For example, Torradi et al. [100] predicted that room temperature superconductivity could be possible in TBCCO with n = 10 and Grant proposed [101] that a $T_c$ up to 200 K might be possible by adding the density of states of all n $CuO_2$-layers. $CuO_2$-layers have been considered to be the essential and most crucial component of cuprate HTSs and the HTS-problem has been treated as a single $CuO_2$-layer problem. Unfortunately, this cannot account for the initial $T_c$-increase of compounds with n > 1. Inter-layer coupling arising from pair-tunneling between layers has been advanced [102]. The subsequent $T_c$-decrease with n > 3 dashes the hope to raise $T_c$ simply by increasing n.

Di Stasio et al. [103] in 1990 found that the charge carrier distribution across the different $CuO_2$-layers within the unit cell of the compound is not uniform with maximum charge carriers in the center layer(s) by considering only the electrostatic interaction between the layers. They then estimated the effect of this non-homogeneous carrier distribution on the electron density of states of a three-dimensional model and found a nonmonotonical dependence of the density of states on n. By assuming a one-to-one correspondence between the density of states and $T_c$ the experimentally observed $T_c$-n curve was qualitatively obtained with a peak-$T_c$ at an n depending on the specifics of the compound. However, the maximum $T_c$ observed occurs always at n = 3.

A rather extensive and systematic investigation was carried out by Q. M. Lin et al. in 1996 [104] on HBCCO concerning this problem. They found that the inhomogeneous charge distribution issue arises naturally from the basic concept of charge redistribution in a conductor, i.e. charges like to move to its surface to minimize the electrostatic interaction. In a homogeneous good conductor, there is no charge residing in its interior. However, when charge is introduced through doping from both sides of the slab (which itself is not a uniform conductor) of the conducting $CuO_2$-layers in a unit cell of the cuprate HTS, a charge distribution gradient will be established with a minimum in the center layer. It has been shown that there exists an optimal doping to obtain the maximum $T_c$. Nature, therefore, prevents us from achieving optimal doping in all $CuO_2$-layers with $n \geq 2$ and thus maximum $T_c$ in all $CuO_2$-layers throughout the cuprate cell. This is because if the outer (inside) layers are optimally doped, the inside layers must be underdoped, leading to a degradation of $T_c$ due to the proximity effect. In such a scenario, the best case for high $T_c$ would be for $n = 2$ where the two $CuO_2$-layers can be doped optimally at the same time without having any underdoped or overdoped layers to drag down the $T_c$. The fact that the $T_c$ is the highest for $n = 3$ strongly suggests that the electrostatic consideration alone cannot be the whole story and that interlayer-coupling has to play a role. A detailed NMR study was carried out by Kitaoka et al. [105] in 2012 on the layer number effect on the electron-pairing glue in the framework of Mott Physics.

Some, including us, have attempted to raise the $T_c$ in cuprates by increasing n but with no avail. Aside from the non-homogeneous charge distribution from layer to layer, crystal instability also sets in as reflected in the increasing difficulty in forming layered cuprates with higher n. To overcome the instability problem, high pressure has been employed to synthesize samples of n higher than 3. Often, even when they are formed, they are not optimally doped. We believe ingenious doping and micro-material-engineering may help relieve this impasse and help further raise the $T_c$ of cuprates.

Hg1223 holds the record-$T_c$ of 134 K at ambient and 164 K under 30 GPa. The drastic $T_c$-enhancement by pressure is unprecedented and remains unexplained, in spite of many attempts. What is equally intriguing is the parallel and similar $T_c$-pressure behavior for the optimally doped Hg1201, 1212, and 1223. The observation suggests that superconductivity in the three members of the homologous series is of the same origin and the interlayer coupling in them does not vary with n. The recent observation of the anomalous $T_c$ –behavior of Bi2223 under high pressure [57] suggests that pressure may be able to enhance the current record $T_c$ of cuprates.

### 3.2. $T_c$ – $\gamma$ (the Sommerfeld coefficient of specific heat)

In 1975, Sleight et al. [106] found superconductivity in the perovskite compound $Ba(Pb_{1-x}Bi_x)O_3$ (BPBO) system with a $T_c$ above 11 K. This $T_c$ was considered to be unexpectedly high at the time, since the material does not consist of any transition metal elements with a high density of states and yet has a $T_c$ above 10 K. Subsequent experiments failed to reveal other anomalies in the compounds. However, Methfessel et al. in 1980 [107] did not find the typical specific heat anomaly $\Delta C_p$ in the BPBO sample. The BPBO sample was detected magnetically to be a bulk superconductor. The absence of $\Delta C_p$ was unexpected. Several possible causes were proposed for the superconductivity ranging from a minute impurity phase to a novel mechanism to a thermodynamic transition of higher order [107]. Later in 1984, Tanaka et al. [108] did observe a small $\Delta C_p/T$ anomaly around $T_c$. The value of $\Delta C_p/T_c$ was only 3.6% of the $C_p/T_c$ background. The observation suggests that the earlier miss of the $\Delta C_p/T$ by Methfessel et al. was due to the small $\gamma$ of BPBO, since according to the BCS theory $\Delta C_p/T_c \sim 1.43\gamma$. It showed that a thermodynamic transition at a higher order in BPBO was not necessary. In addition, it demonstrates that a low $\gamma$ and a low carrier density as in BPBO need not be a deterrent to high $T_c$, in contrast to conventional wisdom.

Soon after a quality $(La,Sr)_2CuO_4$ [(LS)CO] sample was made, Batlogg et al. in 1987 [109] determined the basic properties of the compound. They found that (LS)CO had a low $\gamma$ in spite of its high $T_c$ like

BPBO. By comparing (LS)CO and BPBO with the conventional low temperature superconductors, they found that the two compounds might distinguish them as a separate class of superconductors as shown in the $T_c$-$\gamma$ plot in Fig. 25. For instance, their $T_c$s are about three times those of other superconductors with comparable $\gamma$'s. As more HTSs are discovered, they fall into the same band of the plot.

This class of superconductors is characterized by high $T_c$ and low $\gamma$. They display a wide range of anomalous properties that defy our understanding of ordinary metals. The questions are "do HTSs form a class of materials of their own?" and "does a low $\gamma$ have deeper implications for all these anomalous properties and high $T_c$ associated with cuprate HTSs?"

3.3. $T_c$ – p (carrier density)

As shown in the generic phase diagram in Fig. 1, cuprate HTSs are known to undergo a universal sequence of transitions with doping or carrier concentration (p) from an antiferromagnetic Mott insulator → superconductor → "normal" metal, demonstrating the strong correlation characteristic of electrons in these compounds. Superconductivity occurs only over a limited p-range $p_{min} < p < p_{max}$. The $T_c$ rises to above 0 K at $p > p_{min}$, peaks at an optimal $p = p_{op}$ with a maximum $T_c = T_c^{max}$ and drops to below 0 K at $p > p_{max}$. Compounds with $p < p_{op}$ are known as underdoped, with $p = p_{op}$ optimally doped and with $p > p_{op}$ overdoped. In an attempt to understand the occurrence of HTS, trends of its evolution with p have been sought. Uemura et al. measured the muon-spin relaxation rate, which is proportional to $p_s/m^*$, where $p_s$ is p in the superconducting state and $m^*$ the effective mass of electrons. Uemura et al. [110] were the first to observe in 1988 a linear $T_c$ – $p_s/m^*$ relation in YBCO and (LS)CO, suggesting a high energy scale for the coupling between superconducting carriers. They later found such a linear $T_c$ – $p_s/m^*$ relation holds not only for the underdoped high $T_c$ cuprate but also for the bismuthate, organic, Chevrel-phase systems, and proposed that this feature distinguished these exotic superconductors from the ordinary BCS superconductors as shown in Fig. 26. One possible common thread for these exotic superconductors may be the strong electron correlation of the compounds. If true, there must be other factors to dictate the $T_c$. This is because the linear relation does not fit for the optimal doped or overdoped cuprates and the Chevrel phase systems. In heavily doped cuprate samples, $T_c$ shows saturation and suppression with increasing doping, which may be attributed to the scattering of the nonsuperconducting carriers [111].

Presland et al. found a universal $T_c/T_c^{max}$ – p relation for cuprates of different families and different numbers of $CuO_2$-layers per formula. $T_c/T_c^{max}$ varies parabolically with p, i.e. $T_c/T_c^{max} = 1- 82.6(p-p_0)^2$ with $p_0 = 0.16$, as shown in Fig. 27 [112], p is defined as the number of carriers per $CuO_2$ and is simply obtained from valence balance or the thermoelectric power at room temperature. In contrast to the relation of Uemura et al., this universal relation fits the overdoped samples, although with a slight deviation [113]. The disagreement may be due to the difference between p (at room temperature) and $p_s$ (near 0 K in the superconducting state), i.e. not all carriers in the compounds participate in the superconducting process.

Based on the p-dependent pressure effect of the cuprate HTSs, a qualitative T-p was proposed by Chu et al. in early 1990s as shown in Fig. 28. A $T_c^{max}$ of ~ 150-160 K [114] was estimated by interpolation for cuprate (50s K for Fe-based superconductors) at $p = p_{op}$ by assuming a rigid electron energy band and a positive pressure effect on p. This seemed to be consistent with the $T_c$-record obtained to date in Hg1223, although detailed studies on HBCCO required modifications under pressures to account for the observation. If this is true, the hope to further raise the $T_c$ by pressure may be limited for layer cuprates. However, for 3D compounds, the situation can be different.

Recognizing the inadequacy of the Uemura relation $T_c$ – $p_s/m^*$ for the optimally doped and overdoped cuprates, Homes et al. in 2004 examined the optical properties of a wide range of superconductors and found the scaling $p_s \propto \sigma_{dc}T_c$, where $p_s$ and $\sigma_{dc}$, the dc conductivity, are determined optically far below and at approximately $T_c$, respectively, as shown in Fig. 29 [115]. The relation appears to have been broadened to hold for all of the superconductors investigated, regardless of $T_c$, doping level, nature of dopant, crystal structure, type of disorder, and the measured direction of $\sigma_{dc}$, suggesting a possible common origin of superconductivity in all material systems. Later study by Tallon et al. [116] pointed out possible limitations and conditions for the relation to hold or to break down.

### 3.4. $T_c$ – $T_{sf}$ (a measure of spin-fluctuations)

The actinide-based Heavy Fermion superconductors (HFSs), $PuCoGa_5$, $PuRhGa_5$, and $NpPd_5Al_2$ are all unconventional superconductors with the highest $T_c$ among all HFSs at 18.5 K, 8.7 K, and 4.9 K, respectively. The high $T_c$s have been attributed to the 5f-electrons, which have been shown to sit at the border between itinerancy and localization, signaling some kind of instabilities, and to have a wider band-width and stronger spin-orbit coupling compared with the 4f-electrons. These unusual characteristics place this group of compounds between the 3d-HTSs and the 4f-HFSs and become the bridge between the two, as clearly demonstrated by the $T_c$ - $T_{sf}$ relation in Fig. 30, where $T_{sf}$ is the characteristic temperature of the spin-fluctuations, extracted from their nuclear quadrupole resonance and heat capacity measurements [117]. Such a $T_c$-$T_{sf}$ seems to suggest that the study of HFSs may provide insight into the underlying working mechanism of HTS and may even lead to higher $T_c$ with better performance.

## 4. Future Prospects of Cuprate HTSs

The future of cuprate HTSs can be vast and bright, but challenging. Future HTSs should possess all or some of the following desirable characteristics to enhance our understanding of HTS science and to expedite the commercialization of HTS devices:

• *higher $T_c$ – by modifying the layered cuprates and discovering new cuprate compounds to facilitate the testing of the limit of some existing theoretical models and the development of the comprehensive microscopic theory, and to make applications more practical*

Although the prediction of $T_c$ is beyond the reach of current theories, some do suggest that the $T_c$ achieved to date is far too low [118]. It has been reasoned [119] that the partially filled 3d-shell of Cu in the HTS cuprates may place them in the class of transition metal oxides. Transition metal oxides display various electronic phase transitions at temperatures exceeding 300 K, driven by the strong electron-electron interactions and strong electron-phonon interactions. If HTS arises from similar interactions as suggested by these models, a $T_c$ much higher than the present record appears not to be impossible. To obtain a $T_c$ in cuprates much higher than the present record will therefore allow one to determine the validity and test the limit of current models. The optimization of the various interactions may hold the key [10].

A higher $T_c$ enables one to operate a HTS device at a higher temperature and thus more efficiently. A HTS device with a higher $T_c$ also has a greater safety margin to avoid accidental quenching than its lower $T_c$ counterpart when operated at the same temperature [120]. The ultimate goal is to attain a $T_c$ above the ambient temperature so that no cryogen or cryocooler is needed. When this is accomplished, HTSs may be put on an equal footing with semiconductors in terms of some consumer electronic and other applications.

• *enhanced pinning and reduced anisotropy – by modifying the charge reservoir block of layer cuprates and discoverying new cuprates to provide a material system to test the current models, to enhance the current carrying capacity, and to simplify the preparation of quality materials and high performance devices*

Cuprate HTSs have a layered structure and display a large anisotropy in many physical properties. The significance of the $CuO_2$-layers in these cuprates cannot be overstated in the current theoretical models of HTS [121]. They have two salient features, namely, the 2D-nature and the unique square-planar atomic arrangement of Cu and O. It is known that in a strictly 2D system, fluctuations will prevent any long-range order, and thus a thermodynamic phase transition, from taking place. Inter-layer coupling between the $CuO_2$-layers or inter-block coupling between the active blocks has to play a role in facilitating the occurrence of the bulk superconducting transition observed in cuprates. The question is to what extent the coupling affects HTS. Studying HTS compounds with different anisotropies under various pressures and different doping will help answer this question.

The large anisotropy of the existing HTS materials puts serious constraints on material processing and device fabrication, since HTS materials must be atomically aligned or in single-crystalline form for devices. A material with a reduced anisotropy will relax such a stringent requirement of atomic alignment for devices and thus greatly simplify the material processing and device fabrication. An additional benefit of such materials with a reduced anisotropy will be the improvement in flux pinning [122] and hence the device performance.

• *improved chemical homogeneity, stability, and structural perfection – by doping and micromaterial engineering the cuprate HTSs to make the easy acquisition and correct interpretation of experimental data possible and to improve the performance and lifetime of the devices*

Cuprate HTS materials have a complex chemistry and a layered structure. As a result, impurity phases, albeit minute, are often present, and intra-layer breakage, inter-layer linkage, and non-uniform oxygen distribution frequently occur. The impurity and the oxygen-deficient phases often generate magnetic signals, which can be confused with the intrinsic magnetism assigned to the $CuO_2$-layers as the origin of HTS by many models [121]. The cross-layer linkage in a sample with an imperfect structure can confound the intra-layer properties with the very different inter-layer properties. The attainment of the intrinsic properties of the HTS materials is indispensable for the development of a comprehensive microscopic HTS theory. This requires pure and uniform samples with perfect structures.

The presence of impurities and structurally defected phases of dimensions greater than the coherence length limits the current-carrying capacity and thus degrades the performance of the HTS material. Since many of the impurity phases easily form carbonates and/or hydroxides and thus act as the degradation centers of the HTS material, they shorten the lifetime of the HTS devices. The loss of oxygen and the relatively high chemical reactivity of some HTS compounds, especially in their imperfect form, can further reduce the lifespan and performance of the HTS devices. Therefore, chemically stable and structurally perfect HTS materials are needed to manufacture commercially viable HTS devices.

• *reduced costs of material processing by modifying the cuprate HTSs and possibly their phase diagrams*

The present high cost of HTS devices originates mainly from material processing and device fabrication. Because of the complex chemistry of formation and the stringent quality requirement of the cuprate HTS material, processing of HTS material requires multiple elaborate steps, usually in a well controlled environment and on a highly demanding substrate [123]. The process is tedious and time consuming, and the substrate is costly. Layered cuprates have to be prepared in their atomically perfect forms for scientific study and device application. Melt-texturing at high temperature and epitaxial growth in a vacuum environment are the two techniques most commonly employed to date: the former limits the substrate material only to the expensive pure or doped Ag and Au, and the latter is commercially unattractive for large-current HTS device fabrication. It has been shown that the HTS/Ag interface lowers the texturing temperature [124] and a less oxidizing or reducing atmosphere suppresses the formation temperature of cuprates [125]. Therefore, by modifying the phase diagram of cuprate phase formation via doping and controlling the processing atmosphere, one can lower the compound formation and texturing temperatures to simplify the material processing and device fabrication. It has been demonstrated [126] that a Ca-overdoped layer of YBCO can improve the superconducting current transporting across the gain-boundaries in the mis-oriented YBCO layer underneath it. This suggests that the stringent requirement on atomic alignment can perhaps be relaxed.

## 5. Conclusion

The lure of superconductivity has continued since its discovery 103 years ago, because of its intellectual challenges to scientists and its technological promises. Its impact on science goes far beyond superconductivity to quarks, neutron stars, superfluid He3, Bose-Einstein condensation, Marjorana fermions, etc. One of the major driving forces has been the never-ending search for higher $T_c$ for science and technology. Cuprate HTSs have played an impressive role in advancing the science, materials, and experimental techniques of HTSs and beyond. Nature has been kinder to physicists than to mountaineers: for the mountaineers the real excitement ceased in 1953 when Mt. Everest was conquered, but for the physicists the search for HTSs with higher $T_c$ and better performance will go on and the excitement will continue.

## Acknowledgments

The work in Houston is supported in part by US Air Force Office of Scientific Research Grant No. FA9550-09-1-0656, the T. L. L. Temple Foundation, the John J. and Rebecca Moores Endowment, and the State of Texas through the Texas Center for Superconductivity at the University of Houston.

Figure Captions

Fig. 1. Typical phase diagram for a hole-doped high-$T_c$ cuprate superconductor. AFI = antiferromagnetic insulator, SG = spin glass, black dots-QCP = quantum critical points, UD = underdoped, OP = optimal doped, OD = overdoped. [*"BCS: 50 years", edited by Leon N Cooper and Dmitri Feldman, World Scientific Press, p. 441(2011).*]

Fig. 2. The schematic layered structure of cuprate HTS. $A_mE_2R_{n-1}Cu_nO_{2n+m+2}$ [A-m(2n-1)n or –02(n - 1)n when m = 0] for m = even (a) and odd (b). [*"Future High $T_c$ Superconductors," C. W. Chu, Chapter 5 in Part G, "Emerging Materials," ed. D. Shaw, in Handbook of Superconducting Materials, ed. D. Cardwell and D. Ginley, Vol. 2: Characterization, Applications and Cryogenics (Bristol: IOP, 2003).*]

Fig. 3. The evolution of $T_c$ with time, i. e. low temperature superconductivity (LTS) and high temperature superconductivity (HTS).

Fig. 4. Crystal structures of a. the T'214 phase ($Nd_{2-x}Ce_x CuO_4$); b. the T214 phase ($La_{2-x}Sr_x CuO_4$); c. the T*214 phase ($Nd_{2-x-z}Ce_z CuO_4$). *[Y. Tokura et al., Nature 337, 261(1989).]*

Fig. 5. ρ(T) of La-Ba-Cu-O shows a $T_c$ of 35 K (La214) in 1986 by Bednorz and Müller. *[ J. G. Bednorz and K. A. Müller, Z. Phys. B 64, 189(1986)].*

Fig. 6. R(T) of La-Ba-Cu-O shows a $T_c$ up to 70 K in 1986 by Chu et al. *[C. W. Chu, AIP Conf. Proc. 169, 220 (1988)]*

Fig. 7. χ(T) of La-Ba-Cu-O showed the first Meissner signal above 77 K on 1/12 disappeared on 1/13 in 1987 by Chu et al. *[C. W. Chu, AIP Conf. Proc. 169, 220 (1988)]*

Fig. 8. ρ(T) and χ(T) of Y-Ba-Cu-O shows the $T_c$ above 93 K (Y123) in 1987 by Wu/Chu et al. [*M. K. Wu et al., Phys. Rev. Lett. 58, 908(1987)*]

Fig. 9. Crystal structure of $YBa_2Cu_3O_{6+x}$.

Fig. 10. Woodstock of Physics in New York City Hilton, March 18, 1987.

Fig. 11. Closing Ceremony for the White House Millennium Time Capsule "Honor the Past – Imagine the Future" that includes a YBCO puck, held at the National Archives in Washington, D.C., December 6, 2000.

Fig. 12. Crystal structure of $Ru_1Sr_2RE_1Cu_2O_8$.

Fig. 13. Magnetization measurements of Ru 1222 (a-c) and Ru1212 (d-f). [*I. Felner et al., Phys. Rev. B 55, R3374(1997); D. Ahmad et al., J. Supercond. Nov. Magn. 27, 1807 (2014).*]

Fig. 14. ρ(T) of Bi-Sr-Ca-Cu-O shows a $T_c$ up to 105 K in 1988 by Maeda et al. *[H. Maeda et al., Jpn. J. Appl. Phys. 27, L209 (1988)].*

Fig. 15. Crystal structures of the Bi phases of general formula $Bi_2Sr_2Ca_{n-1}Cu_nO_y$ with n = 1-4.

Fig. 16. R(T) for a $Tl_2Ba_2Cu_3O_{8+x}$ sample with a $T_c$ up to 90 K in 1988 by Sheng and Herman. *[Z. Z. Sheng and A. M. Hermann, Nature 332, 55 (1988)].*

Fig. 17. ρ(T) of Tl-Ba-Ca-Cu-O shows a $T_c$ up to 120 K (Tl2223) in 1988 by Sheng and Hermann. *[ Z. Z. Sheng and A. M. Hermann, Nature 332, 138(1988*).

Fig. 18. ρ(T) of Hg-Ba-Ca-Cu-O shows a $T_c$ up to 134 K (Hg1223) in 1993 by Schilling et al. *[A. Schilling et al., Nature 363, 56 (1993)].*

Fig. 19. Crystal structures of the Hg phases of general formula $HgBa_2Ca_{n-1}Cu_nO_y$ with n=1-5.

Fig. 20. ρ(T) of Hg12(n-1)n for n = 1, 2, 3 under pressures up to 45 GPa shows a $T_c$ up to 164 K in 1993 by Gao et al. [*L. Gao et al., Phys. Rev. B 50, 4260(R)(1994)].*

Fig. 21. Infinite-layer structure.

Fig. 22. a. M/H vs. T and b. R(T) for a $(Ca_{0.3}Sr_{0.7})_{0.9}CuO_2$ shows a $T_c$ up to 110 K in 1992 by M. Azuma et al. *[M. Azuma et al., Nature 356, 775 (1992)].*

Fig. 23. χ(T) of $Ba_2Ca_{n-1}Cu_nO_x$ shows a $T_c$ up to 126 K in 1997 by Chu et al. *[ C. W. Chu et al., Science 277, 1081(1997)].*

Fig 24. $T_c$ as a function of the number of $CuO_2$ layers for different superconductor families. [See Table. I]

Fig 25. $T_c$ vs. $\Upsilon$ for various superconductors. The experimental quantity $\Upsilon$ in heat capacity measurements is proportional to the electronic density of states. [*B. Batlogg et al., Phys. Rev. B 35, 5340(R)(1987); B. Batlogg, Solid State Commun. 107, 639(1988); J. S. Urbach et al., Phys. Rev. B 39, 12391(R) (1989); R. A. FISHER et al., Physica C 162,503(1989); A. JUNOD et al., Physica B 165&166, 1335 (1990); B. F. Woodfield et al., Physica C 235-240,1741 (1994); R. A. Fisher et al., Phys. Rev. B 38, 11942(R)(1988).*]

Fig. 26. Plot of $T_c$ vs σ (T→0) □ $n_s/m^*$ of cuprates, BKBO, Chevrel-phase, BEDT, Nb and HF systems (inset). The straight line in the inset corresponds to the linear relation found for the cuprates. *[Y. J. Uemura et al., Phys. Rev. Lett. 66, 2665(1992)].*

Fig. 27. Schematic phase diagram for cuprate superconductors showing the parabolic superconducting domain for $La_{2-x}Sr_xCu0_4$. [*M.R. Presland et al., Physica C 176, 95(1991)*]

Fig. 28. Schematics of the universal dependence of $T_c$ on pressure proposed in 1991 by Chu et al. *[J.G. Lin et al., Physica C 175, 627(1991)].*

Fig. 29. Plot of the superfluid density ($\rho_s$) versus the product of the d.c. conductivity ($\sigma_{dc}$) and $T_c$ for copper oxides. [*C. C. Homes et al., Nature 430, 539(2004)*]

Fig. 30. $T_c$-$T_0$ where $T_0$ measures the strength of spin-fluctuations. [*C. Pleiderer, Rev. Mod. Phys. 81, 1551(2009)].*

**Table I**

| Compounds | Symmetry | symbol | No. of $CuO_2$ planes | Max $T_c$ (ambient pressure) | Max $T_c$ (high pressure) | Year of discovery | Ref. |
|---|---|---|---|---|---|---|---|
| ***LMCO-214 type*** | | | | | | | |
| $(La,M)_2CuO_4$ | T | 0201 | 1 | 36.2K | 52.5K | 1986 | 1-4 |
| $La_2CuO_{4+x}$ | O | 0201 | 1 | 45K | ~48K | 1988 | 5-8 |
| $La_2CuO_4F_x$ | T | 0201 | 1 | ~35K | -- | 1988 | 9 |
| $(Nd,Ce)_2CuO_4$ | T | 0201 | 1 | ~25K | ~25(-$dT_c/dP$) | 1989 | 10-12 |
| (Nd, Ce, Sr)$_2CuO_4$ | T | | | 28K | -- | 1988 | 13-14 |
| $Sr_2CuO_{3+x}$ | T | 0201 | 1 | 94K | --- | 1993 | 15-16 |
| $Sr_{2-x}Ba_xCuO_{3+\delta}$ | T | 0201 | 1 | 98K | -- | 2009 | 17 |
| $(A,A')_2CuO_{2-x}B_{2+x}$: A=Ca, Sr, A'=Na; B=Cl or F | T | 0201 | 1 | ~46K | -$dT_c/dP$ | 1994 | 18-20 |
| | | | | | | | |
| ***YBCO-related type*** | | | | | | | |
| $YBa_2Cu_3O_{7-x}$ | O | 1212 | 2 | 93K | 93K | 1987 | 21-22 |
| $RBa_2Cu_3O_{7-x}$ (R= La, Nd, Sm, Eu, Gd, Ho, Er, and Lu) | O | 1212 | 2 | 93K | 93K | 1987 | 23 |
| $YBa_2Cu_4O_8$ | O | 2212 | 2 | 81 | 108 | 1988 | 24-26 |
| $Y_2Ba_4Cu_7O_{15-\delta}$ | O | 1212 +2212 | 2 | 92? | -- | 1988 | 27-29 |
| $YSr_2Cu_3O_7$ (under high oxygen pressure) | T | 1212 | 2 | 60K | -- | 1988 | 30-32 |
| $(Cu,M)Sr_2(Y,Ca)Cu_2O_7$ M stabilized Sr 123 M = Li, Al, Ga, Fe, Co, Ti, Ge, V, Cr, Mo, W, Re, Pb, B, $SO_4$, $CO_3$, $PO_4$, (Bi + Cd) | T | 1212 | 2 | ~73K | +$dT_c/dP$ | 1994 | 33-37 |
| $LaBaCaCu_3O_7$ | T | 1212 | 2 | ~78K | | 1988 | 38-39 |
| $(La,Sr)_2CaCu_2O_6$ | T | 0212 | 2 | 60 | | 1990 | 40 |
| $(Ca,Na)_2CaCu_2O_4Cl_2$ | T | 0212 | 2 | 49K | | 1995 | 41-42 |
| $(Sr,Ca)_3Cu_2O_{4+\delta}Cl_{2-y}$ | T | 0212 | 2 | 80K | | 1995 | 43 |
| $NbSr_2(Nd,Ce)_2Cu_2O_{10}$ | T | 1222 | 2 | 28K | | 1992 | 44-45 |
| $GaSr_2(Y,Ca)Cu_2O_7$ | O | 1212 | 2 | 70K | | 1991 | 46-48 |
| $PbBaSr(Y,Ca)Cu_3O_8$ | T | | | 65K | | 1991 | 48-51 |
| $Ru_1Sr_2Gd_1Cu_2O_8$ | T | 1212 | 2 | 26K | | 1995 | 52, 53 |
| $RuSr_2(Ln_{1+x}Ce_{1-x})Cu_2O_{10}$ (Ln=Nd, Sm, Eu and Gd) | T | 1222 | 2 | 40K | | 1995 | 52, 54 |
| | | | | | | | |
| ***Bi, Pb-based type*** | | | | | | | |
| $Bi_2Sr_2Ca_{n-1}Cu_nO_{2n+4}$ (n = 1, 2, 3, 4): | | | | | | | |
| $Bi_2Sr_2CuO_6$ | T | 2201 | 1 | <10 | | 1987 | 55-57 |
| $Bi_2Sr_2CaCu_2O_8$ | T | 2212 | 2 | ~84 | + | 1988 | 58-62 |
| $Bi_2Sr_2Ca_2Cu_3O_{10}$ | T | 2223 | 3 | ~110 | 135 | 1988 | 57, 63, 64 |
| $Bi_2Sr_2Ca_3Cu_4O_{12}$ | T | 2234 | 4 | 90 | | 1988 | 65 |
| | | | | | | | |
| $Bi_2Sr_2(Ln,Ce)_2Cu_2O_{10}$ | T | 2222 | 2 | 34 | | 1990 | 66 |
| $(Bi,M)Sr_2(R,Ce)Cu_2O_y$ (M=Cu, Cd; R=Nd, Gd) | T | 1212 | 2 | ~68 | | 1992 | 67 |
| $(Bi,Sr)_2YCu_2O_{6+x}$ | T | 0212 | 2 | 75 | | 1994 | 68 |
| $(Bi,Cu)Sr_2(Ln,Ce)_2Cu_2O_z$ | T | 1222 | 2 | | | 1992 | 69 |
| | | | | | | | |
| $(Bi_2Sr_2CuO_6)_n(Sr_2CuO_2CO_3)_{n'}$ (n=1, n'=1 or 2) | T | Intergrowth | 1 | 40 | | 1993 | 70 |
| | | | | | | | |
| $Pb_2Sr_2(Ln,Ca)Cu_3O_8$ | O | 3212 | 2 | 70 | | 1988 | 71 |
| $Pb_2(Sr,La)_2Cu_2O_6$ | O | 3201 | 1 | 26 | | 1989 | 72 |
| $(Pb,Cu)\ Sr_2(Y,Ca)Cu_2O_x$ | O | 1212 | 2 | ~60 | | 1990 | 73, 74 |
| $PbSr_2Ca_2Cu_3O_x$ | T | 1223 | 3 | 115 | | 1995 | 75 |
| $PbSr_2CuO_{5+\delta}$ | T | 1201 | 1 | 40 | | 1999 | 76 |
| $Pb_2Sr_2Ce_2Cu_3O_{10+Y}$ | T | 3222 | 2 | <24 | | 1990 | 77 |
| $Pb(Sr,La)_2Ln_2Cu_2O_z$ | T | 1222 | 2 | <30 | | 1995 | 78 |
| | | | | | | | |
| ***Tl-based type*** | | | | | | | |
| $Tl_2Ba_2Ca_{n-1}Cu_nO_{2n+4}$ (n = 1, 2, 3, 4): | | | | | | | |
| $Tl_2Ba_2CuO_6$ | T | 2201 | 1 | 90 | | 1987 | 79, 56 |
| $Tl_2Ba_2CaCu_2O_8$ | T | 2212 | 2 | 110 | | 1988 | 80-83 |
| $Tl_2Ba_2Ca_2Cu_3O_{10}$ | T | 2223 | 3 | 125 | 131 | 1988 | 80-85 |
| $Tl_2Ba_2Ca_3Cu_4O_{12}$ | T | 2234 | 4 | 104 | | 1988 | 86 |
| | | | | | | | |
| $TlBa_2Ca_{n-1}Cu_nO_{2n+3}$ (n = 1, 2, 3,4): | | | | | | | |
| $TlBa_2Cu_1O_5$ | T | 1201 | 1 | <40 | | 1988 | 87, 88 |
| $TlBa_2\ Ca_1Cu_2O_7$ | T | 1212 | 2 | 82 | | 1988 | 87 |
| $TlBa_2\ Ca_2Cu_3O_9$ | T | 1223 | 3 | 116 | | 1988 | 87 |
| $TlBa_2\ Ca_3Cu_4O_{11}$ | T | 1234 | 4 | 122 | | 1988 | 89 |
| $TlBa_2\ Ca_4Cu_5O_{13}$ | T | 1245 | 5 | <100 | | 1989 | 90 |
| | | | | | | | |
| $TlBa_2(Eu,Ce)_2Cu_2O_9$ | T | 1222 | 2 | <40 | | 1992 | 91 |
| | | | | | | | |
| ***Hg-based type*** | | | | | | | |
| $HgBa_2Ca_{n-1}Cu_nO_{2n+2}$ (n = 1, 2, 3, 4, 5, 6, 7): | | | | | | | |
| $HgBa_2CuO_{4+x}$ | T | 1201 | 1 | 95 | 118 | 1993 | 92, 95 |
| $HgBa_2Ca_1Cu_2O_{6+x}$ | T | 1212 | 2 | 114 | 154 | 1993 | 93, 95 |
| $HgBa_2Ca_2Cu_3O_{8+x}$ | T | 1223 | 3 | 133 | 164 | 1993 | 93-95 |
| $HgBa_2Ca_3Cu_4O_{10+x}$ | T | 1234 | 4 | 125 | 143 | 1993 | 96-97 |
| $HgBa_2Ca_4Cu_5O_{12+x}$ | T | 1245 | 5 | 110 | | 1994 | 98-99 |
| $HgBa_2Ca_5Cu_6O_{14+x}$ | T | 1256 | 6 | 107 | | 1994 | 98-99 |
| $HgBa_2Ca_6Cu_7O_{16+\delta}$ | T | 1267 | 7 | <90 | | 1994 | 98 |
| $Hg_2Ba_2(Y,Ca)Cu_2O_8$ | T | 2212 | 2 | ~70 | | 1994 | 100 |
| $(Hg_{0.7}W_{0.3})Sr_2(Ce_{0.58}Eu_{0.42})_3 Cu_2O_{11+\delta}$ | T | 1232 | 2 | 33 | | 2004 | 101 |
| | | | | | | | |
| ***Infinite layer and related*** | | | | | | | |
| $Sr_{1-x}Nd_xCuO_2$ | T | 0011 | infinite | 40 | | 1991 | 102-104 |
| $Sr_{1-x}Ba_xCuO_2$ | T | 0011 | infinite | 90 | | 1992 | 105 |
| $Sr_{1-x}CuO_2$ | T | 0011 | infinite | 110 | | 1992 | 106 |
| $Sr_{1-x}Ca_xCuO_2$ | T | | infinite | ~100 | | 1993 | 107 |
| $Sr_{n+1}Cu_nO_{2n+1+\delta}$: ($n$ = 1, 2, 3,...) | T | | n | 100 | | 1993 | 108 |
| $Ca_{13.5}Sr_{0.5}Cu_{24}O_{41}$ (6 MPa applied pressure only) | T | | | ~12K | | 1996 | 109-111 |
| | | | | | | | |
| ***Other distinct types*** | | | | | | | |
| $Sr_2Ca_{n-1}Cu_nO_{2n+\delta}F_{2\pm y}$ ($n$ = 2; $T_c$ = 99 K, $n$ = 3; $T_c$ = 111 K) | T | n | n | 111K | | 1996 | 112 |
| $(Sr,Ca)_2(Sr,Ca)_{n-1}Cu_nO_x$ (n = 2, 3, 4) | T | 0212<br>0223<br>0234 | n | ~85 | | 1993 | 113-114 |
| $CuBa_2Ca_{n-1}Cu_nO_{2n+2+x}$: | T | 12(n-1)n | n | 117 | | 1994 | 115-116 |
| $GaSr_2Ca_{n-1}Cu_nO_x$ (n = 3, 4) | O | 12(n-1)n | n | 107K | | 1994 | 48 |
| $(Cu_{1-x}(CO_2)_x)_m(Ba,Sr)_2Ca_{n-1}Cu_nO_y$<br>m = 1; n = 2, 3, 4, 5; x ≠ 0 or x = 0<br>m = 2; n = 3, 4, 5; x ≠ 0 | T | m2(n-1)n | n | 117 | | 1992 | 117-127 |
| $Ba_2Ca_{n-1}Cu_n(Ca_xCu_y)O_{8+\delta}$ | T | 02(n-1)n | n | 126 | | 1996 | 128-132 |
| | | | | | | | |
| | | | | | | | |

Table I: Classification of different hole-doped cuprates by family, including years of discovery and representative milestone work with references.

Ref:

**Table II**

| Compounds | Coherence length ab; c (Å) | Penetration depth ab;c(nm) | Gap ratio | $\Delta C/T_c$ ($mJ/K^2$ mol) | $H_{c1}$ ab;c (mT) | $H_{c2}$ ab;c (T) | Ref. |
|---|---|---|---|---|---|---|---|
| ***LMCO-214 type*** | | | | | | | |
| $(La,Sr)_2CuO_4$ | 32; 0.55-3<br>32-38;- | 254-440;- | 8.9<br>4.3 | 5.2-10.2<br>1.75-17.5 | 7;30 | 64;- | 1-8 |
| $(Nd,Ce)_2CuO_4$ | 70.2; 3.4<br>72-80;- | 360;-<br>130;- | 3.1<br>4.1 | | | -;6.2 | 9-12 |
| ***YBCO-related type*** | | | | | | | |
| $YBa_2Cu_3O_{7-x}$ | 16;3<br>15.4-32.3;- | 1360;318<br>140-240;- | 13 | 30-44<br>40-60 | 7;21<br>-;12-37 | 674;122<br>-;32-140 | 12-16 |
| $EuBa_2Cu_3O_{7-x}$ | 27;6<br>35;3.8 | -;170 | | | | 190;45<br>245;28 | 17-19 |
| $YBa_2Cu_4O_{8-x}$ | 19.5;- | 196;-<br>198;- | | 16 | -;19.7 | -;87 | 20-22 |
| $RuSr_2GdCu_2O_8$ | 5;5 | 340;2400 | 2 | 55.1 | 0.4;2.8 | 30;-<br>133;133 | 23-25 |
| $RuSr_2(Gd,Ce)Cu_2O_{10}$ | 140;28 | | 4-7 | 66.8 | | 39;8 | 24,26,27 |
| $Y_2Ba_4Cu_7O_{15-x}$ | | | | 42 | | | 28 |
| ***Bi-based type*** | | | | | | | |
| $Bi_2Sr_2CuO_6$ | 55;32<br>35-45;15 | | 5.4-7.4 | | | 17;10<br>16-27;43<br>22;6.6 | 29-32 |
| $Bi_2Sr_2CaCu_2O_8$ | 27;0.45<br>19;-<br>21.5;2.8 | 250;-<br>-;300<br>210;- | 10-12<br>9 | 3.9<br>20 | -;7<br>6.3;-<br>19;- | 90;-<br>542;71.5 | 8,29,<br>33-39 |
| $Bi_2Sr_2Ca_2Cu_3O_{10}$ | 29;0.93<br>9.7;0.2 | 200;1000 | 9-11 | 12 | | -;57.9 | 29,36,38<br>40-42 |
| ***Pb-based type*** | | | | | | | |
| $Pb_2Sr_2(Ca,Y)Cu_3O_{8+x}$ | 18.5;3 | 257.5;642.5 | | | 9.5;50.5 | 590;96 | 43,44 |
| $(Bi,Pb)_2Sr_2CaCu_2O_8$ | 19;20 | 178;- | | | | -;89 | 45 |
| $(Bi,Pb)_2Sr_2Ca_2Cu_3O_{10}$ | -;2<br>13;- | 232;- | | 16<br>37 | | 39;198 | 46-50 |
| $(Bi,Pb)_2(Sr,La)_2CuO_{6+x}$ | 26;- | 358;- | | | | | 51 |
| ***Tl-based type*** | | | | | | | |
| $TlBa_2Ca_2Cu_3O_9$ | 14;1 | 150;- | | | | | 40 |
| $TlBa_2Ca_3Cu_4O_{11}$ | 12.6;- | 156;- | | | -;1020 | -;207 | 52 |
| $Tl_2Ba_2CuO_6$ | 50;2<br>52;3 | 72;2000<br>170;- | 3.0-5.6 | 0.4 | -;6 | | 34,36,<br>53-55 |
| $Tl_2Ba_2CaCu_2O_8$ | 31;6.8 | 125;-<br>221;- | 8.0 | 35 | 60;- | 99;- | 36, 56-60 |
| $Tl_2Ba_2Ca_2Cu_3O_{10}$ | 12.4;-<br>28;0.8 | 103;3250<br>210;-<br>196;- | 4.6 | 20 | -;18.6 | 760-1400;<br>42-69<br>-;212 | 57,59<br>61-64 |
| $Tl_2Ba_2Ca_3Cu_4O_{12}$ | 45;10 | | | | | | 36 |
| (Tl,Pb)$Sr_2Ca_2Cu_3O_9$ | 11.7;- | 180;-<br>158;- | | | -;23.1 | -;240 | 8,62 |
| ***Hg-based type*** | | | | | | | |
| $HgBa_2CuO_{4+x}$ | 21.1;12.1<br>20;-<br>21;- | 260;454<br>145;-<br>117;- | 7.9<br>6.4 | | 8.2;12.9 | 125;72<br>-;100 | 65-69 |
| $HgBa_2CaCu_2O_{6+x}$ | 16.6;4.0<br>13.9;- | 205;825<br>191;-<br>166.5;- | 9.5 | | 20.8;50 | 455;113<br>-;170 | 58,60-62 |
| $HgBa_2Ca_2Cu_3O_{8+x}$ | 4.4;19.3<br>12;-<br>-;1.5<br>1.9;15<br>18;-<br>13;- | 154;680<br>174;-<br>210;6100<br>1300;170<br>206;-<br>165;2805 | 13 | 57 | 10;33.9<br>-;45<br>5;28<br>-;29 | 390;88<br>-;201.6<br>1150;149<br>-;108<br>-;190 | 65,67,68,<br>70-74 |
| $HgBa_2Ca_3Cu_4O_{10+x}$ | 12.7;- | 157;- | | | | -;205 | 52 |
| $HgBa_2Ca_4Cu_5O_{12+x}$ | 17.2;- | 170.7;- | | | | -;111.2 | 68 |
| $HgBa_2$(Ca,Sr)$Cu_2O_{6-\delta}$ | 13.9;2 | 191.3;- | | | | -;170.4 | 75 |
| (Hg,Pb)$Ba_2Ca_2Cu_2O_y$ | 16;- | 191;- | | | | -;128 | 76 |
| (Hg,Pb)$Sr_2Ca_2Cu_2O_y$ | 12;- | 169;- | | | | -;226 | 76 |
| (Hg,Pb)$(Ba,Sr)_2Ca_2Cu_3O_y$ | 15;- | 183;- | | | | -;145 | 77 |
| ***Infinite layer and related*** | | | | | | | |
| $Sr_{1-x}Nd_xCuO_2$ | | | | | 1.5;- | | 78 |

**Table II: Representative physical parameters for selected hole-doped cuprates with references.**

Ref.

**Figure 1**

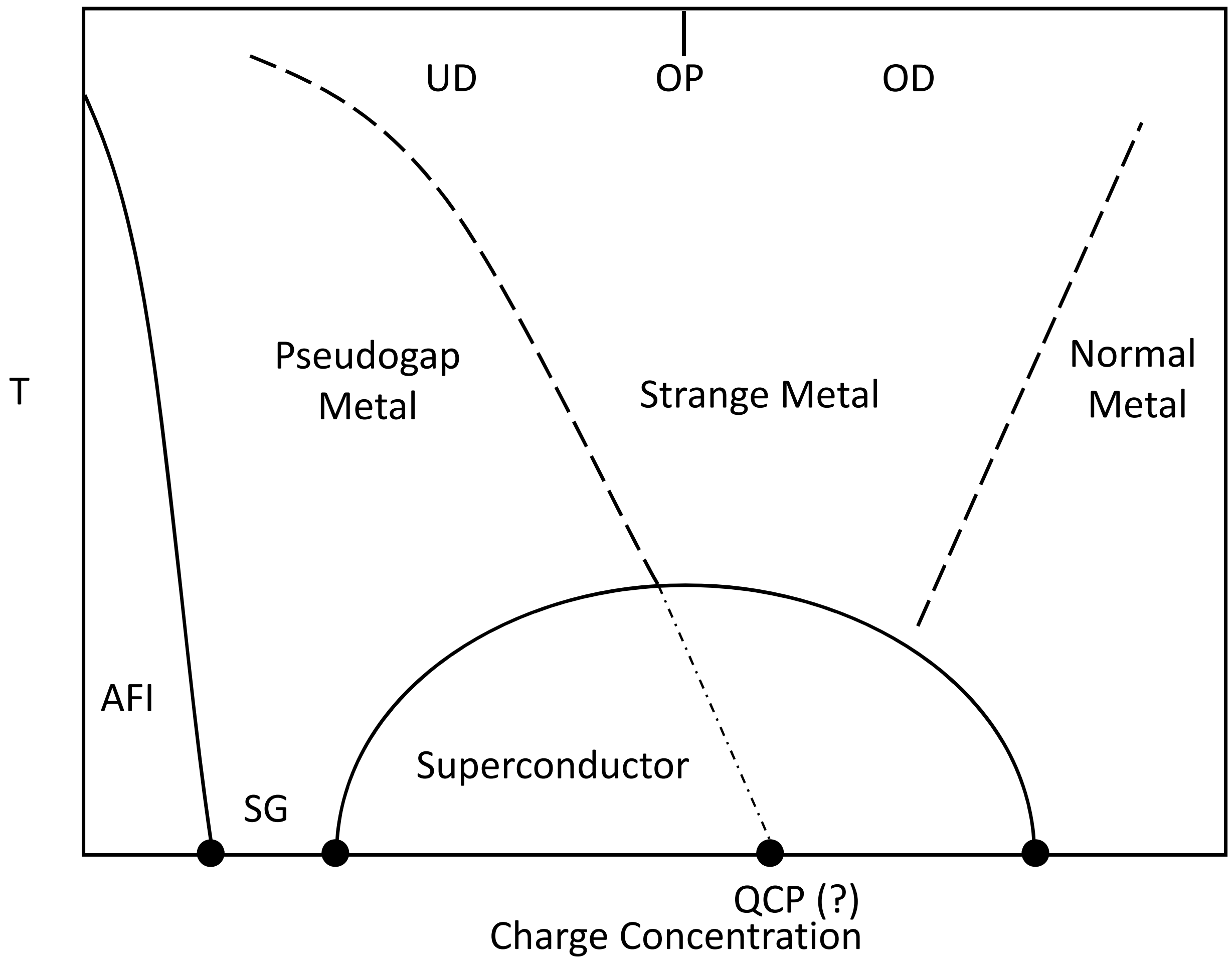

Figure 2

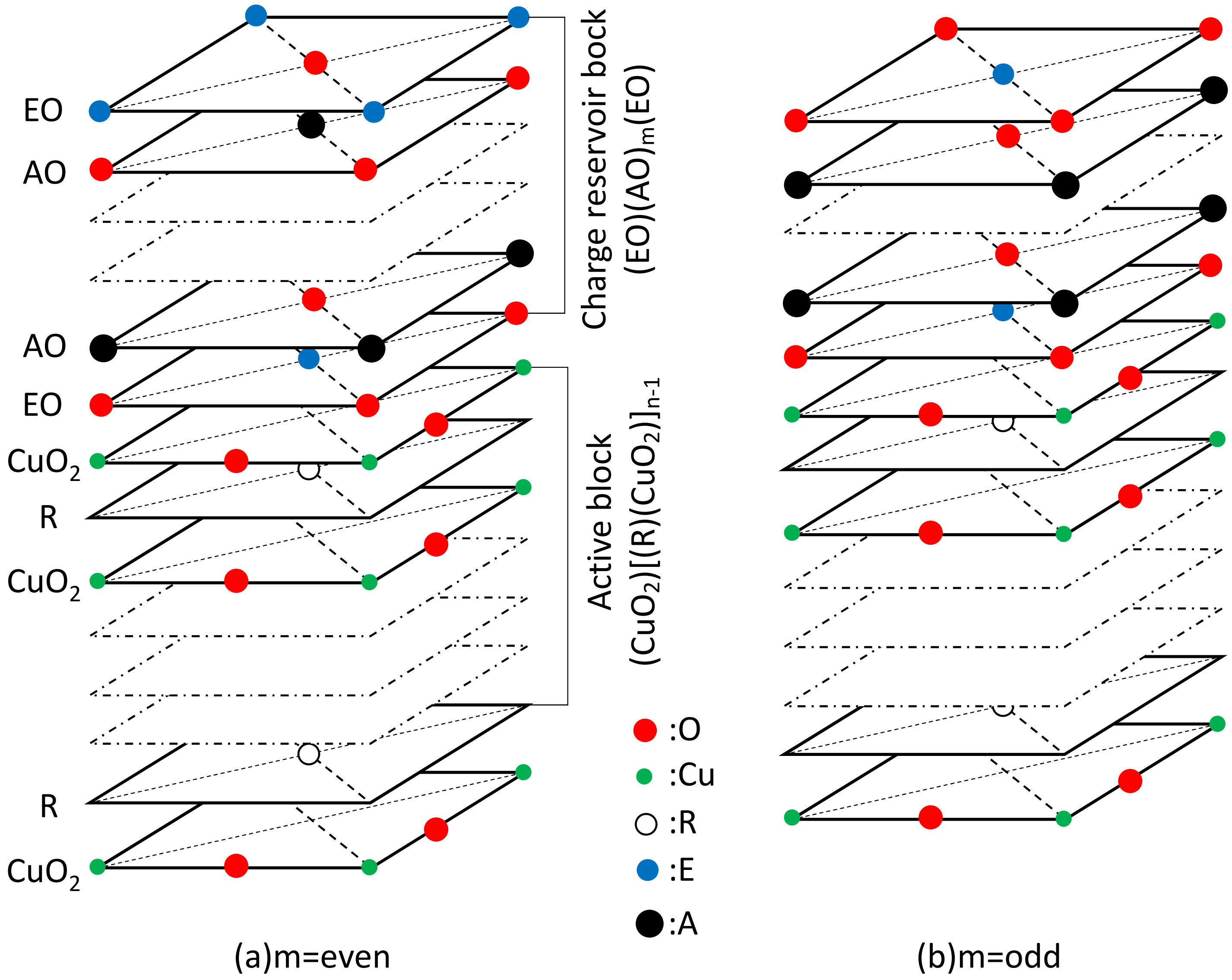

Figure 3

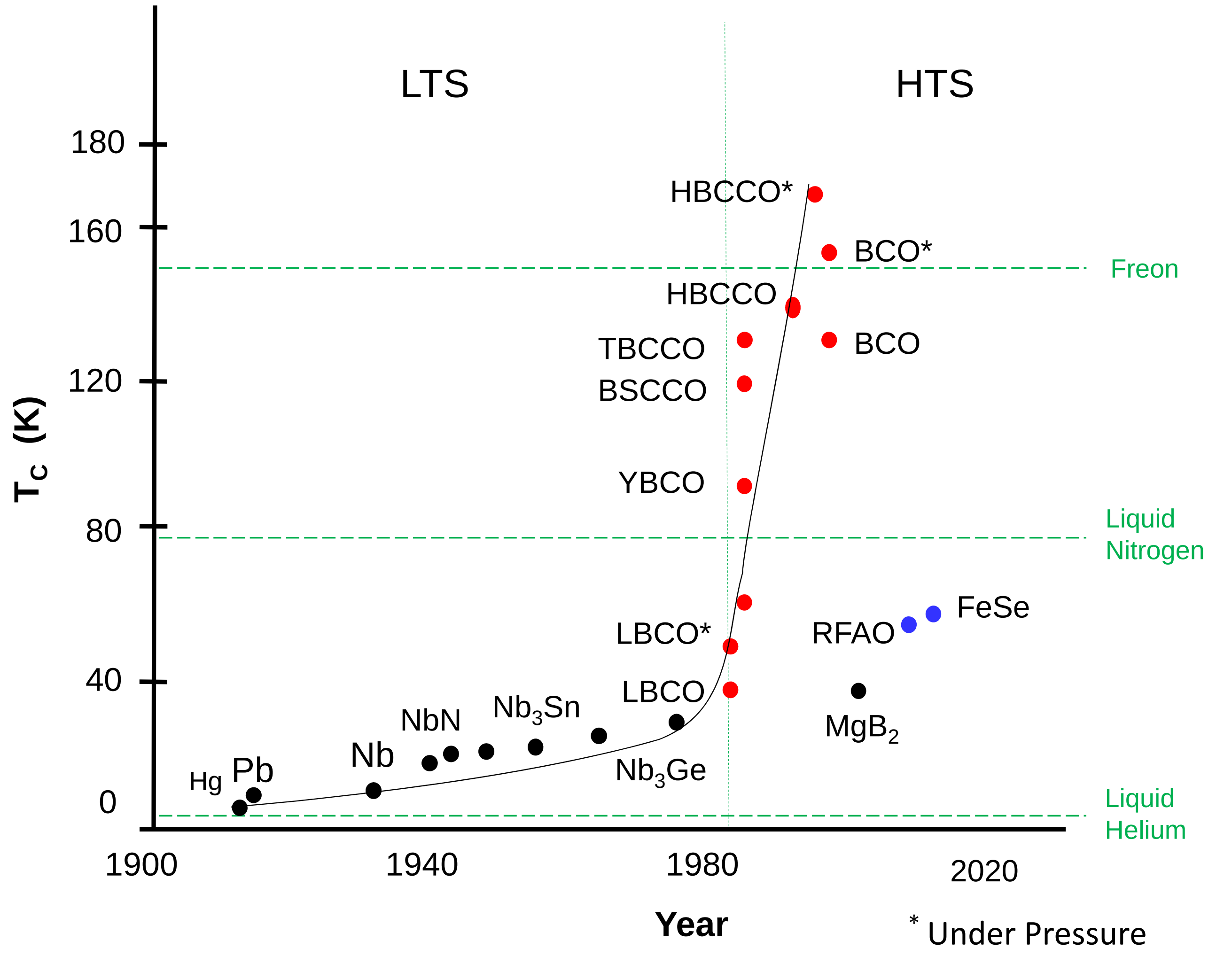

**Figure 4**

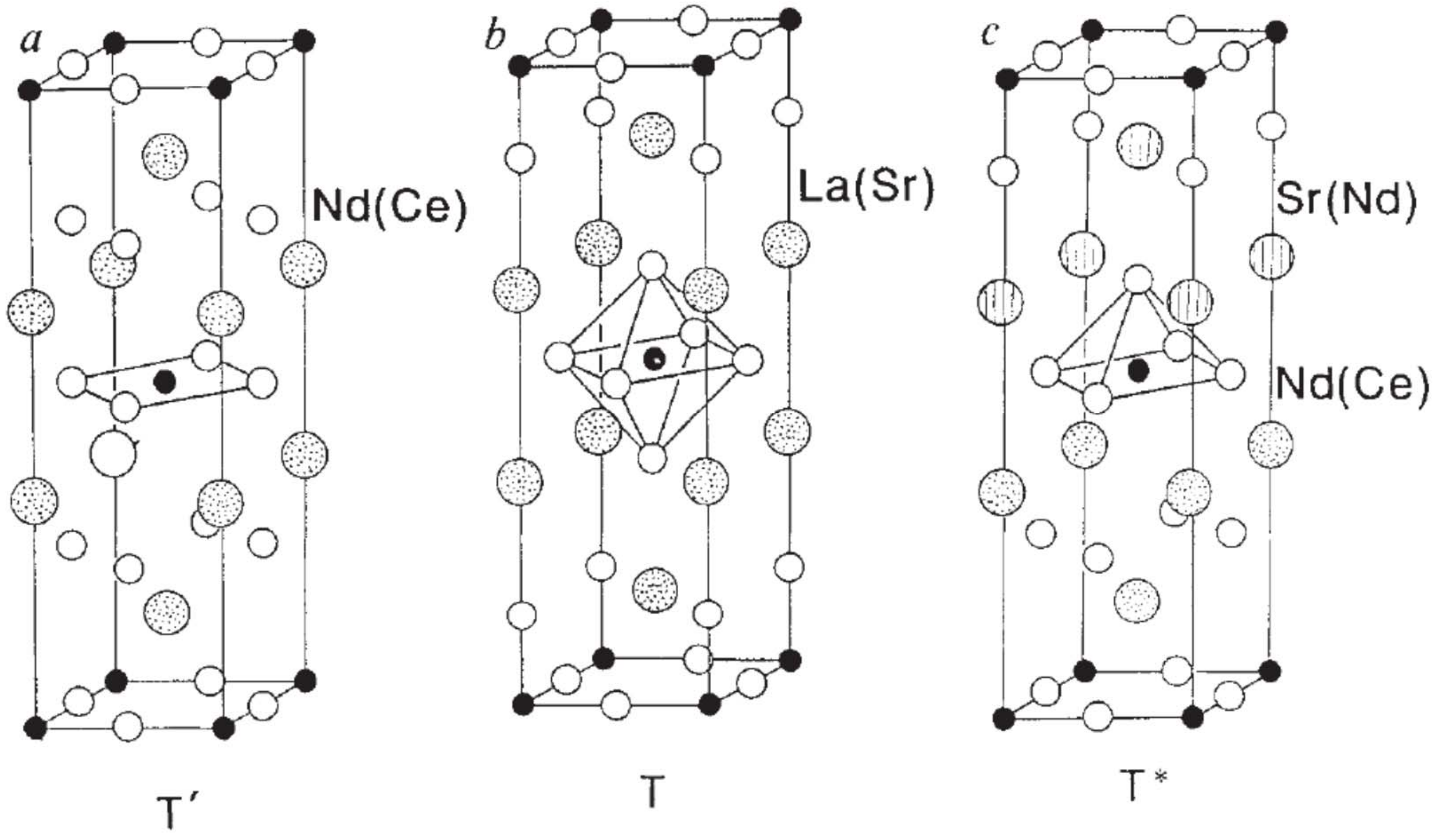

Figure 5

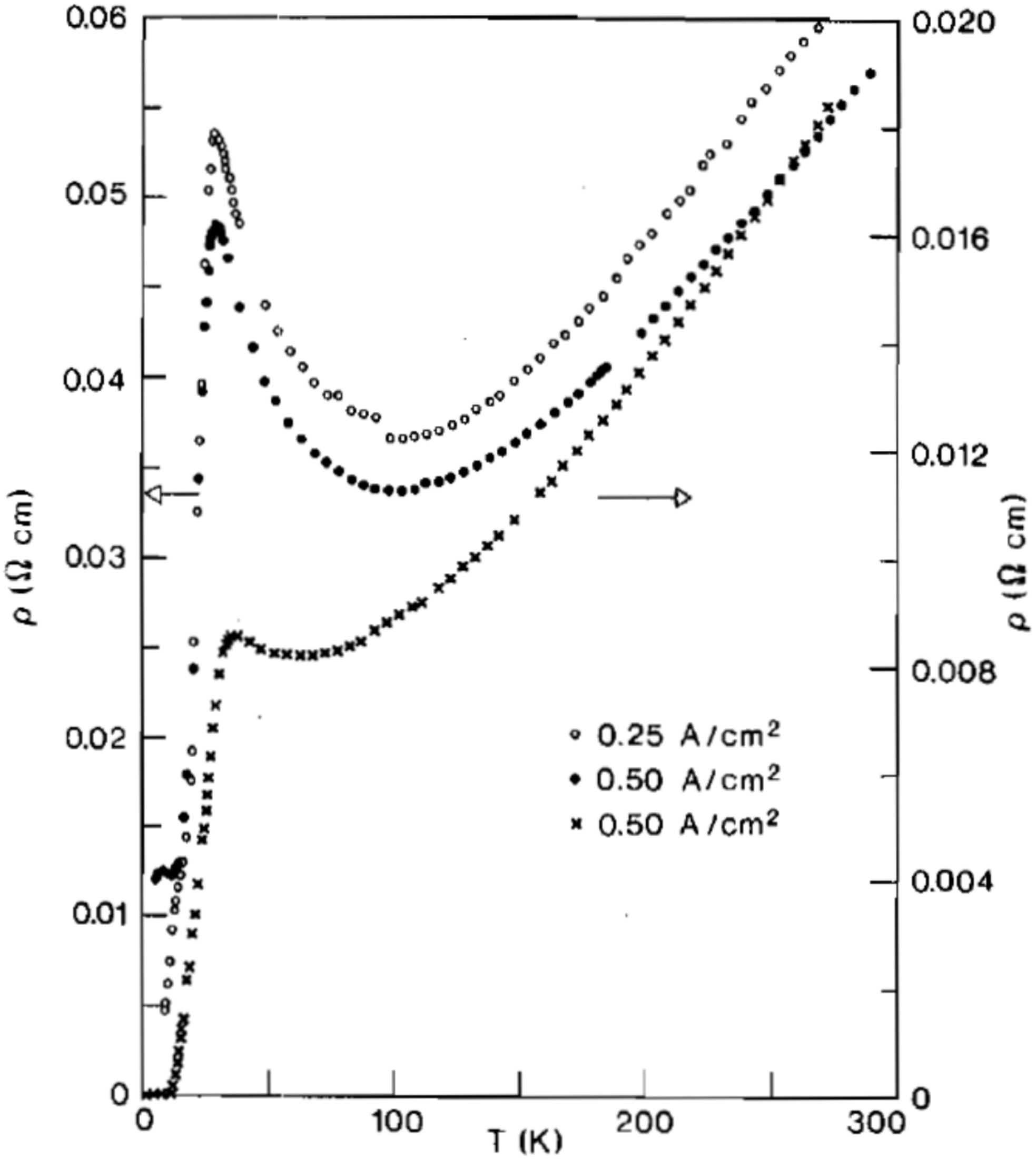

Figure 6

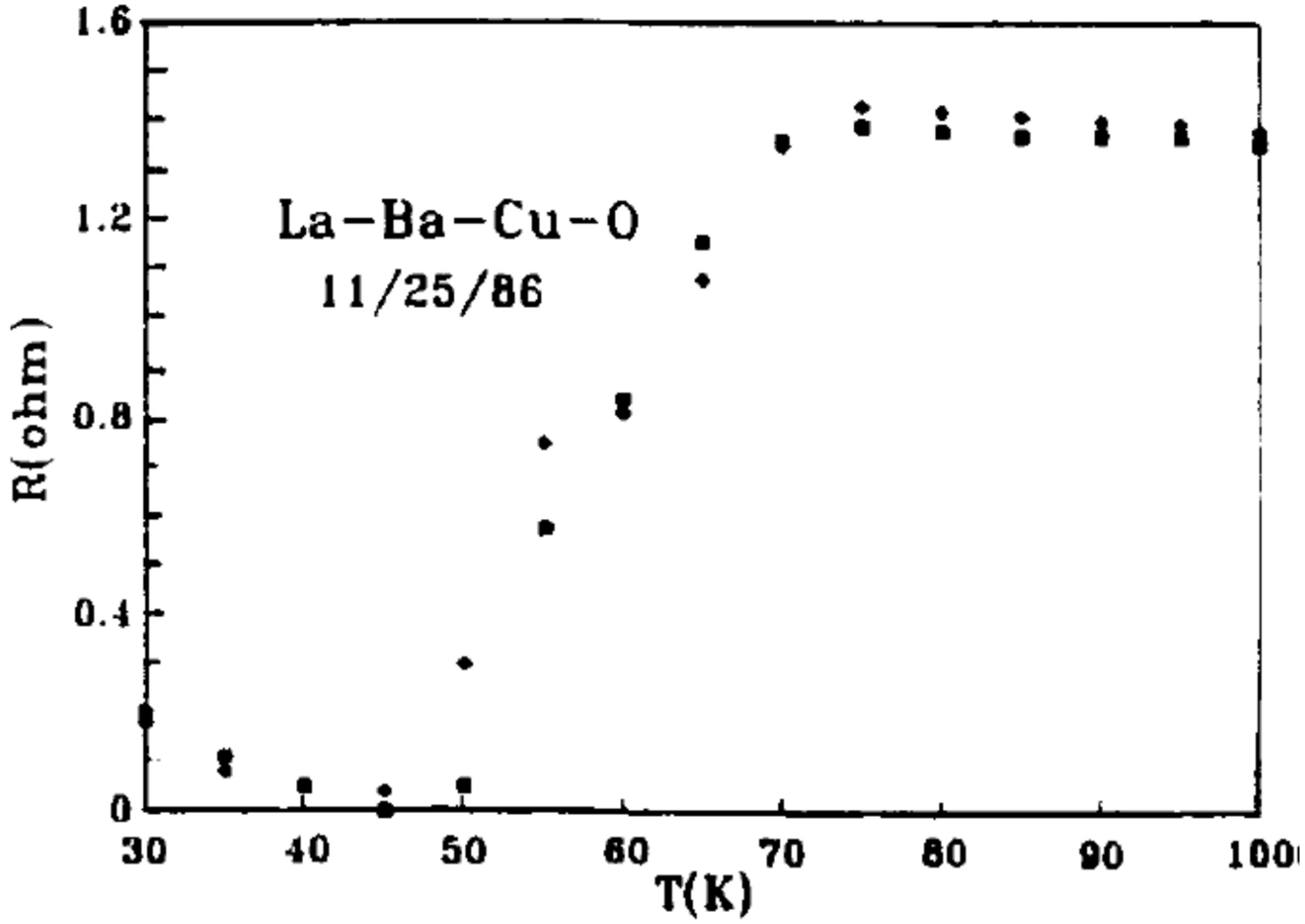

Figure 7

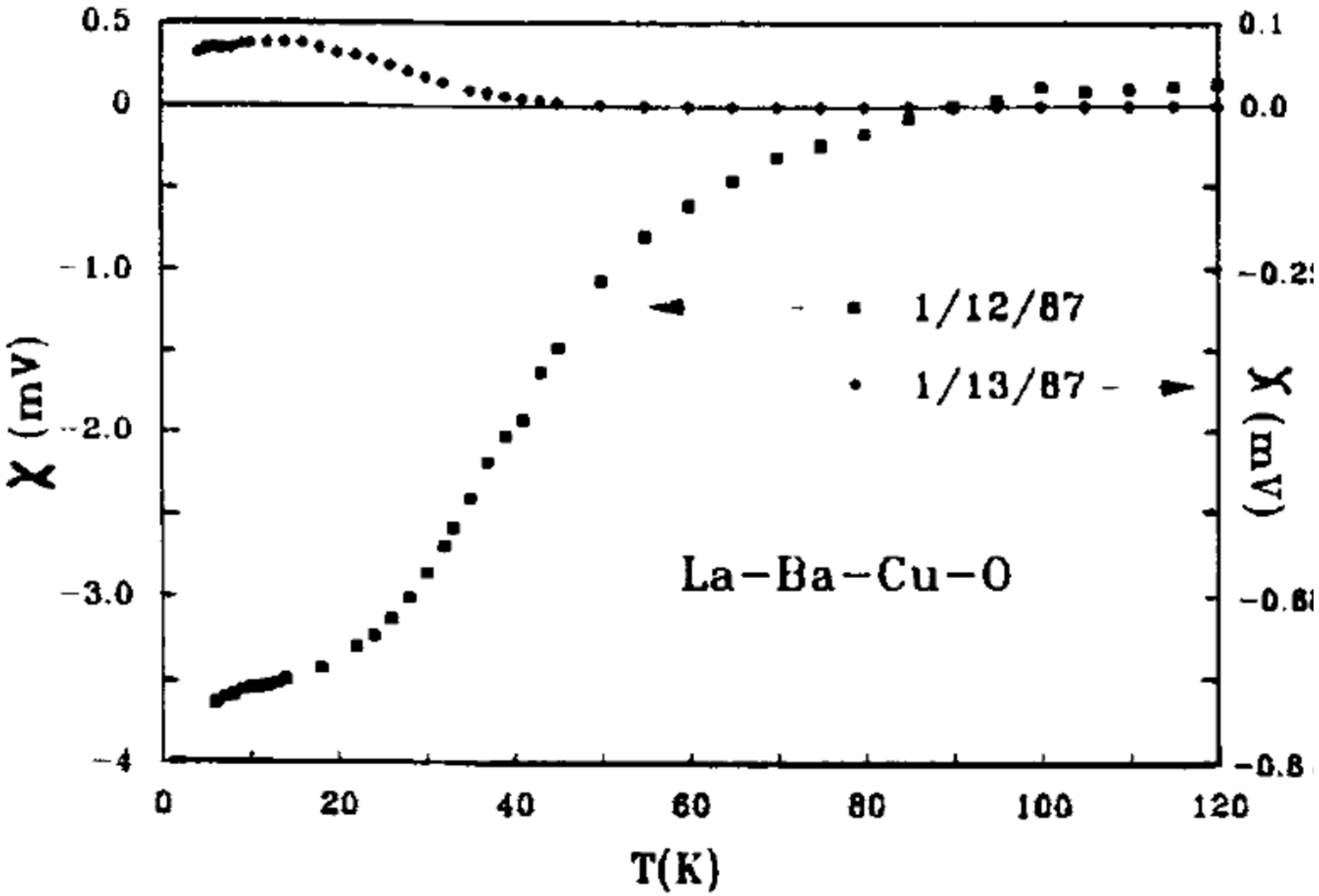

Figure 8

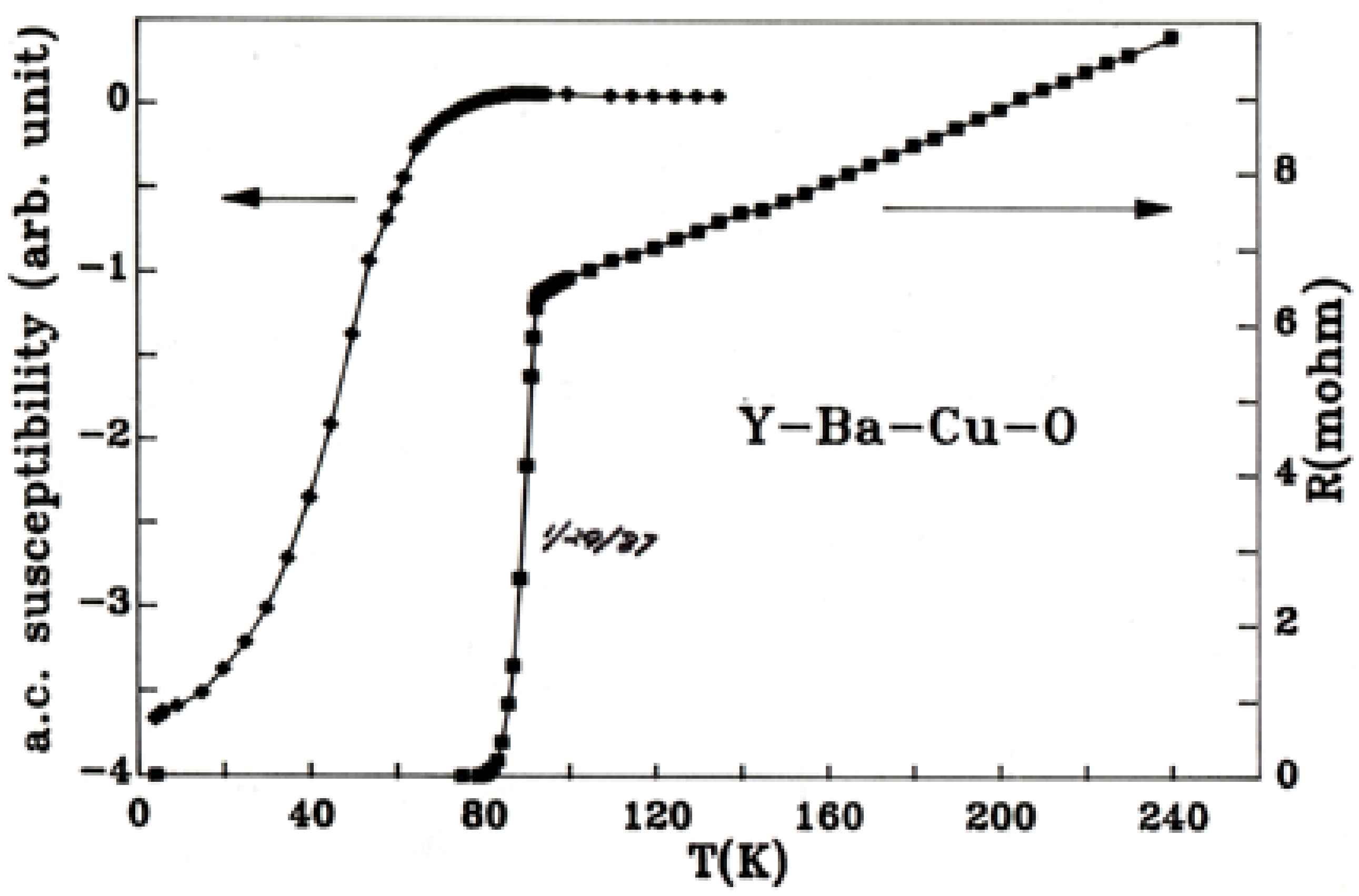

**Figure 9**

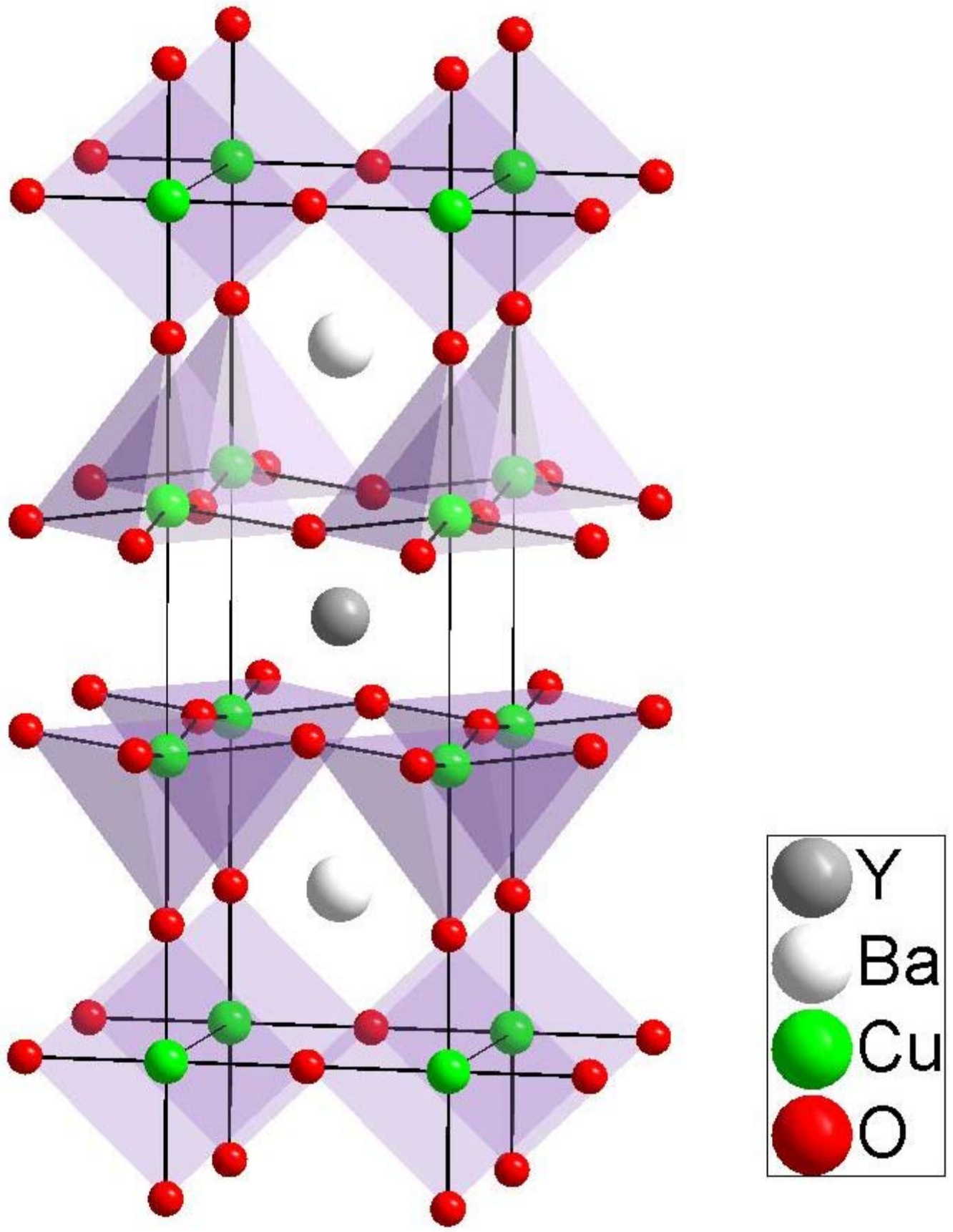

**Figure 10**

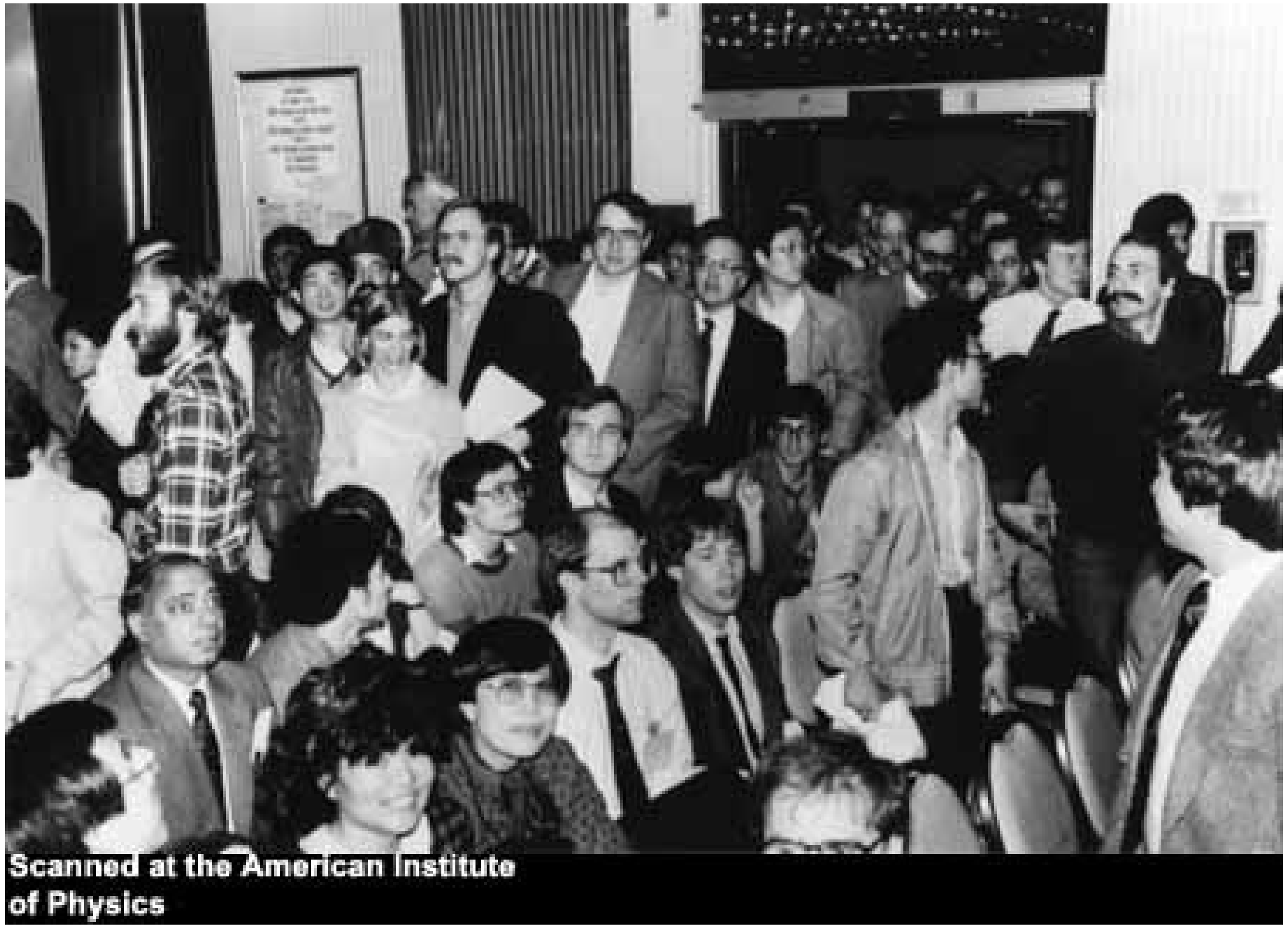

**Figure 11**

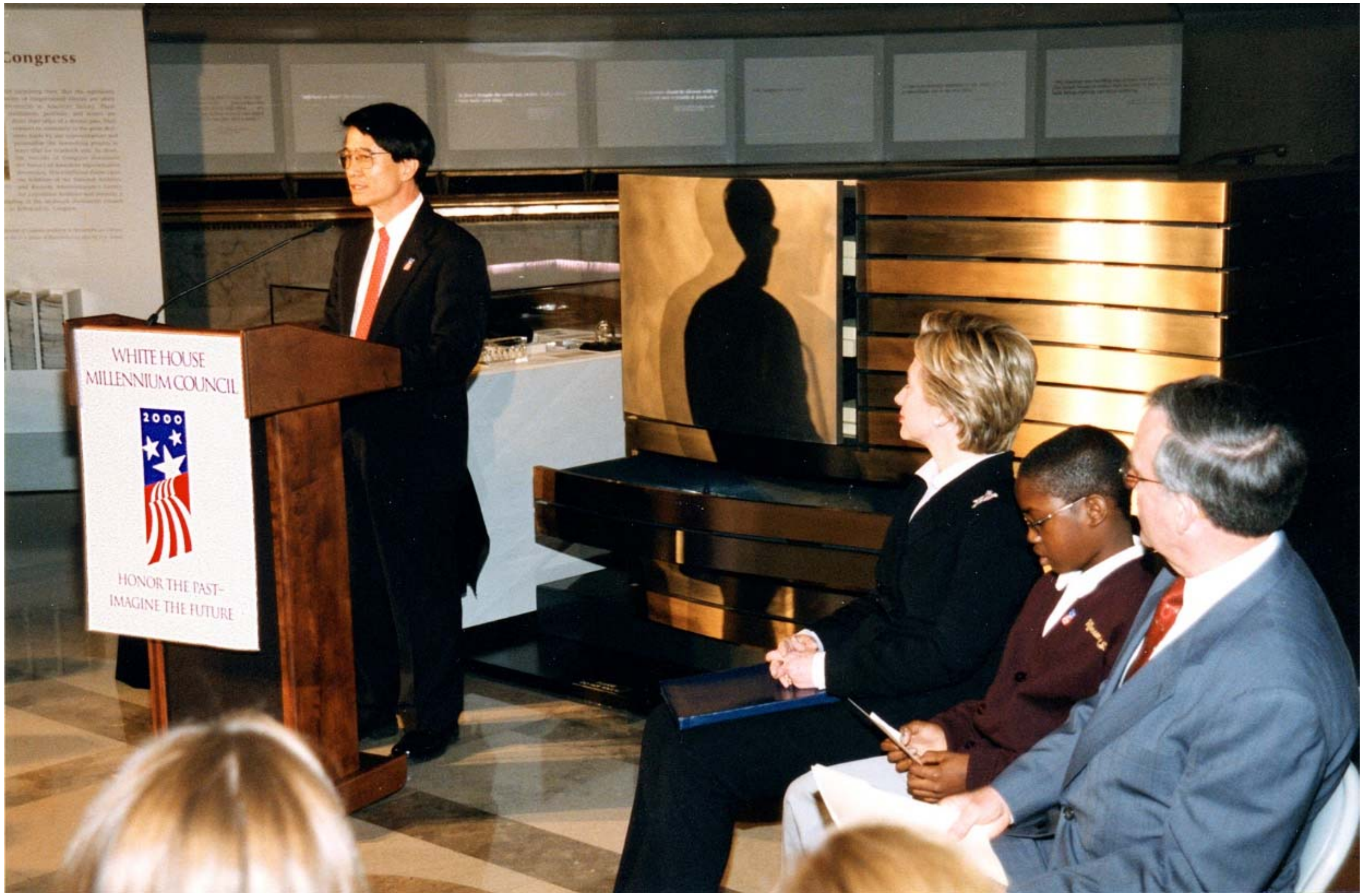

**Figure 12**

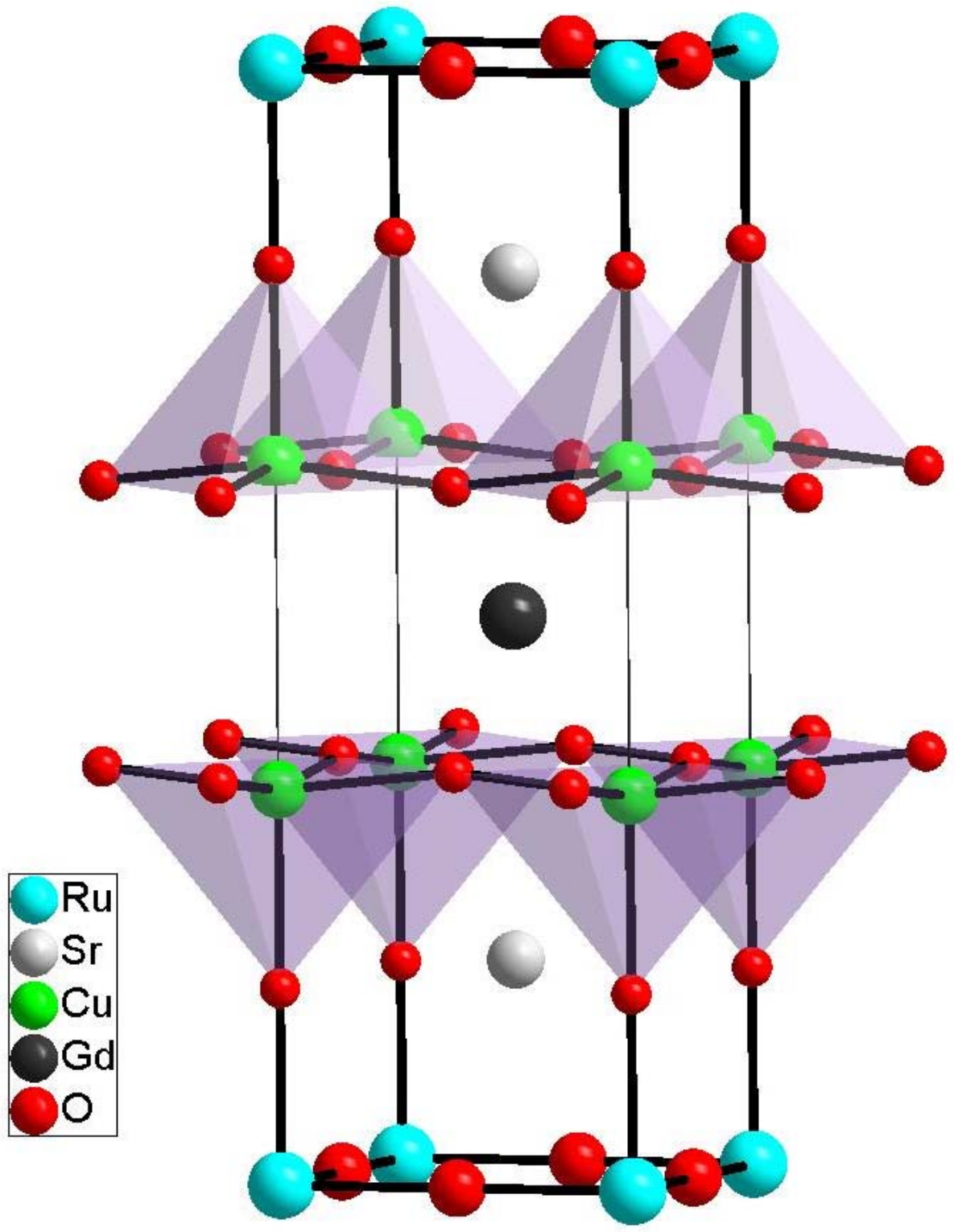

Figure 13

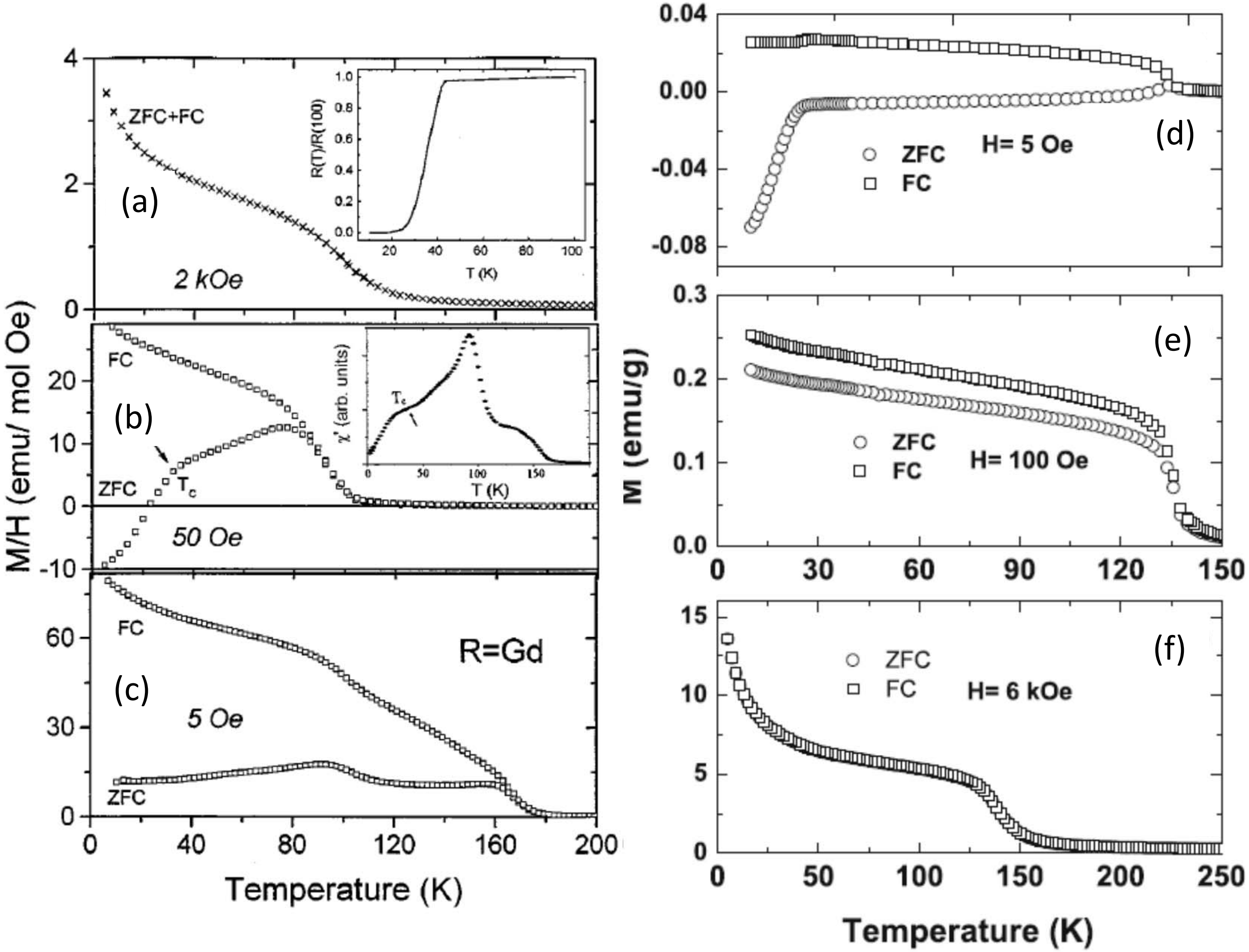

Figure 14

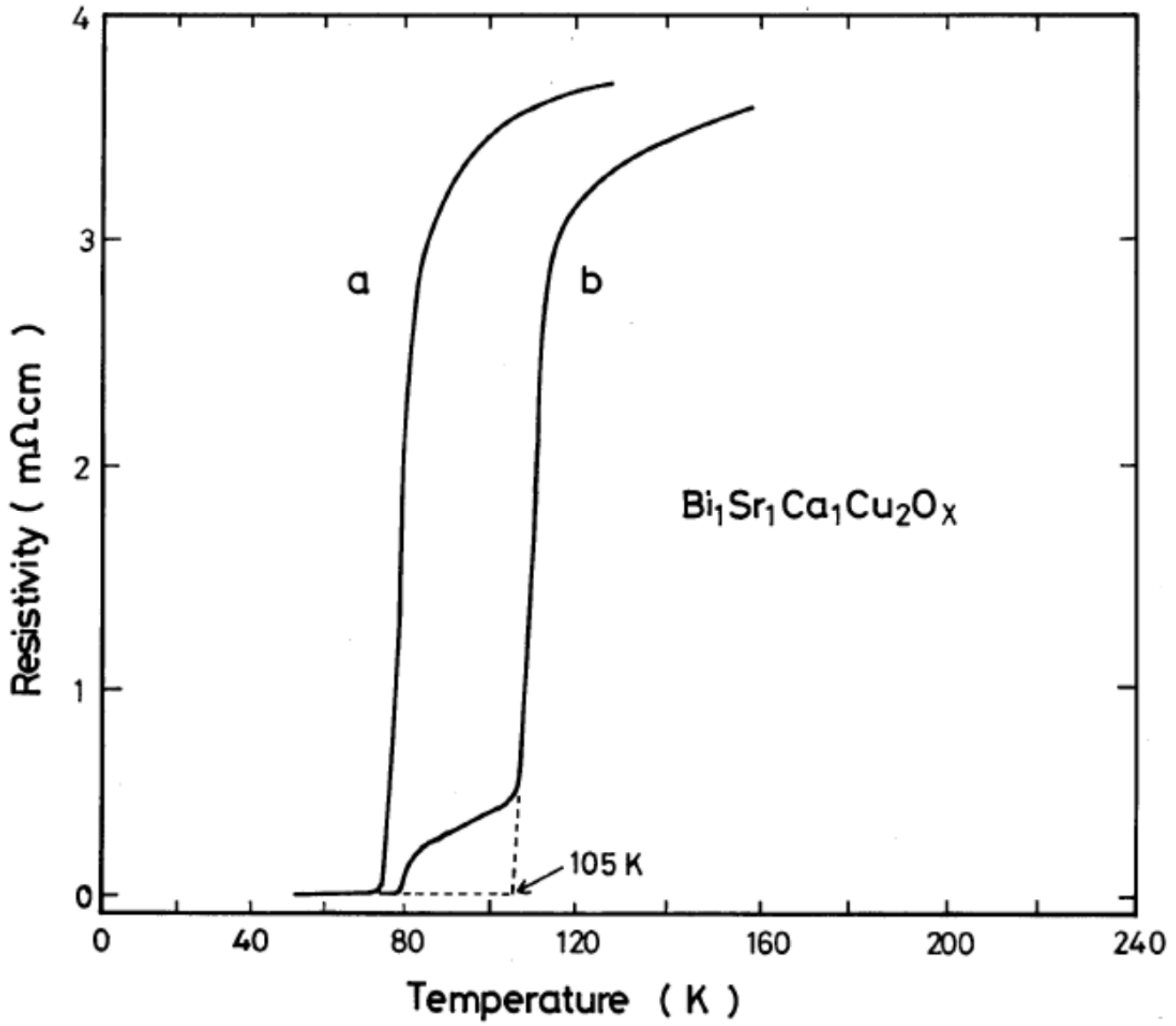

Figure 15

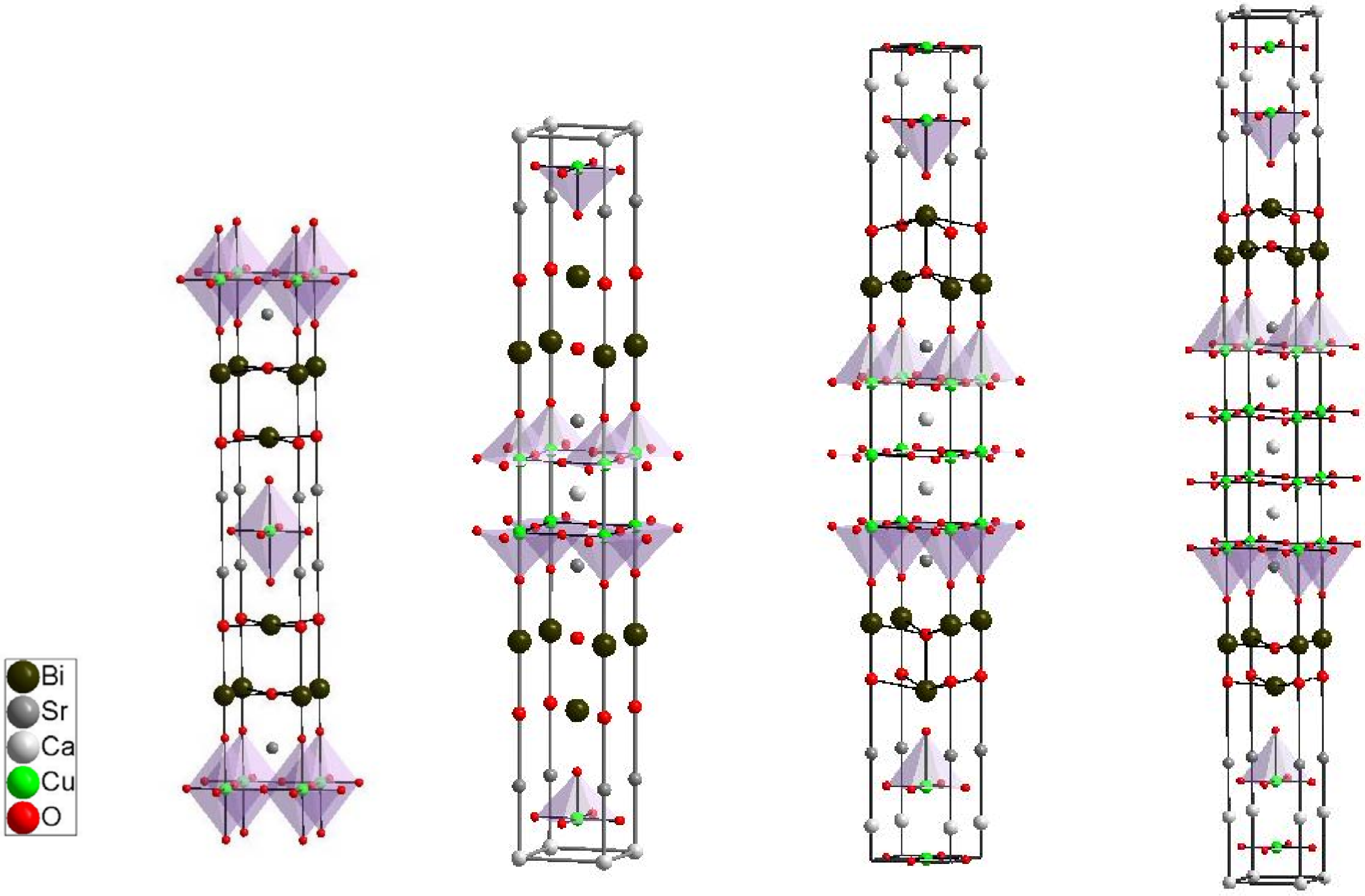

Figure 16

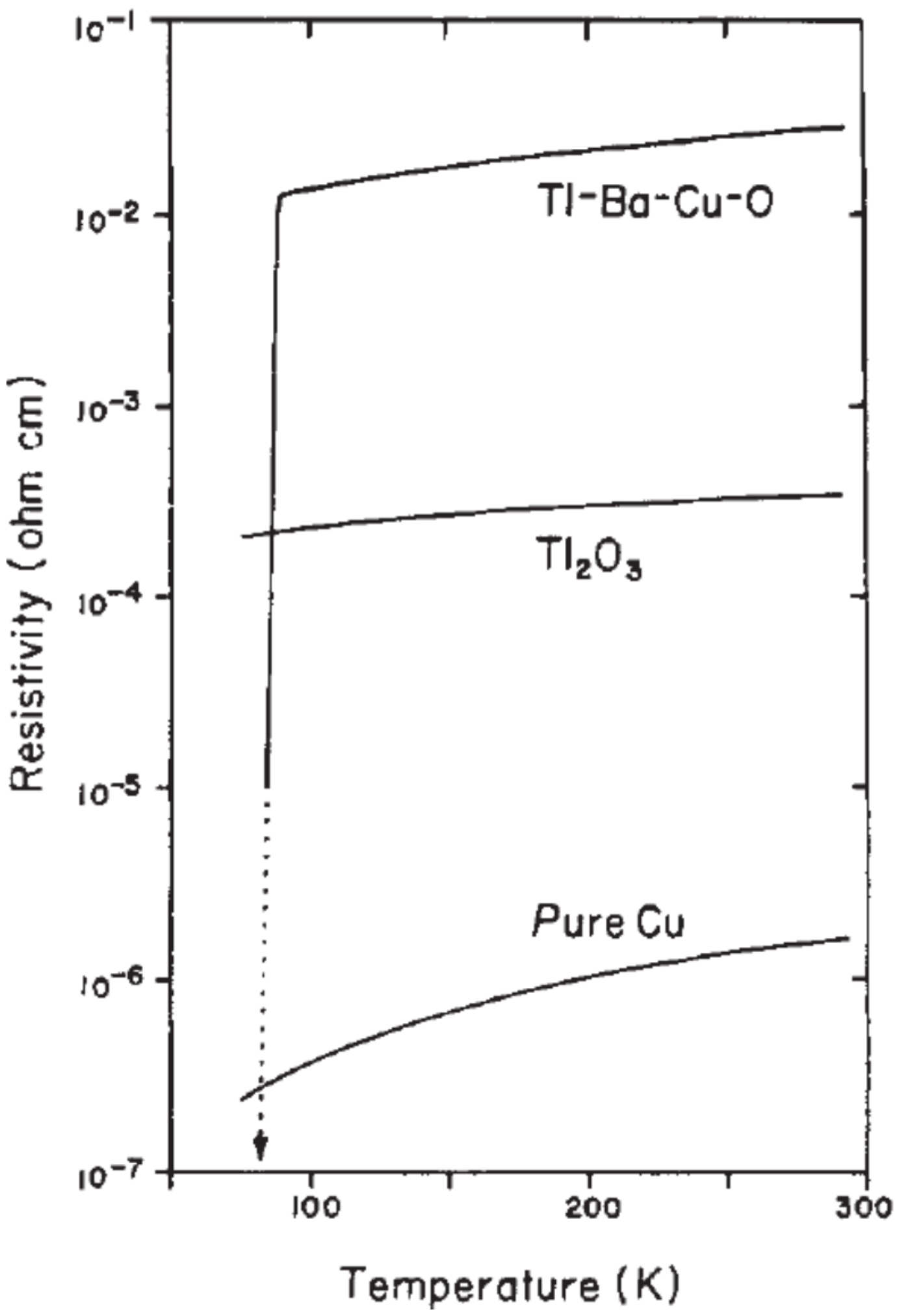

**Figure 17**

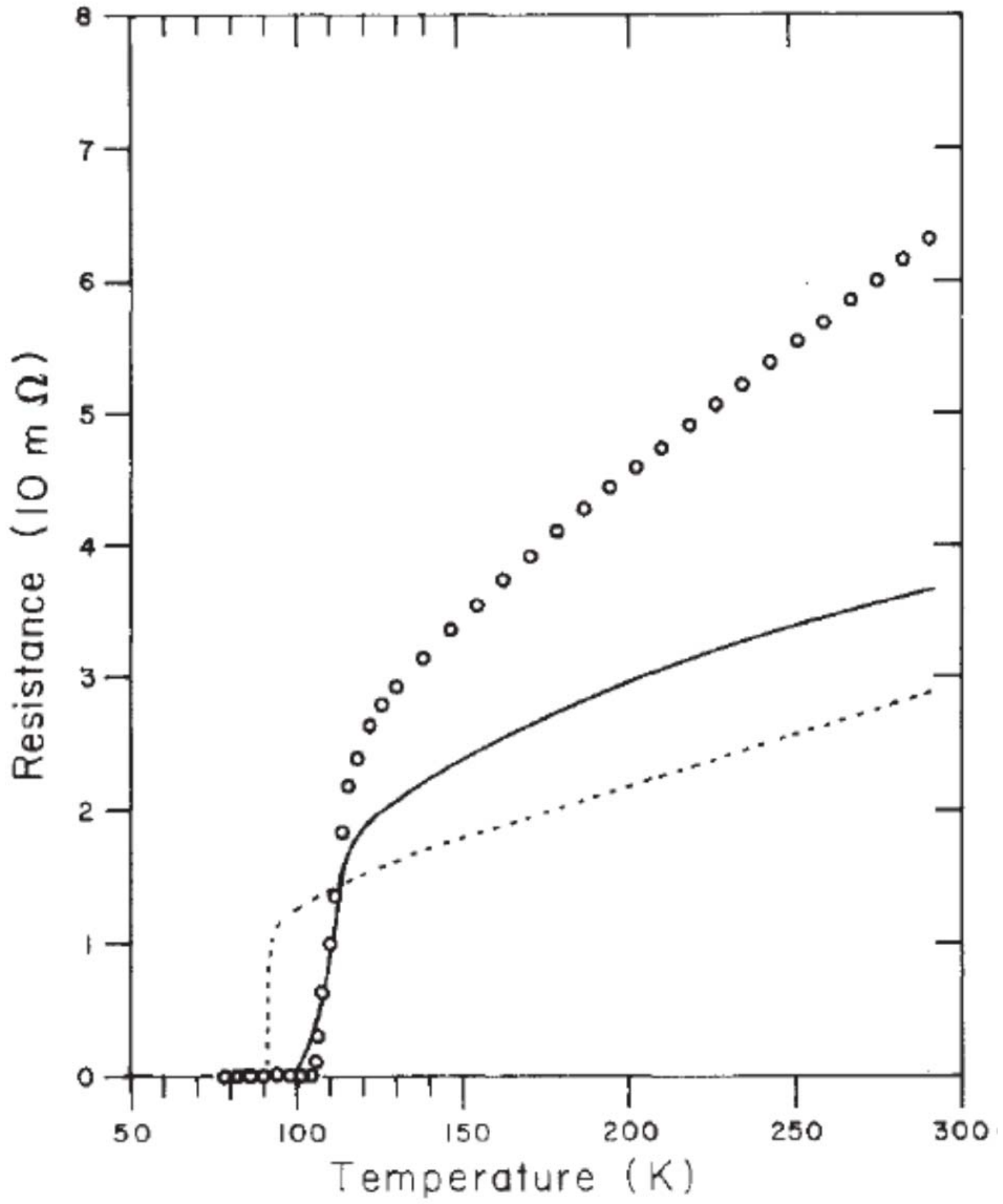

Figure 18

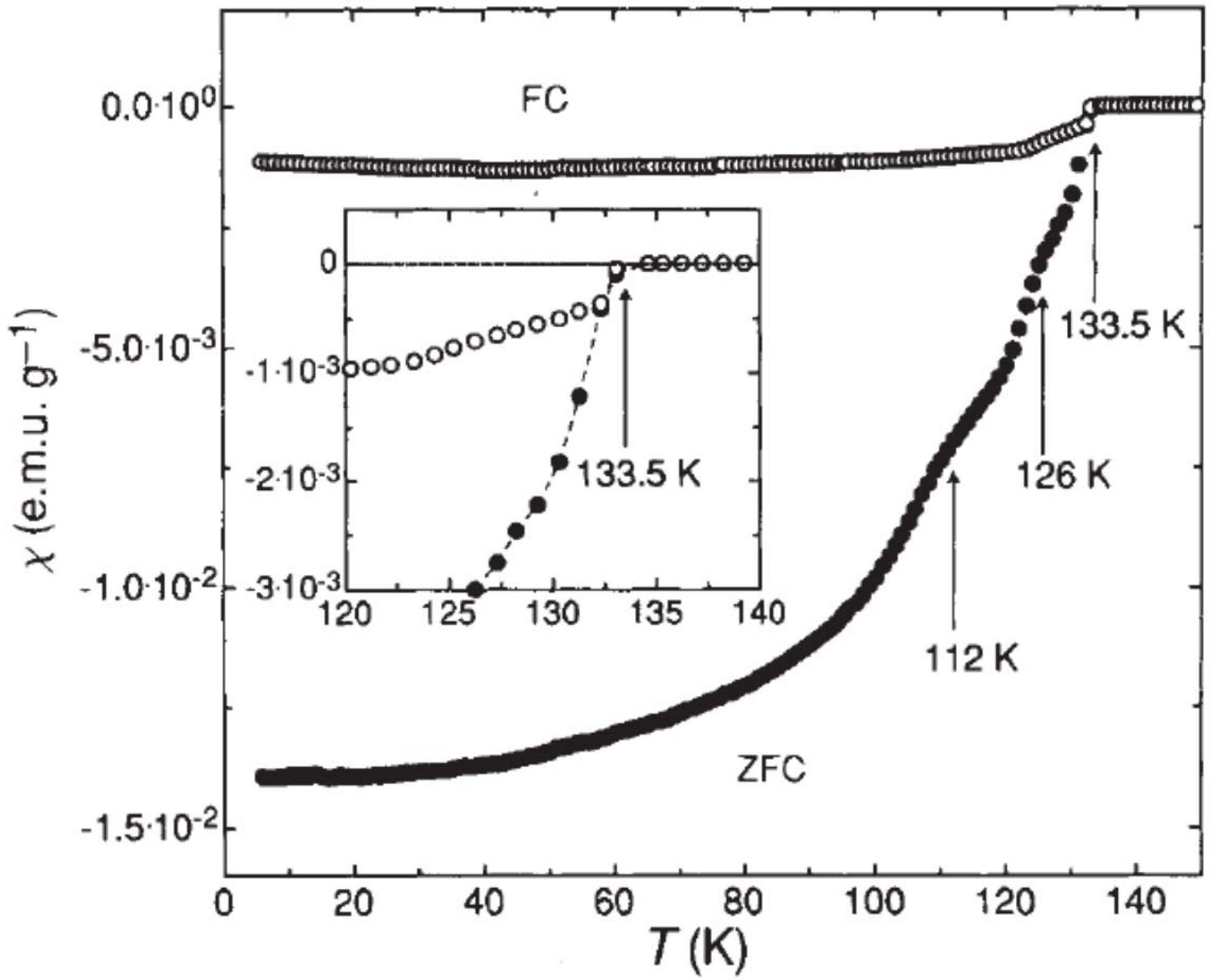

Figure 19

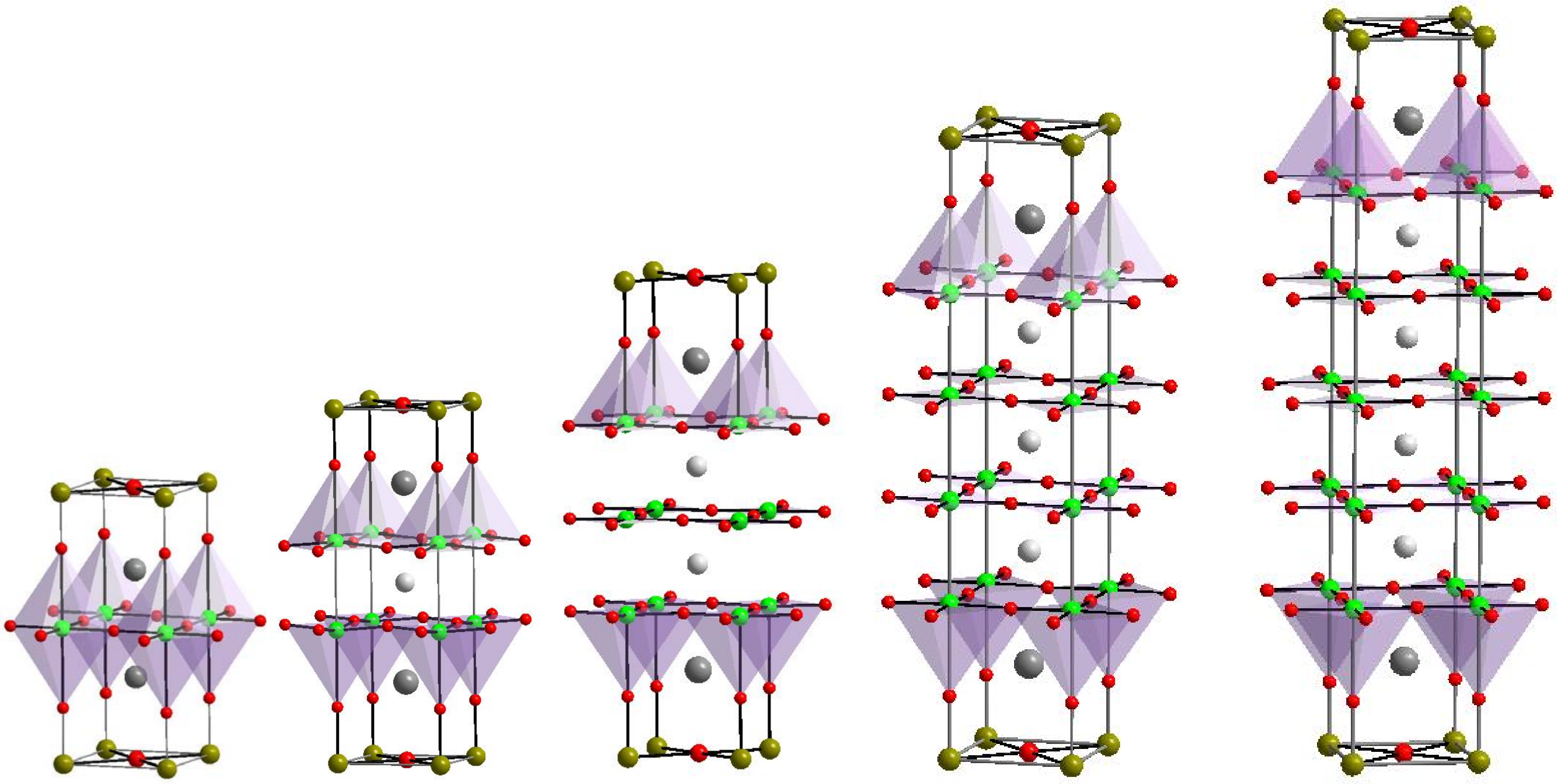

**Figure 20**

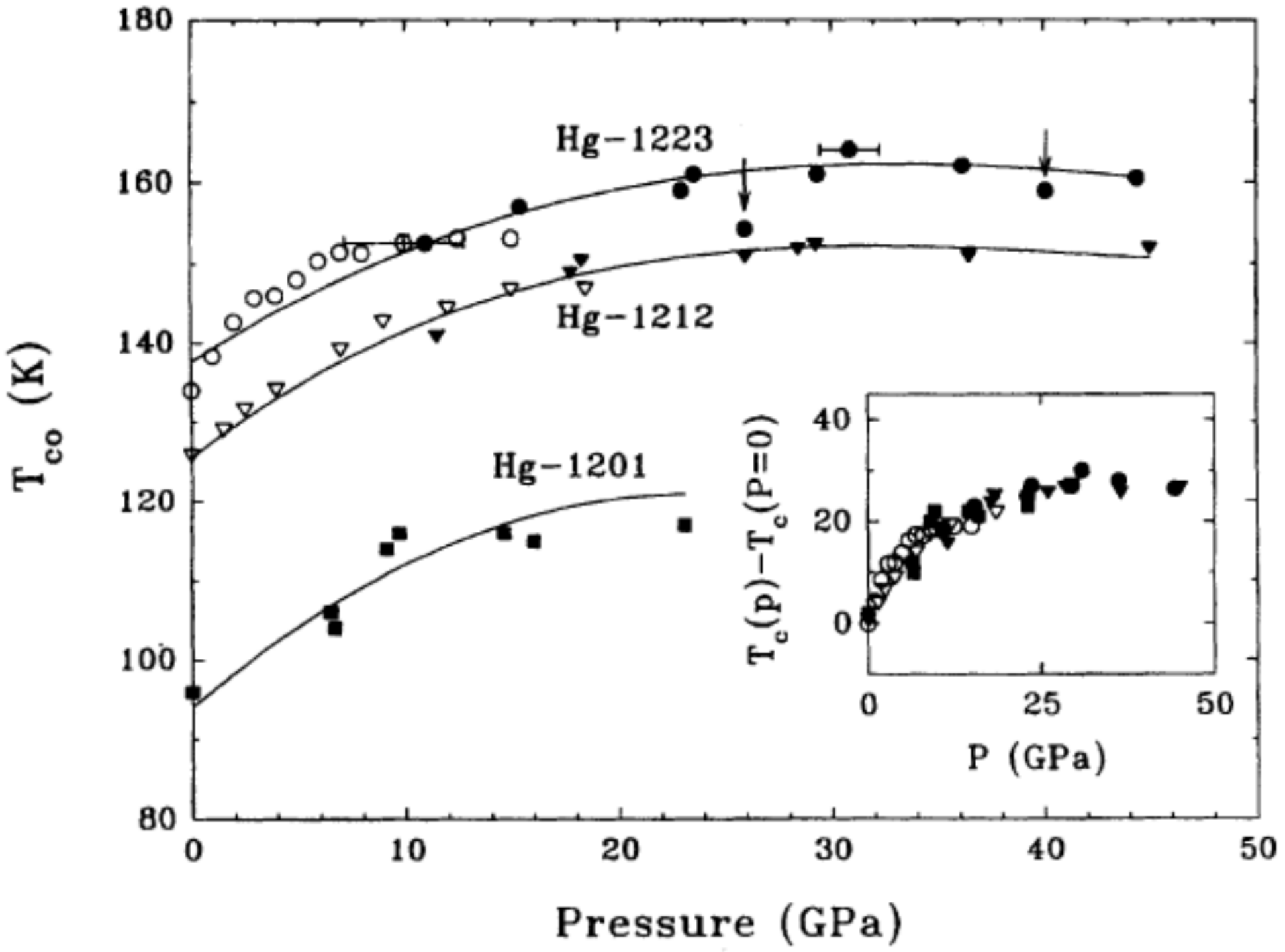

**Figure 21**

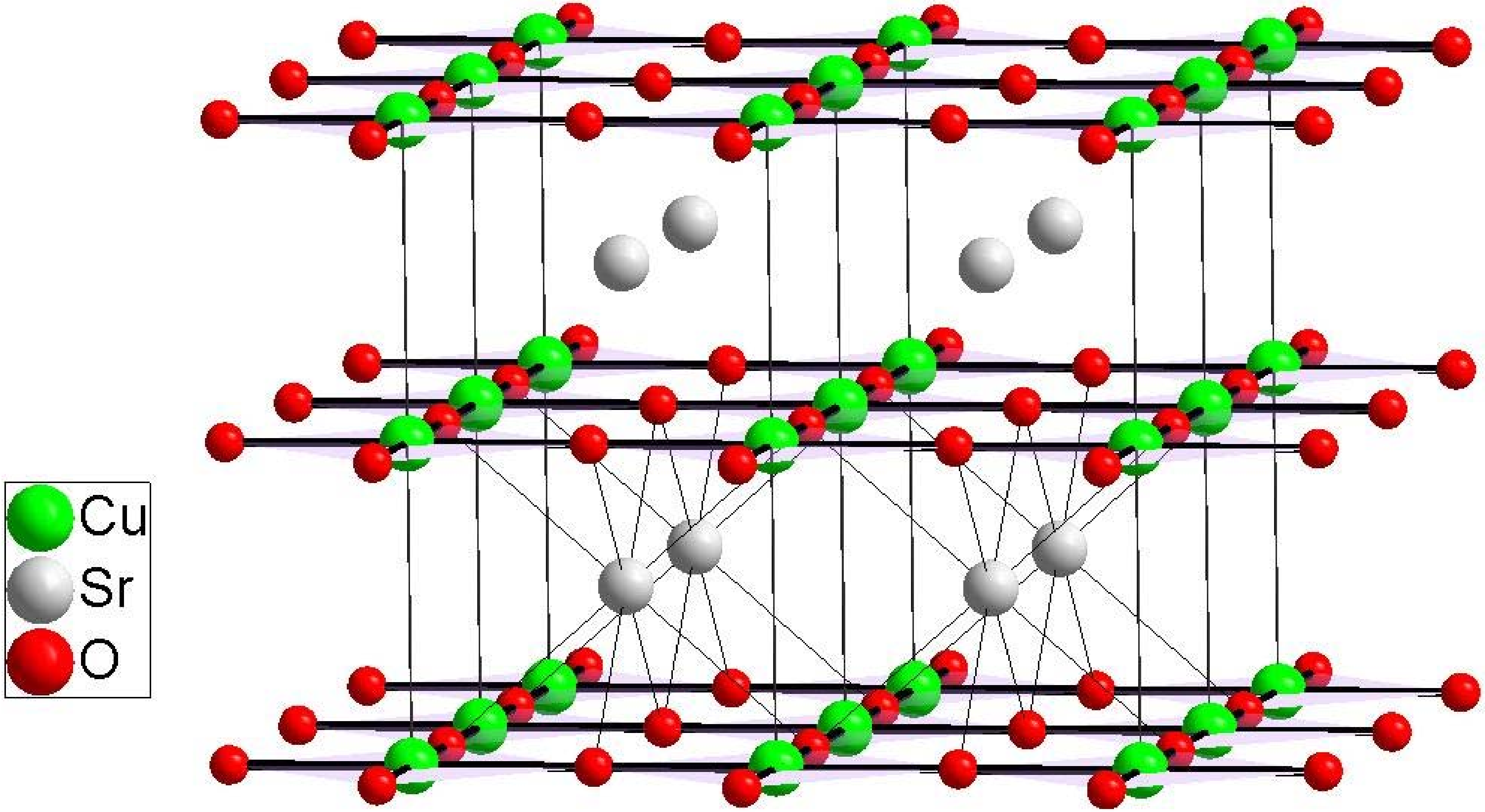

**Figure 22**

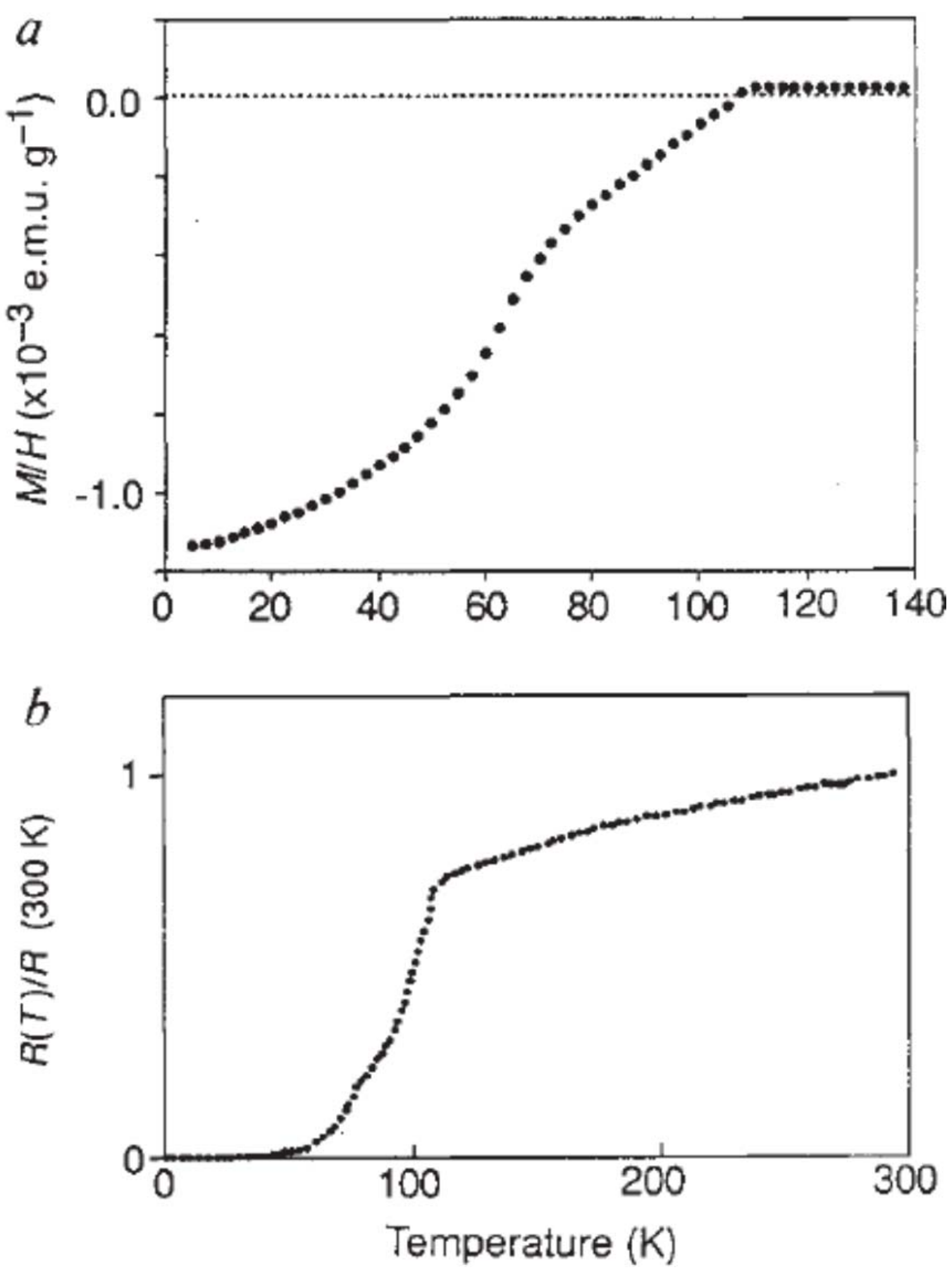

Figure 23

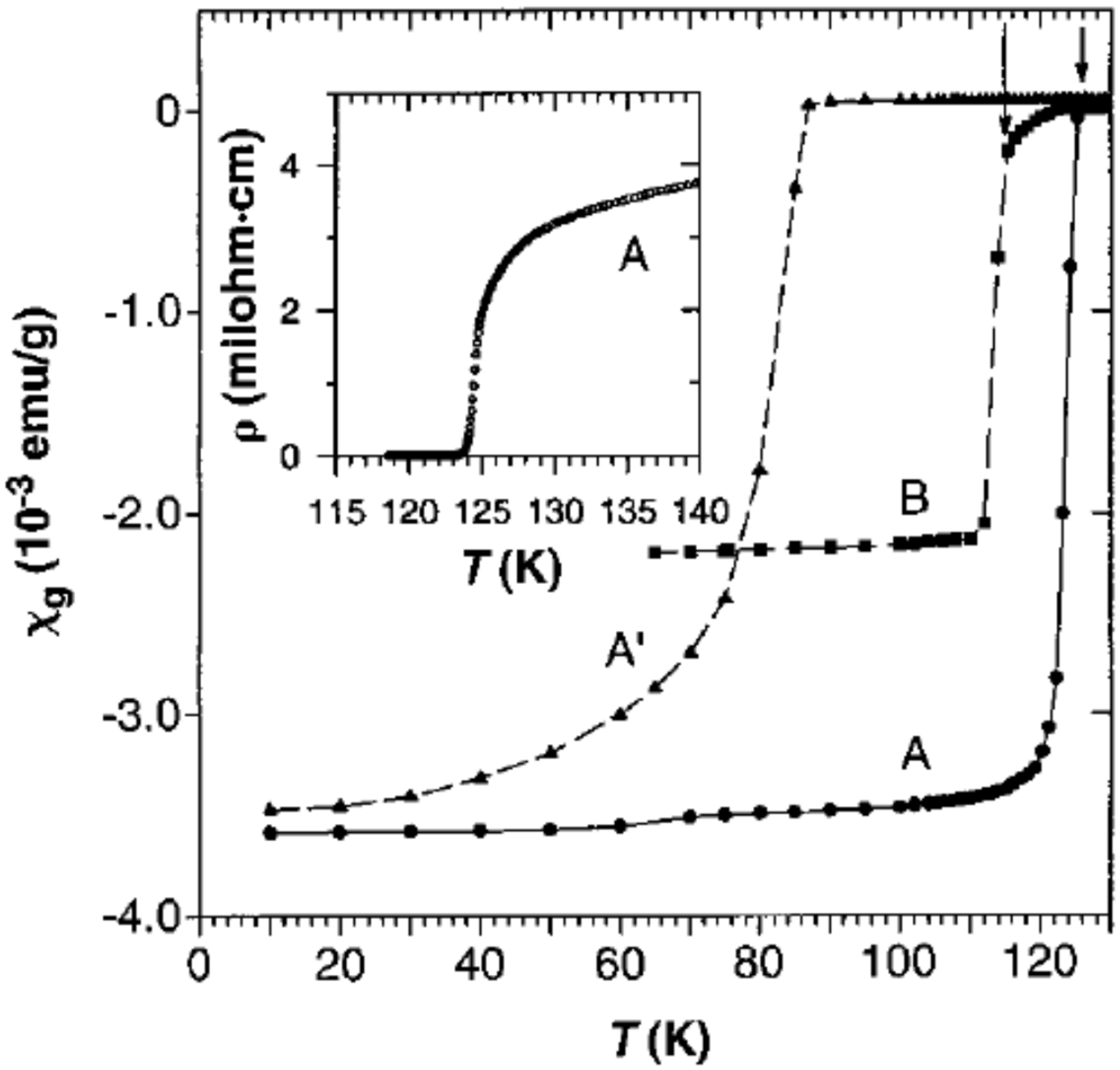

**Figure 24**

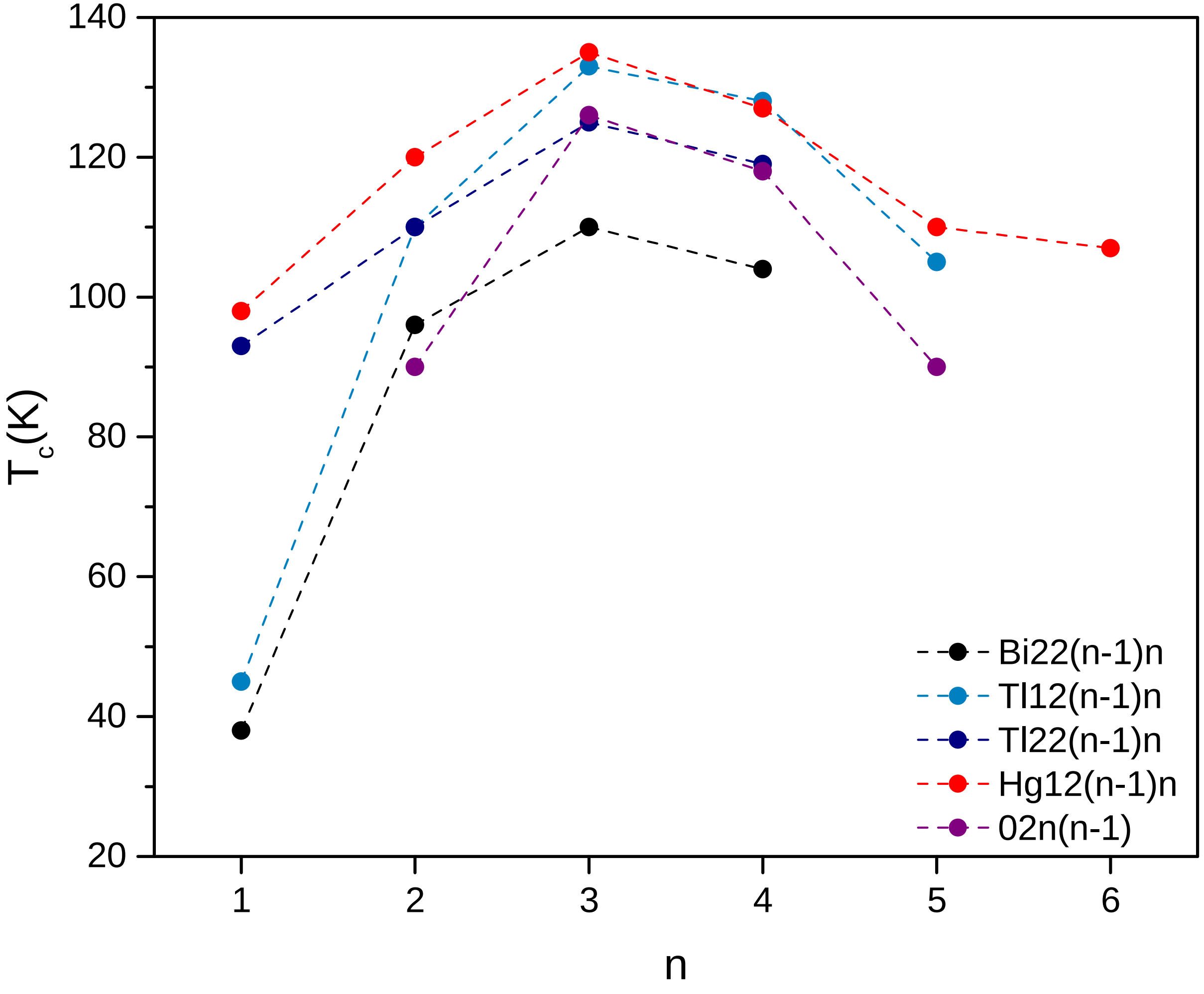

**Figure 25**

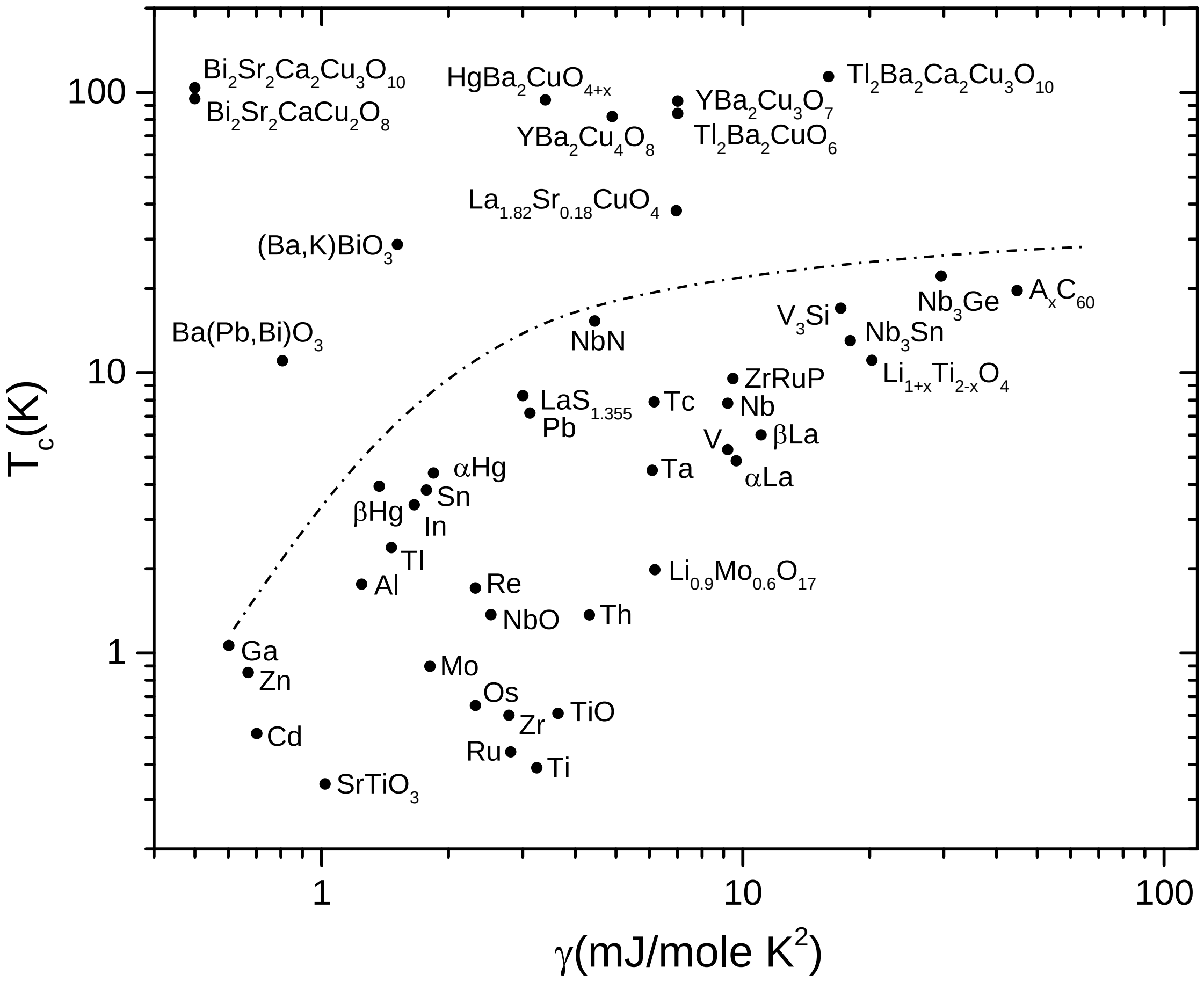

**Figure 26**

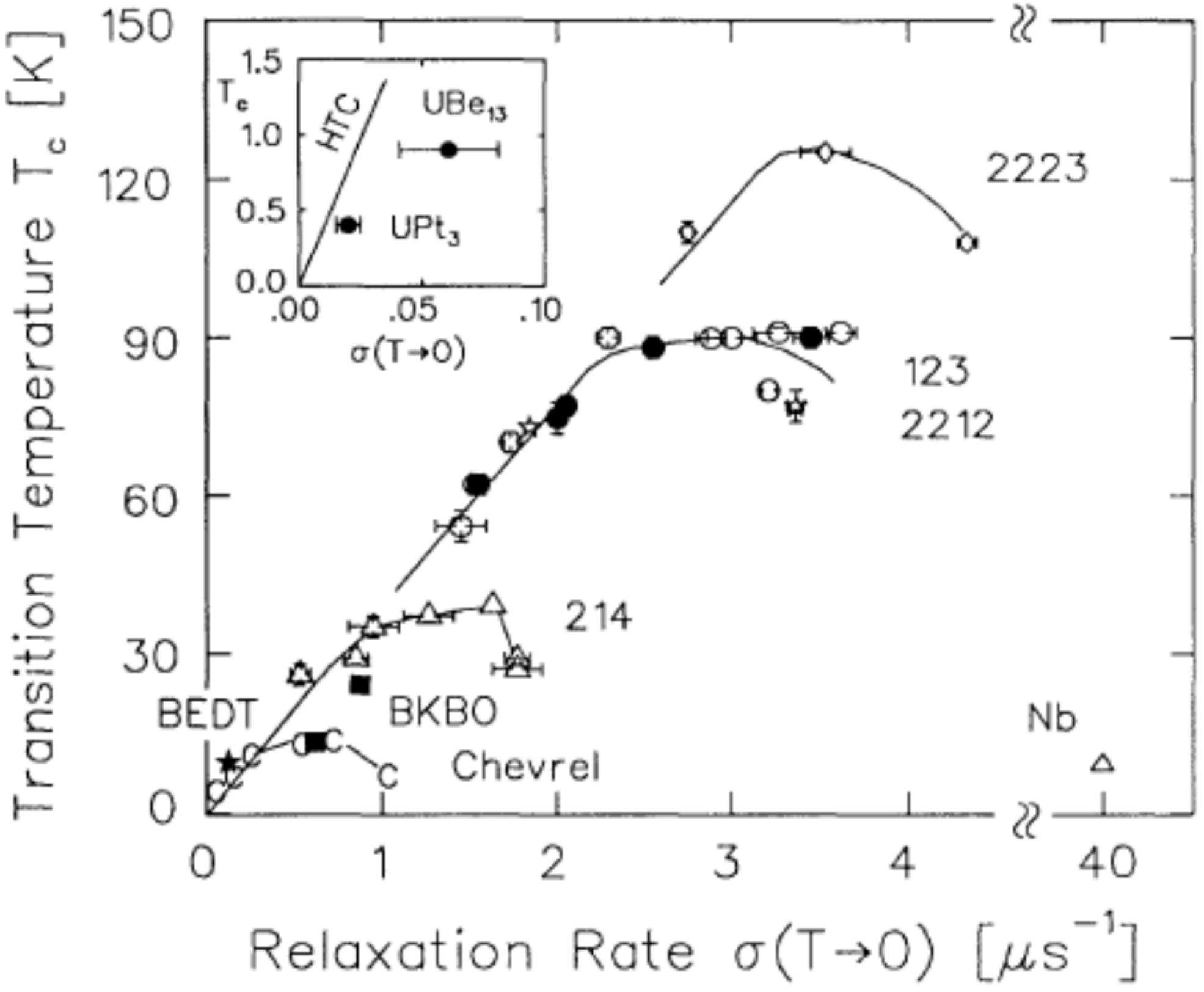

Figure 27

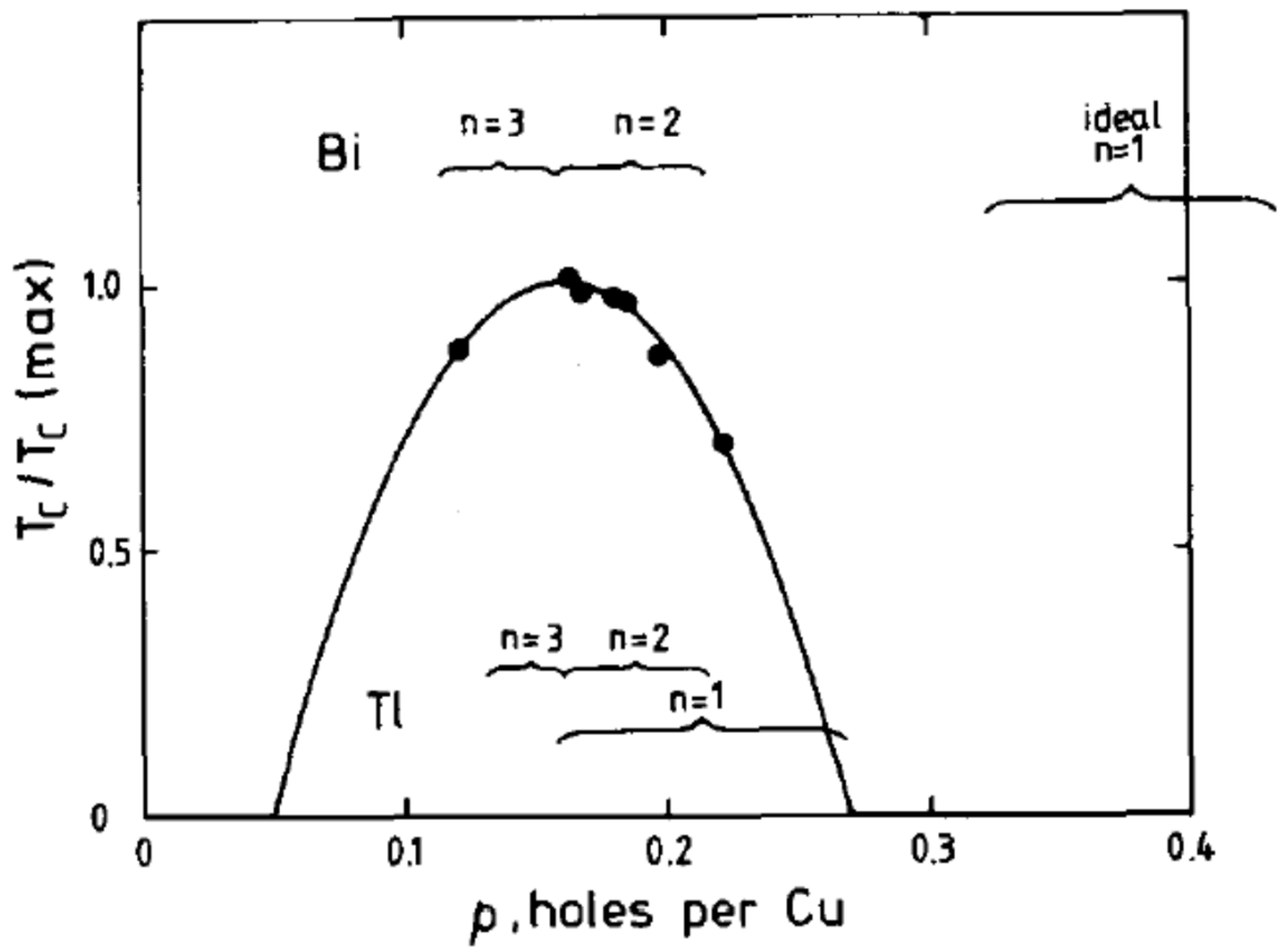

Figure 28

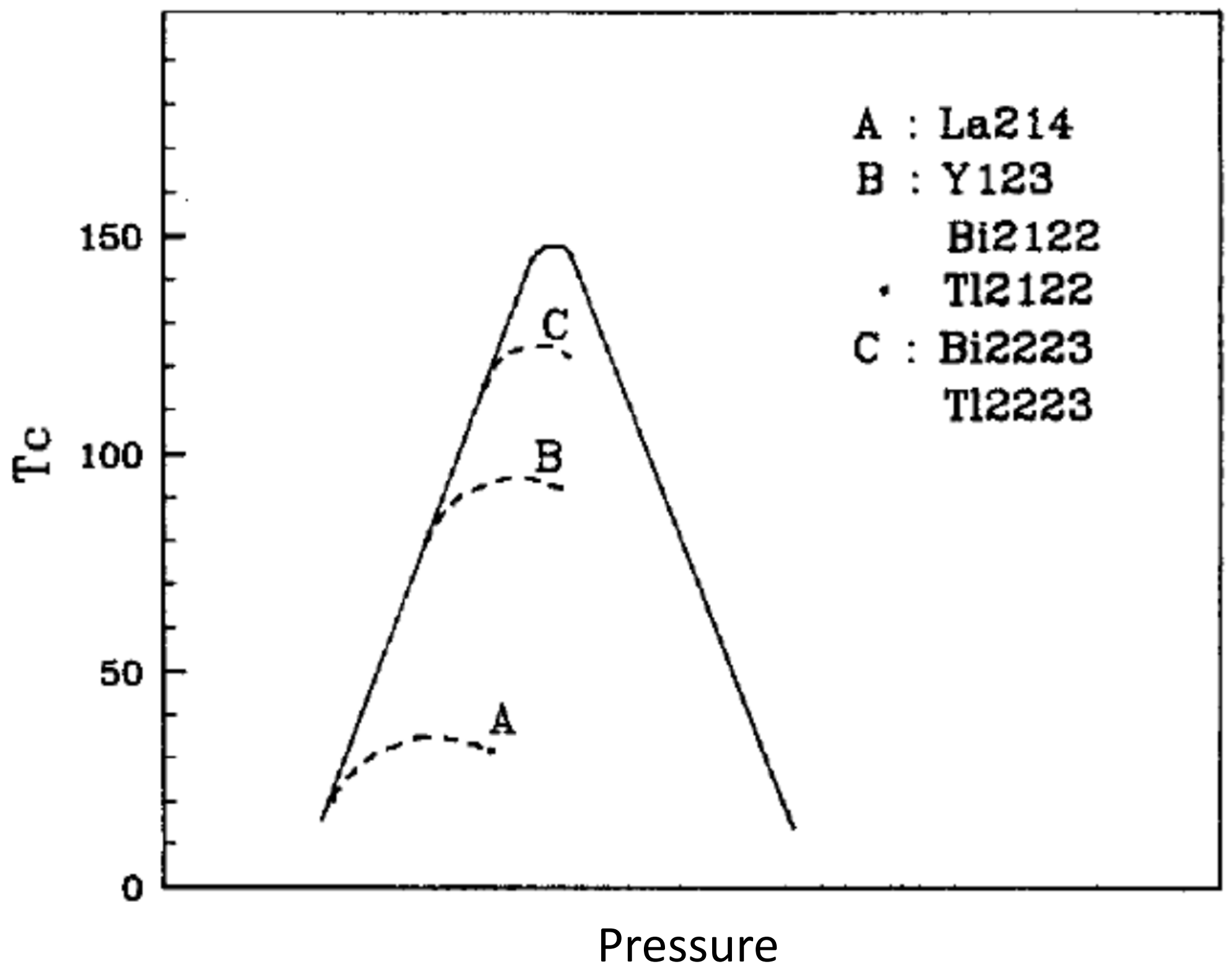

**Figure 29**

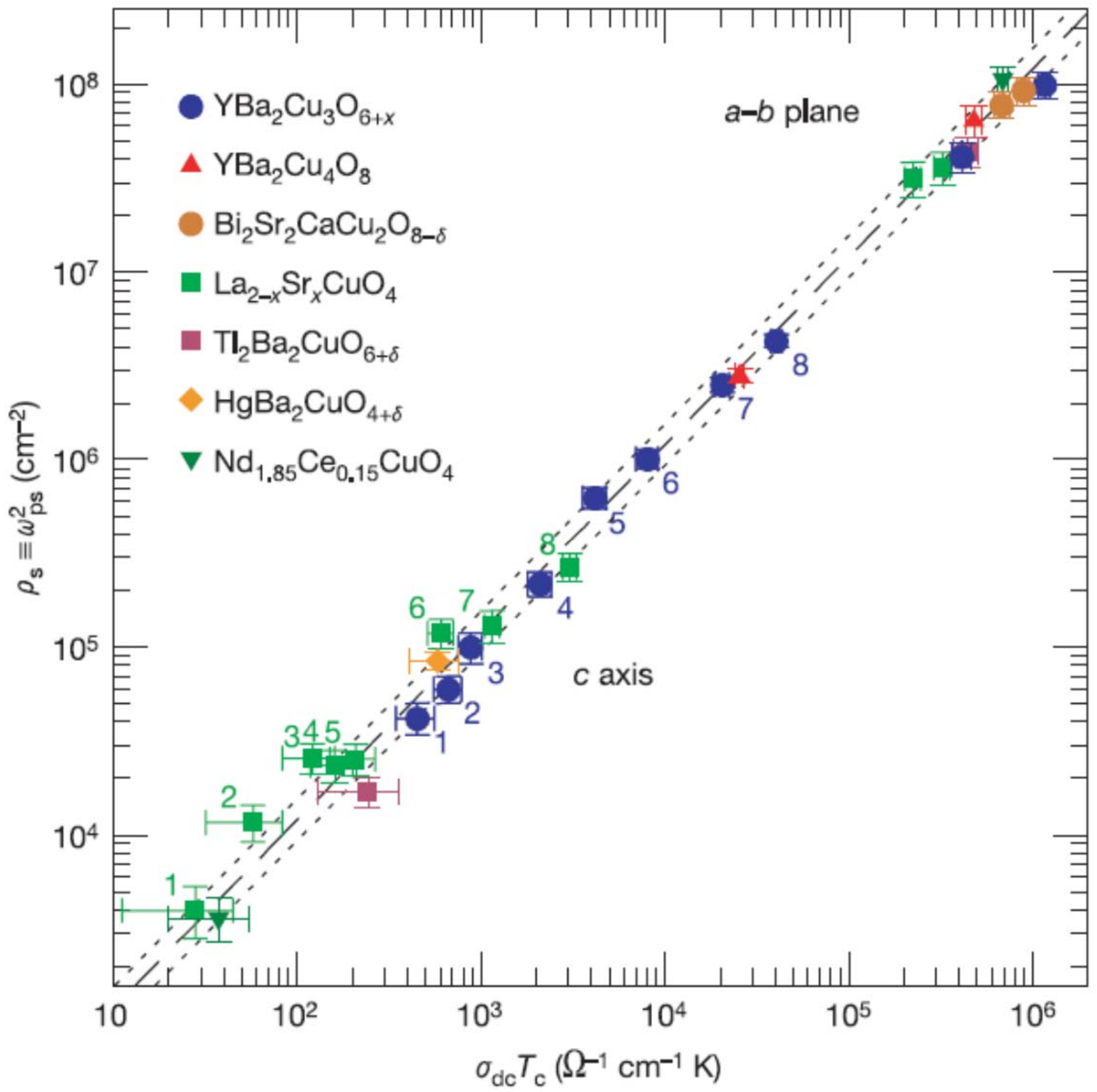

**Figure 30**

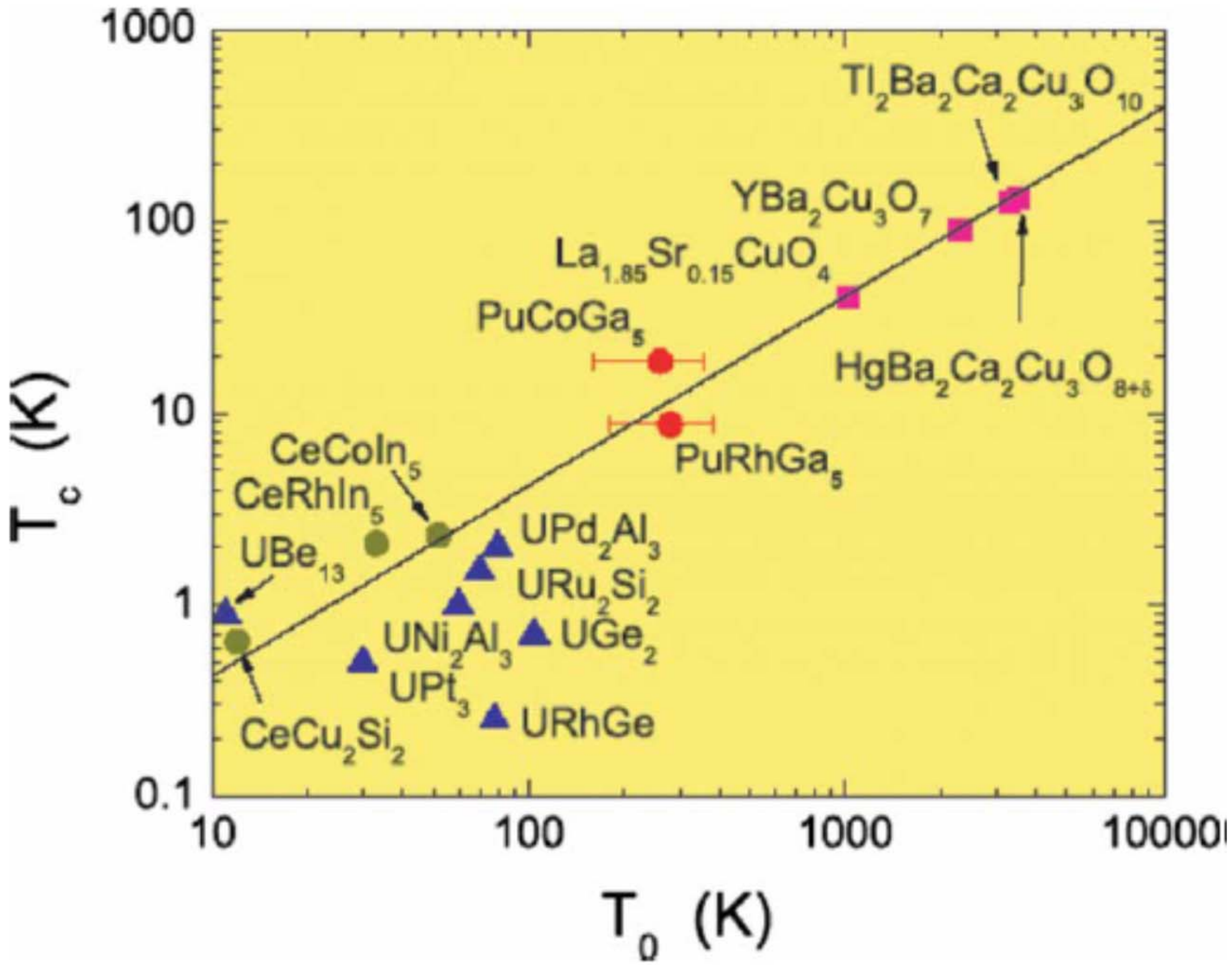